\PassOptionsToPackage{table,dvipsnames,x11names}{xcolor} 
\documentclass{neutralclass}

\usepackage[utf8]{inputenc}
\usepackage[T1]{fontenc}
\usepackage[english]{babel}
\usepackage[autostyle]{csquotes}
\usepackage{lmodern}

\usepackage{subcaption}
\usepackage{wrapfig}
\usepackage{multirow}

\usepackage{amsmath}
\usepackage{amssymb}
\usepackage{amsthm}
\usepackage{dsfont}
\usepackage{bm}

\usepackage{mathtools}
\usepackage{enumerate}
\usepackage{enumitem}

\theoremstyle{definition}
\newtheorem{definition}{Definition}
\newtheorem*{definition*}{Definition}

\theoremstyle{plain}

\newtheorem{algorithm}[definition]{Algorithm}

\newcommand{\R}{\mathbb{R}}
\newcommand{\E}{\mathbb{E}}
\newcommand{\N}{\mathbb{N}}

\newcommand{\1}{\mathds{1}}

\newcommand{\X}{\mathcal{X}}
\newcommand{\x}{\mathbf{x}}

\renewcommand{\hat}{\widehat}
\renewcommand{\tilde}{\widetilde}
\renewcommand{\d}{\ensuremath {\,\mathrm{d}}}

\begin{document}

\title{Modeling inhomogeneous spatial point configurations with applications to replicated patterns in real-life waiting crowds}

\author{Lars Sickert Karam$^{1}$, Rui M. Castro$^{1}$, Maarten Schoukens$^{2}$ and Alessandro Corbetta$^{3}$}

\affil{$^1$Department of Mathematics \& Computer Science and Eindhoven Artificial Intelligence Systems Institute (EAISI), Eindhoven University of Technology, Eindhoven, The Netherlands}

\affil{$^2$Department of Electrical Engineering and EAISI, Eindhoven University of Technology, Eindhoven, The Netherlands}

\affil{$^3$Department of Applied Physics \& Science Education and EAISI, Eindhoven University of Technology, Eindhoven, The Netherlands}

\email{l.sickert.karam@tue.nl}

\begin{abstract}
    In this article, we connect statistical inference for spatial point processes with the analysis of waiting pedestrian crowds through two interconnected contributions. First, on the methodological side we develop an inference procedure for semiparametric spatial point process models leveraging replicated spatial patterns, i.e., multiple approximately independent realizations from the same process. Second, we show that spatial point processes provide a suitable modeling framework for waiting pedestrians, capturing two key aspects: spatial inhomogeneity driven by location attractiveness and repulsive interactions between pedestrians. These two components are central to the inference problem itself, since spatial point process modeling hinges on disentangling background intensity from interaction. Although replicated spatial patterns are rare in point process literature, they are available here through a unique real-life pedestrian dataset, thereby directly linking the methodological development to the physical application. We use the proposed methods to fit and evaluate determinantal and Gibbs point processes in a simulation study and a real-world case study. Despite persistent challenges in decoupling the influences of inhomogeneity from interaction, these models are able to reproduce key empirical features of waiting pedestrians.

    \keywords{active matter, statistical inference, traffic and crowd dynamics, agent-based models}
\end{abstract}

\section{Introduction}\label{sec:intro}

\begin{figure}[ht]
    \centering
    \begin{subfigure}[b]{\textwidth}
        \centering
        \includegraphics[width=\textwidth]{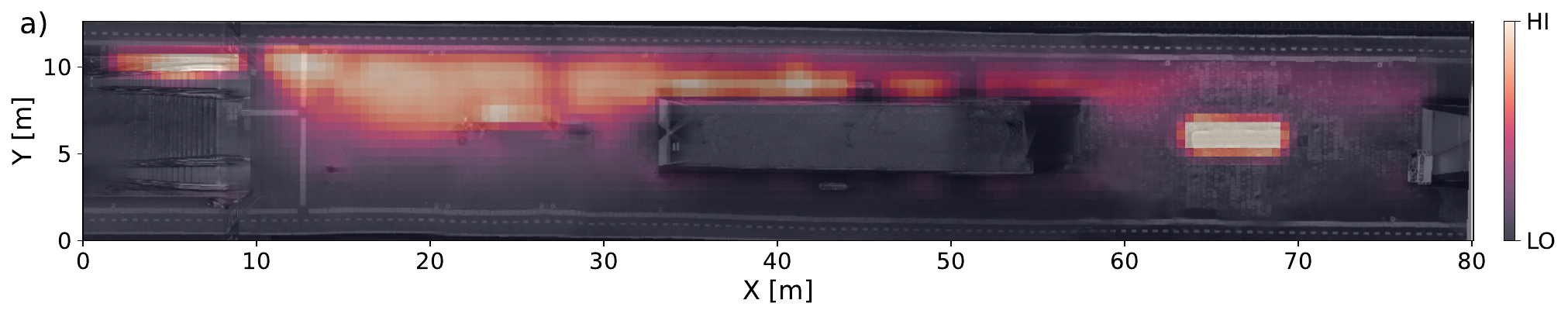}
    \end{subfigure}

    \begin{subfigure}[b]{\textwidth}
        \centering
        \includegraphics[width=\textwidth]{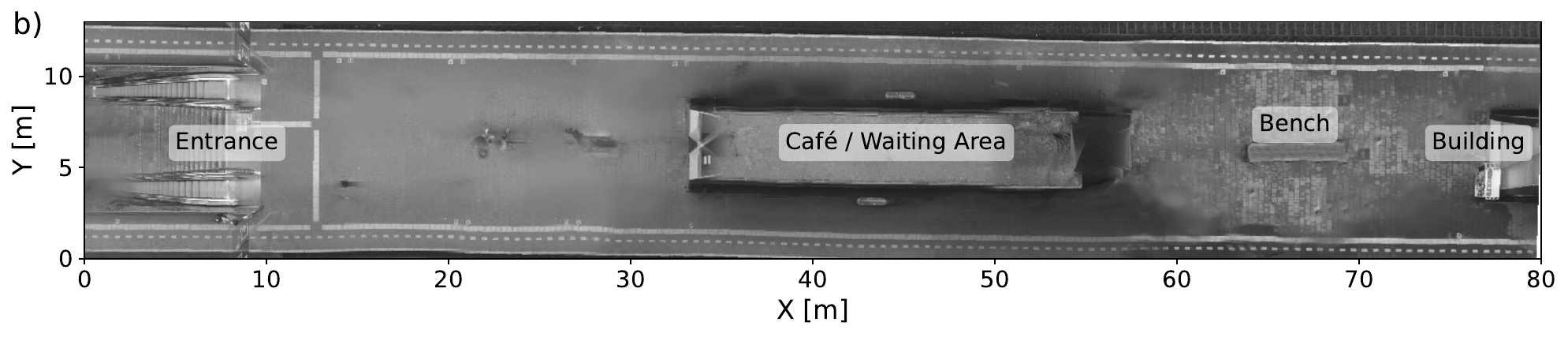}
    \end{subfigure}
    \caption{(a) Overhead view of the estimated one-particle marginal density of pedestrians (see Section~\ref{subsec:data_features}\eqref{eq:one-particle_marginal} for a definition of this quantity, and Section~\ref{subsec:poisson_inference}\eqref{eq:kde_intensity} for the estimate) on the platform, estimated from an aggregate of all snapshots. A clear preference for certain areas is obvious, such as the bench on the right, highlighting the presence of an inhomogeneous background intensity $\beta$ as in \eqref{eq:spp_density}. Thresholding of the color gradient is done at the $99^{\text{th}}$ percentile for visualization purposes. Panel (b) shows an overhead view of the platform (Eindhoven Central Station Platform 2, tracks 3 and 4) with points of interest marked.}
    \label{fig:agg_heatmap}
\end{figure}

\begin{figure}[t]
    \centering
    \begin{subfigure}[b]{\textwidth}
        \centering
        \includegraphics[width=\textwidth]{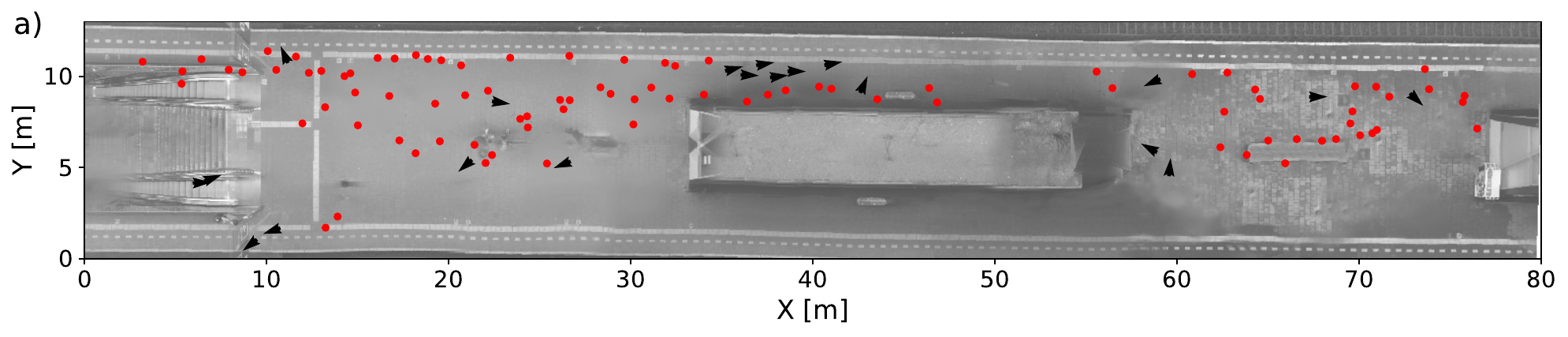}
    \end{subfigure}
    \begin{subfigure}[b]{\textwidth}
        \centering
        \includegraphics[width=\textwidth]{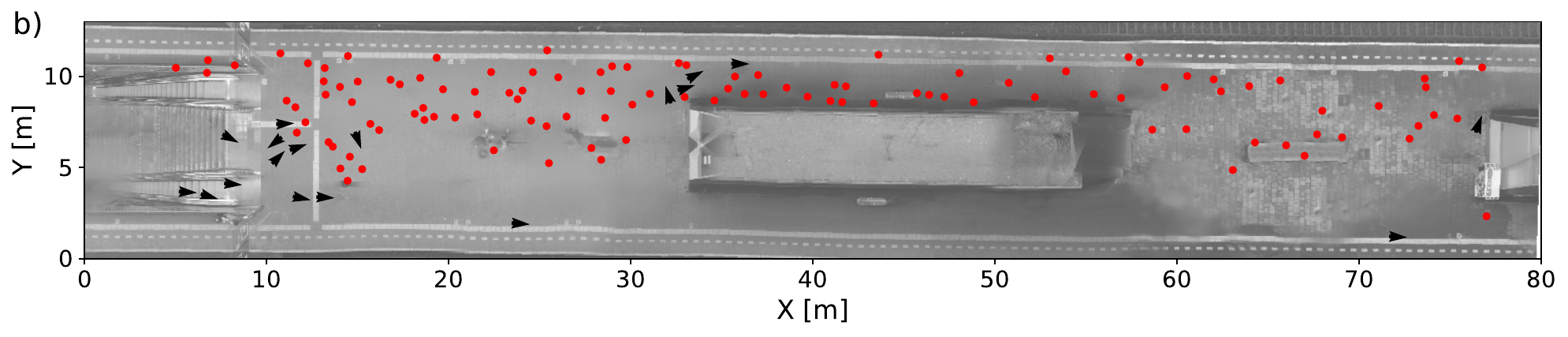}
    \end{subfigure}

    \caption{Example snapshots, with positions of waiting pedestrians in red, passing pedestrians (see definition in Section~\ref{sec:data}) as black arrowheads showing direction of movement. (a) 110 pedestrians, 82\% waiting, (b) 133 pedestrians, 86 \% waiting.}
    \label{fig:samples}
\end{figure}

Imagine arriving at a train station platform to wait for your next connection. Collectively, how have the people on the platform arranged themselves — are they waiting in the shade, on a bench, or near the information board? How close are they standing to other passengers? By abstracting the locations of pedestrians as points in two-dimensional space, the central questions arising from modeling the collective behavior of these individual pedestrians align closely with classical challenges in statistical inference for spatial point processes, stochastic models for spatial point arrangements~\cite{Baddeley_2015,vanLieshout_2000,Moller_2004}. Whether the focus is modeling waiting pedestrian location distributions or inferring the distribution of a configuration of points in space, the primary goal is to estimate both an inherent attractiveness of different locations, which we call the \emph{spatially inhomogeneous intensity}, and the \emph{interaction} between pedestrians (or alternatively, points in space). Simultaneous inference of these two phenomena is challenging, yet decoupling the two is equally difficult~\cite{Baddeley_2015,Kulesza_2012}. In this article we investigate the suitability of several spatial point process models by taking advantage of a wealth of pedestrian data from train stations (see Section~\ref{sec:data}, and Figures~\ref{fig:agg_heatmap} and~\ref{fig:samples}). Notably, this dataset features multiple snapshots of comparable settings, each of which can be seen as an (approximately) independent realization of a spatial point process. Within the spatial point process literature, this is a rather uncommon feature, as most research has so far relied on single-sample inference (see~\cite[Chapter 16.2]{Baddeley_2015} and references therein). Access to these multiple realizations of the same process, referred to as \emph{replicated spatial patterns}, unlocks the use of averaging to control for high stochastic variability and makes nonparametric techniques such as kernel smoothing for intensity estimation more reliable. In this manner, using the inferred distributions, we can construct practically useful models of a generative nature. Beyond the methodological contributions of this approach, accurate quantitative spatial point process models of waiting pedestrian behavior can serve a crucial role in designing transportation infrastructure and understanding agent-based physical systems~\cite{Kupper_2023a}.

Samples from a spatial point process are in general random configurations of a (possibly also random) number of points, which we will refer to as both \emph{configurations} or \emph{snapshots} interchangeably. The latter term aligns more closely with our considered practical application, in which configurations are birds-eye snapshots of pedestrian positions on a train station platform (see Figure~\ref{fig:samples}). Formally, a configuration of $n$ points may be written as
\[
\x = \{x_1,\dots,x_n\}, \qquad x_i \in \X \text{ for all } i \in \{1,\dots,n\},
\]
where $\X$ is a connected subset of the plane (or, more generally, of $d$-dimensional space). For each $n \in \mathbb{N}_0$, denote the space of finite point configurations of size $n$ by
\[
\X_n^* \coloneqq \{\x \subset \X : N(\x)=n\},
\]
where $N(\x)$ denotes the number of points in the configuration $\x$. The space of all finite point configurations in $\X$ is then given by the countable union
\[
\X^* \coloneqq \bigcup_{n \in \N_0} \X_n^*.
\]
We regard a configuration $\x$ as a realization of a spatial point process, that is, as a random variable taking values in $\X^*$. In general, the number of points $N(\x)$ may itself be random; however, in our setting it is useful to consider models conditionally on $N(\x)$, as this allows us to separate inference of the spatial distribution of the points from inference of the distribution of the total count. For practical convenience, we will also use the notation $\x$ for random configurations taking values in $\X_n^*$, in a slight abuse of notation. While conditioning on $N(\x)$ has little to no effect in settings where inference is performed using single samples~\cite{Moller_2022}, in settings of replicated spatial patterns, likelihood-based inference involves summing up contributions from configurations of varying sizes. This necessitates a careful conditional setup to ensure contributions are weighted appropriately.

A spatial point process (conditional or otherwise, as this only affects the normalization) is conveniently described by a probability density function $f$, which we factorize as
\begin{equation}\label{eq:spp_density}
    f(\x) \propto \left(\prod_{x \in \x} \beta(x)\right) \Gamma(\x),
\end{equation}
where $\beta$ embodies the inherent spatial heterogeneity, and $\Gamma$ captures the interaction between points. In the above, $\propto$ denotes proportionality (i.e., the two sides of the expression above are equal up to a constant factor, that depends on $\beta$, $\Gamma$ and the number of points $n$ in the configuration, but not on $\x$ itself). The normalizing constant (also known as partition function) of the density is in general intractable, and its formulation varies slightly depending on whether the model is conditional (see Section~\ref{subsec:gpp}). The function $\beta \colon \X \to [0,\infty)$ assigns an intensity value to each point $x_i$ in the configuration $\x$, taking into account the background likelihoods of different positions in the observation area. Further, $\Gamma$ describes the interactions between points in $\x$, either by penalizing or boosting the likelihood of certain configurations, and is typically assumed to be invariant to both translations and rotations of points (see e.g.~\cite[Chapter 13.12.2]{Baddeley_2015}). While in the remainder of this article we adhere to formulation \eqref{eq:spp_density}, which is most common in mathematical and statistical literature~\cite{Baddeley_2015,vanLieshout_2000,Moller_2004}, we can equivalently define the density in a form more common in physics literature~\cite{Dereudre_2019,Georgii_1979}:
\begin{equation}\label{eq:spp_density_physics}
    f(\x) \propto \exp \left(- U(\x) \right), 
\end{equation} 
where $U$ is a potential of the form 
\[ U(\x) = \sum_{x \in \x} \phi_1(x) + \sum_{\{x, y\} \subset \x} \phi_2(x, y) + \sum_{\{x, y, z\} \subset \x} \phi_3(x, y, z) + \dots \]
Here $\phi_k$ embodies the $k$-th order interaction. Clearly, the first term in the right-hand-side above corresponds to $\beta$ via $\beta(x) = \exp({-\phi_1(x)})$, and the remaining terms correspond to $\Gamma$. In this notation it becomes clear that spatial point processes are energy-based models, where intensity values (here corresponding to the first-order terms $\phi_1$) and interaction values (described by the remaining higher-order terms) can be seen as energetic costs (see e.g.~\cite[Example 15.5(a)]{Daley_2008}). The use of terms of order $> 2$ is less common in practice, as the added complexity complicates such models, and in particular makes statistical inference more difficult, even beyond the intractability of the partition function (a complication already present in order-2 models). Some examples not considered here include the area-interaction model~\cite{Baddeley_1995} (also known as the Widom-Rowlinson model~\cite{Widom_1970}) and the Geyer saturation model~\cite{Geyer_1999}. For concreteness, in this article we focus on Gibbs processes with pairwise interactions (a.k.a. pairwise Markov processes, these are models with only first and second-order interactions)~\cite{vanLieshout_2000,Moller_2004,Daley_2003}, as well as determinantal processes (which are higher-order models)~\cite{Kulesza_2012}. For these two model classes, it is easy to formally separate the intensity and interaction components of the process, even if it remains challenging to decouple them during inference. In practice, $\beta$ (or $\phi_1$) has to embody many environmental factors, limiting the use of very stringent parametric models. On the other hand, for $\Gamma$ (equivalently $\phi_2$, $\phi_3$, etc.) one is generally willing to make simplifying assumptions, such as that it follows a relatively simple parametric structure. In our context, we might be primarily interested in estimating and understanding the interaction between points, rather than the first-order term, which can be seen as an unavoidable nuisance parameter distorting estimation of the higher-order terms. While one component is not inherently more important or interesting than the other, the interaction is less influenced by local effects, and can therefore often be considered to be a more global property. This is true for many physical settings, from condensed matter~\cite{Schwabl_2006} to  collective animal behavior~\cite{Ouellette_2022}. 

Although the mathematical theory of spatial point processes is vast and well-developed, statistical inference in inhomogeneous settings still remains an outstanding challenge. Much of the existing literature on repulsive processes relies on the simplifying, though often unrealistic, assumption of constant spatial intensity, $\beta \equiv \text{const}$, referred to as stationary settings~\cite{Baddeley_2015}. However, in real-world situations spatial inhomogeneity is oftentimes a central feature: environmental influences such as obstacles, terrain, sun exposure, and so on can introduce strong and localized variations in intensity~\cite{Baddeley_2015,Ba_2023}. In the context of waiting pedestrians, this corresponds to location-dependent preferences for, e.g., benches, shade, or simply easy access from the entrance. For such non-stationary settings, it is crucial to disentangle the effects of the background intensity from the effects of interpoint interactions. The majority of approaches for estimation of inhomogeneous models depend on a strict parametrization of the intensity, such as considering the intensity as a linear combination of a set of spatial covariates~\cite{Ba_2023} or as a parametric transformation of a homogeneous process~\cite{Nielsen_2004}. These methods, while tractable, place strong structural assumptions on the process, limiting the flexibility necessary to model many real-world scenarios. In contrast, inference under minimal assumptions, such as nonparametric estimation of the intensity in the presence of interaction, remains an open and largely underdeveloped challenge, and motivates the development of more flexible inference approaches tailored to settings of high spatial variation.

The waiting pedestrian setting offers a notable and unique advantage over many domains in which inference for spatial point processes has been considered. As mentioned above, in most cases estimation of the different components of these processes relies on single-sample data. In forest ecology settings, for instance, this is quite natural, as the configuration of trees in a forest~\cite{Baddeley_2015,Ba_2023,Guan_2008} or insect nest sites~\cite{Lavancier_2015} can only be observed a single time in each spatial region. This makes statistical inference quite challenging, and in particular makes it very difficult to give statistical guarantees without imposing strong and unrealistic assumptions. By comparison, waiting pedestrians can be observed multiple times per day on the same platform, with similar behavior repeating indefinitely throughout the year. Under some considerations these can be meaningfully regarded as multiple independent samples from the same process. In particular, this is the case for this paper, where we construct and use large datasets of replicated patterns extracted from measurements recorded at Eindhoven Central Station~\cite{Pouw_2024a}. This dataset comes within a growing effort towards a  quantitative physical understanding of pedestrian behavior~\cite{Corbetta_2023}, via large-scale real-life data (e.g,~\cite{Pouw_2024a,Gu_2025,Minartz_2025,Gregorj_2023,Murakami_2021}). 

From a pedestrian dynamics perspective spatial point processes offer a natural framework for modeling waiting behavior, as the two main drivers of waiting crowd organization are environmental preference and interaction. These correspond directly to the two core model components in spatial point processes. The physical interpretability of these models makes them appealing building blocks for digital twins. Beyond the train station setting studied in this article, we expect the framework to extend to other scenarios in which waiting behavior arises, provided some knowledge of the local spatial inhomogeneity is obtainable, with the interaction mechanism being likely more universal. More broadly, this work establishes a connection between pedestrian dynamics and spatial point processes, opening methodological avenues for studying waiting crowds that have remained largely unexplored.

This article focuses on the challenge of likelihood-based statistical inference for spatial point processes in inhomogeneous settings with replicated spatial patterns, motivated by configurations of waiting pedestrians. The availability of replicated spatial patterns is used to construct a semiparametric inference procedure not readily available in the usual single-pattern inference setting. More specifically, we investigate how repeated observations can be used to stabilize nonparametric intensity estimation and support the fitting of parametric repulsive interaction models in the presence of spatial heterogeneity. On this basis, we study whether determinantal and Gibbs point process models can capture the empirical structure of waiting crowds, both in controlled simulations and in a real-world train station dataset. Our main finding is that replicated patterns indeed facilitate semiparametric estimation, and that this can lead to models that reproduce key physical features of pedestrian configurations. At the same time, our analysis confirms that separating spatial inhomogeneity from interaction continues to be a crucial challenge in statistical inference, and one that merits further investigation.

\paragraph{Outline:} We begin by reviewing previous literature on the modeling of waiting pedestrians in Section~\ref{subsec:lit}. In Section~\ref{sec:data}, we present the dataset used in our analysis, highlighting key quantitative features that effective models should capture. As a benchmark, we introduce inhomogeneous Poisson-type models without interaction formally in Section~\ref{sec:poisson}, discuss the limitations of disregarding interactions, and highlight these limitations by applying these models to the data. Subsequently, we introduce two main classes of spatial point processes with repulsive interactions: determinantal point processes (DPPs) in Section~\ref{subsec:dpp}, and Gibbs point processes (GPPs) in Section~\ref{subsec:gpp}\footnote{
    While strictly speaking DPPs can be considered to be a special case of GPPs~\cite{Georgii_2005}, we follow the standard practice of treating them separately. It is worth noting that these are essentially the only two repulsive classes of point processes that are widely used, as Cox processes, the third main class of spatial point process models, are more suited to modeling clustering behavior, i.e. attractive interactions. We do not further discuss these models here, for more details see original introduction of these processes in~\cite{Cox_1955}, or see~\cite[Chapter 12]{Baddeley_2015} for an overview.
}. We briefly review the literature on each of these processes and introduce them formally, after which we discuss strategies used for statistical inference in semiparametric settings. In Section~\ref{subsec:cv}, we propose a novel estimation method for such semiparametric models, after which we assess inference procedures for these models using synthetic data in Section~\ref{subsec:numerics}, investigating likelihood-based methods for DPPs and pseudolikelihood-based approaches for GPPs. We apply these models to the pedestrian dataset in Section~\ref{sec:application}, comparing their respective strengths and limitations for modeling waiting behavior. A final discussion in Section~\ref{sec:discussion} closes the paper, reviewing our findings and presenting avenues of further research. 

This article interfaces between methodological contributions and a pedestrian dynamics application, and we therefore recommend a linear reading for a complete picture. Readers interested primarily in the application may skip Sections~\ref{sec:poisson} and~\ref{sec:inference}, which develop the inference machinery behind the results but are not needed to interpret them, and read Sections~\ref{sec:data},~\ref{sec:models}, and~\ref{sec:application} in sequence.

\section{Modeling waiting pedestrians: literature review}\label{subsec:lit}

This section reviews the most relevant literature on modeling waiting crowds and identifies the specific gap addressed in our work: the lack of a quantitative, granular modeling framework for replicated waiting crowd patterns. We give a brief overview of behavioral models and empirical studies from pedestrian dynamics, followed by probabilistic and rule-based models of aggregate passenger distributions on railway platforms. Finally, we discuss a related approach from density-functional theory that is methodologically close to our own. 

While the field of pedestrian behavior modeling has focused more on pedestrian dynamics than on static pedestrians, Johansson et al.~\cite{Johansson_2015} treat the topic of waiting pedestrians as an addition to the social force model~\cite{Helbing_1995}, one of the foundational models for pedestrian dynamics. Modeling is based on the existence of points of interest to which pedestrians are attracted. These induce a waiting area around the point of interest. In those models, as a pedestrian enters this area its activity changes to ``waiting,'' for a pre-determined waiting time. As the social force model is inherently a dynamic model, the authors focus on how pedestrians arrive at a waiting position and then react to other pedestrians in the area. Three different sub-models are proposed. In the preferred velocity model, as a pedestrian enters the area, it is compelled to reduce its velocity to $0$ exponentially quickly. In the preferred position model, the pedestrian reaches a preferred waiting position within the waiting area, and tries to remain here. Finally, in the adaptive preferred position model, the preferred position is modified as a pedestrian's position is disturbed by passing pedestrians, moving toward the new position just as the disturbed pedestrian moves toward it in turn. However, little focus is given to the choice of these preferred positions, as well as to the choice of interest points or waiting times. In addition, this is primarily a qualitative model, as statistical inference is not explored in depth.

Complementary to this approach, several studies have empirically characterized waiting behavior at railway stations. Seitz et al.~\cite{Seitz_2016} discuss waiting as a distinct form of pedestrian behavior, drawing on concepts such as personal-space requirements and the selection of waiting positions. Based on field observations at a railway platform, they introduce measures and hypotheses for waiting behavior and discuss its implications for pedestrian simulations. While this work highlights the importance of spatial arrangements of waiting pedestrians, it is primarily concerned with characterizing waiting behavior empirically, rather than statistical modeling.

More recently, a series of articles by K\"{u}pper and co-authors has examined waiting pedestrians at train stations using both field observations and experiments~\cite{Kupper_2023a,Kupper_2020,Kupper_2023b}. In~\cite{Kupper_2020}, space usage on train platforms is analyzed. In particular, differences between spaces used by waiting and passing pedestrians are considered, and identification methods for waiting pedestrians are developed. In~\cite{Kupper_2023b}, the authors more thoroughly investigate the behavior of waiting pedestrians based on the same platform data by analyzing grouping behavior and the influence of waiting times on the preferred waiting positions. Finally,~\cite{Kupper_2023a} concerns a series of laboratory experiments to study influencing factors on waiting behavior under more controlled circumstances. Tradeoffs between different environmental factors are classified and analyzed. The entire series of papers takes an empirical stance, analyzing behavior both in the field and in laboratory experiments, seeking to gain a physical understanding. However, the majority of results concern behavior at an ensemble level, relying mainly on density profiles and heatmaps, but do not directly define a probability distribution over configurations of individual pedestrian positions. 

Other studies have developed probabilistic or rule-based models for passenger distributions on railway platforms~\cite{Krstanoski_2014,Zhou_2019}. In~\cite{Krstanoski_2014}, boarding and alighting passengers on a platform were studied using multinomial distributions to describe how passengers distribute among train doors. While this analysis is relatively coarse in terms of spatial resolution, it supports the intuition that train platforms are spatially heterogeneous in terms of location attractiveness. This notion is also supported by the analysis in~\cite{Zhou_2019}, in which waiting pedestrian behavior at a Beijing subway station is studied. Using a fuzzy-logic model, different modes of waiting behavior are introduced, and it is shown that distributions of waiting pedestrians are not uniform over the platform. In addition, as density increases, distributions become more uniform, consistent with a regularizing effect, which supports our claim of repulsive interactions. The approach in~\cite{Zhou_2019} is based on behavioral heuristics and designated waiting areas, and is then calibrated against field observations. In contrast, our work takes a slightly more agnostic approach, without assuming any behavioral rules beforehand.

An approach with methodological similarities is~\cite{Mendez_2018}, which applies density-functional theory to crowds, describing their probability distribution through a pair of functions for environmental preference (``vexation'') and social response (``frustration''). While the framework is validated experimentally using walking fruit flies, a very different setting than our own, the methodological contribution is relevant. The framework makes no behavioral assumptions about individuals, relying only on their spatial distribution across replicated observations, and shares our aim of disentangling spatial inhomogeneity from interactions. Nevertheless, a key difference is that~\cite{Mendez_2018} works with local density fields through a binning of the space, which can blur exact nearest-neighbor structures, whereas our point process models describe individual positions directly, thereby retaining microscopic properties such as interpersonal distances.

In our study, we seek to model waiting behavior at a more granular level, thereby mirroring ensemble level patterns while retaining microscopic characteristics. Rather than trajectories, we model the distribution of individual pedestrian locations in static snapshots using spatial point processes, which have so far not been applied in the waiting pedestrian setting. In particular, we propose generative models unifying spatial heterogeneity and interactions that leverage replicated spatial patterns for statistical inference, and verify our models quantitatively on a large, high-resolution dataset of real-life pedestrian crowds. In contrast to the studies above, this yields an explicit probability distribution over configurations of waiting pedestrians estimated from data rather than prescribed by behavioral rules. Such models can generate realistic waiting configurations for pedestrian simulations, and can quantify the attractiveness of platform locations separately from interactive effects, informing platform design and crowd management.

\section{Waiting pedestrian dataset of replicated spatial patterns from Eindhoven central train station}\label{sec:data}

In this section, we introduce a unique dataset of pedestrian configurations at a train station, which can be considered replicated spatial patterns, i.e. (approximately) independent realizations of a spatial point process. We present the dataset by discussing the collection of the data in~\ref{subsec:data_collection}, how configurations are distilled from the raw data in~\ref{subsec:peaks}, and some general simplifying assumptions in~\ref{subsec:waiting_passing}. Finally, we introduce key quantitative features of the pedestrian behavior observable in the dataset in~\ref{subsec:data_features}, features which we would like effective models to capture.

Settings such as this one present a difficult estimation challenge, as the combination of spatial inhomogeneity and interaction between points makes inference quite difficult. It is challenging to cleanly decouple the two phenomena, yet estimating both aspects simultaneously is rather challenging as well. The pedestrian setting seems to have underlying inhomogeneous intensities that vary greatly, without a clear dependence on any spatial covariates, unlike many of the settings studied by previous work concerning inhomogeneous spatial point processes~\cite{Ba_2023,Baddeley_2000a}. This leads to the near necessity of using nonparametric methods to estimate the environmental influence, making existing theory on which to base our efforts even more scarce. On the other hand, in the waiting pedestrian setting described we have a unique feature: one can observe multiple realizations of the same spatial point process. These replicated spatial patterns allow for the use of statistical methods previously not applicable to spatial point process settings, such as cross-validation approaches for the choice of hyperparameters (such as a bandwidth for nonparametric intensity estimation via kernel estimators that we propose), see Section~\ref{subsec:cv}. They also enable future study and validation of statistical guarantees for a variety of established estimation methods. This not only makes this particular setting and dataset worth investigating, but also sets the scene for future investigation of similar settings for the emergence of point process phenomena, perhaps with different or more desirable characteristics for specific statistical questions.

\begin{figure}[t]
    \centering
    \includegraphics[width=\textwidth]{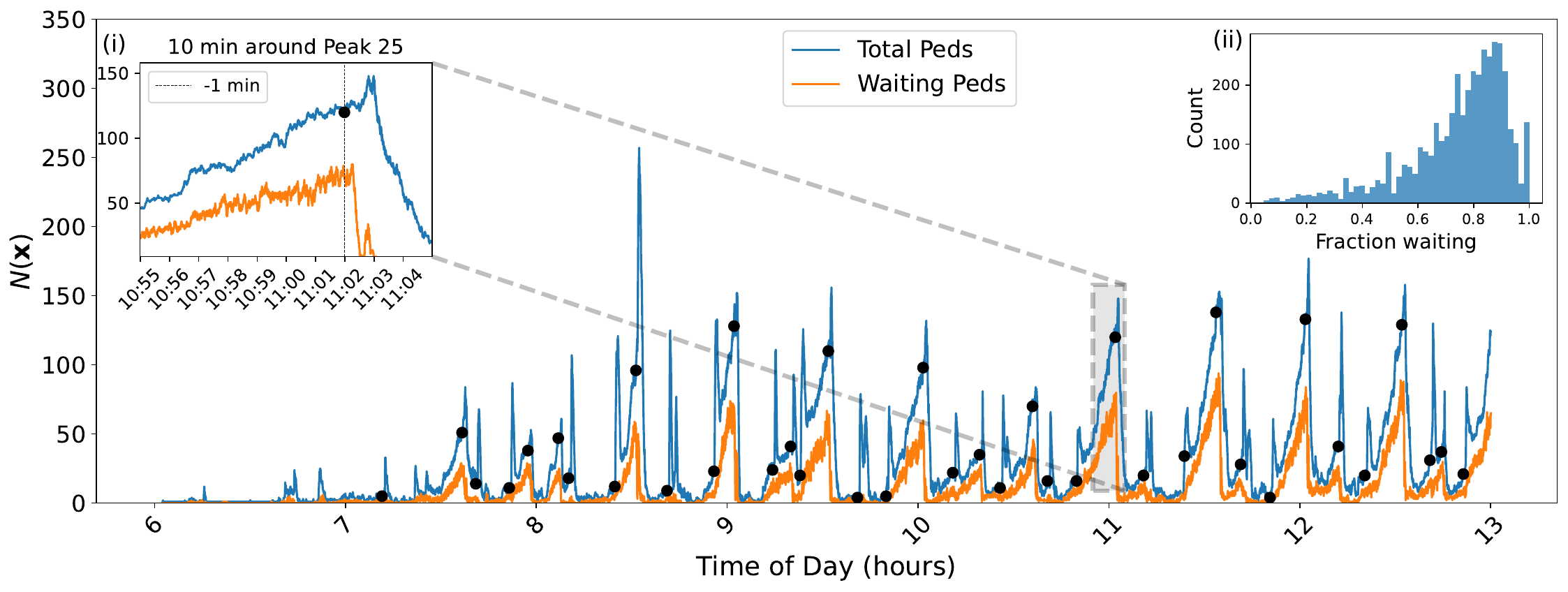}
    \caption{Time series showing the number of pedestrians on the platform from 6:00h to 13:00h for 1 April 2022---total number of pedestrians (passing and waiting) in blue, number of waiting pedestrians in orange. Time of snapshot recording marked in black. Inset (i) shows zoomed in time series around a peak. Inset (ii) shows frequencies of ratios of waiting vs. passing pedestrians in the full dataset.}
    \label{fig:peaks}
\end{figure}

\subsection{Data collection and pre-processing}\label{subsec:data_collection}

The data we use for this article was gathered at Eindhoven Central Station in the Netherlands. Pedestrian location is recorded as two-dimensional positional coordinate data at a sub-centimeter resolution and decisecond frequency ($10\,$Hz temporal resolution) by overhead sensors on a full platform---all platform depictions in this paper are a \emph{birds-eye-view}, as presented in Figure~\ref{fig:agg_heatmap}(b). The platform has a staircase lined by escalators on the left side as its only regular entrance and exit, as well as elevators further on the left side (not visible). Note that in the following, orientations (top, bottom, left, right) correspond to the view seen in Figure~\ref{fig:agg_heatmap}(b). Trains arrive at the top and bottom. In the center of the platform, a caf\'e takes up a closed covered area with entrances on the upper, lower, and left side of the building as seen from above. Similarly, on the far right of the platform, another building obscures the view. Between the two buildings, there is a bench, which serves as a significant attraction point for waiting pedestrians (see e.g. the hotspot of activity in Figure~\ref{fig:agg_heatmap}). The data is collected automatically and anonymously, enabled by the use of specialized sensors tracking pedestrian positions. For more details, see~\cite[Appendix A]{Corbetta_2018}. An excerpt of the full dataset ($60$ consecutive days in spring of $2022$) is publicly available~\cite{Pouw_2024b}. To ensure reproducibility of the results and to restrict the computational complexity of the analysis, this article uses the same timeframe contained in this excerpt. From the raw trajectory data, we use a velocity-based filtering method to identify approximately static pedestrians, i.e. pedestrians that barely change location within a certain window in time (but could potentially turn, or adapt different postures). We classify these pedestrians as \emph{waiting}, and the remaining pedestrians as \emph{passing}. See Appendix~\ref{app:data_prep} for more details about the algorithm for detecting static pedestrians and filtering methods. After filtering the data appropriately, we generate snapshots of the platform, each containing a configuration $\x \in \X$ of coordinate positions of waiting pedestrians, where $\X$ now corresponds to the two-dimensional observation area. 

\begin{figure}[t]
    \centering
    \begin{subfigure}[ht]{0.56\textwidth}
        \centering
        \includegraphics[width=\textwidth]{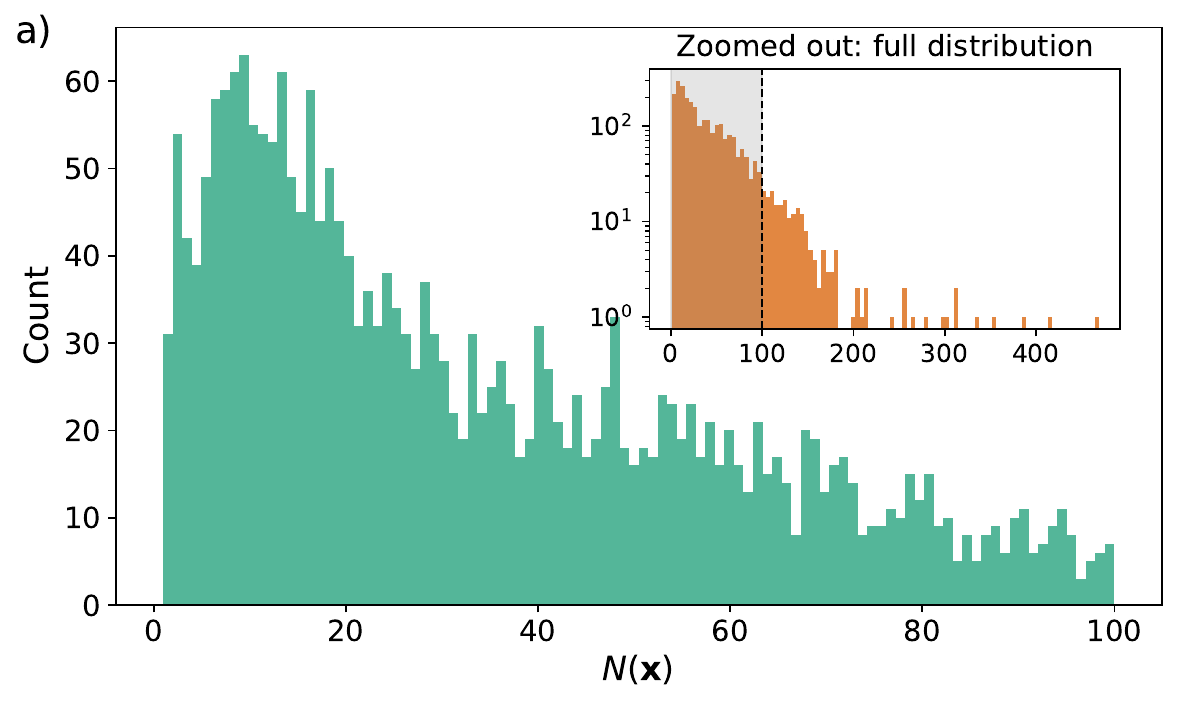}
    \end{subfigure}
    \begin{subfigure}[ht]{0.43\textwidth}
        \centering
        \includegraphics[width=\textwidth]{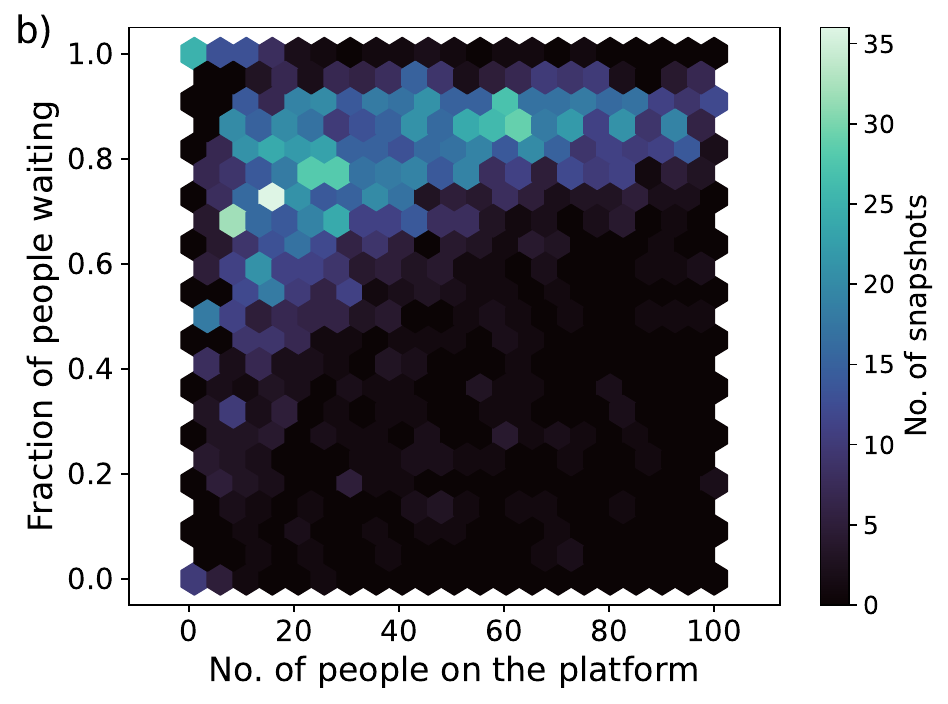}
    \end{subfigure}

    \caption{(a) Histogram of the number of (waiting) people per snapshot $N(\x)$ in the dataset for waiting crowds of less than 100 pedestrians, corresponding to the bulk of the samples. Inset showing the same histogram for the full dataset, counts in log scale showing a clear exponential progression. (b) 2D Histogram of the crowd size (waiting and passing) vs. the fraction of people waiting for snapshots of less than 100 total pedestrians. While at low densities most ratios are represented, a higher occupancy correlates with a higher fraction of waiting pedestrians.}
    \label{fig:summary_stats}
\end{figure}

\subsection{Obtaining snapshot data from peaks in crowd size.}\label{subsec:peaks}

To obtain a set of snapshots of comparable scenarios, we identify peaks in the number of pedestrians on the platform over the entire observed time period, which generally correspond to train arrivals. This is clearly observable by the sharp drops after a certain time (see Figure~\ref{fig:peaks}), and are more reliable than pre-determined train timetables, as delays and slight real-time adjustments to train planning are frequent. To obtain snapshots with large numbers of waiting pedestrians, we choose a moment in time one minute before the peak (i.e. arrival of the train). This choice is based on observing the peaks in the number of waiting pedestrians $N(\x)$ compared to peaks in the total number of pedestrians, as well as physical considerations. For instance, when a train enters the platform, waiting people begin to move towards expected points of entry, thereby reducing the number of candidate points. Figure~\ref{fig:peaks} also shows that while this may not be obvious from a visual inspection of the plot, settings in which the ratio of waiting to passing pedestrians is high are quite common at these moments in time. Thereby, with this choice, we not only ensure that typical waiting scenarios are captured (i.e. before a train arrival), but we also ensure that the influence of passing people on the waiting behavior is small. To secure that the inhomogeneous intensity is comparable, we split the data, focusing only on snapshots consisting of settings with a majority of pedestrians in the top half of the platform (as seen from above in Figure~\ref{fig:agg_heatmap}(b)). This ensures that we separate settings in which trains arrive at different sides of the platform, which can be seen as different phenomena (for more detail, see Appendix~\ref{app:data_prep}). These pragmatic choices ensure, from a modeling perspective, that snapshots can be considered approximately independent samples of a point process, some examples of which are in Figure~\ref{fig:samples}. While independence is not entirely guaranteed, as the same person could be included in several snapshots\footnote{
    For instance, a person may have arrived to wait for a train with an hour to spare, and thus may appear in more than one subsequent snapshot. Alternatively, the same person may take the same train each morning, or may take a train from the same platform more than once per day. In any case, these scenarios are not particularly common, and are assumed to have insignificant effects.
}, these effects are assumed to be negligible. The dataset generated using these steps consists of $2629$ snapshots, with in total $108 348$ total unique pedestrian positions, where each snapshot can be considered to be a replicated spatial pattern, i.e. independent realizations from a spatial point process. Figure~\ref{fig:summary_stats} shows some general summary statistics for the dataset. We can see that a wide variety of different types of configurations are covered, with different crowd densities being well represented, as well as different ratios of waiting to passing pedestrians. In particular, Figure~\ref{fig:summary_stats}~(b) shows that while low fractions of waiting pedestrians are common in lower density snapshots, at higher densities they are quite rare. The dataset of snapshots obtained from the public data~\cite{Pouw_2024b} is also publicly available~\cite{Sickert_2026a}\footnote{
    Due to filtering and anonymization before release, the public dataset~\cite{Sickert_2026a} (including snapshots where the majority is in the bottom half of the platform) contains $32/4175$ fewer snapshots (ca. $0.77\,$\%) than the dataset used in this article. This does not lead to any notable changes in the results.
}.

\begin{figure}[t]
    \centering
    \begin{subfigure}[ht]{\textwidth}
        \centering
        \includegraphics[width=\textwidth]{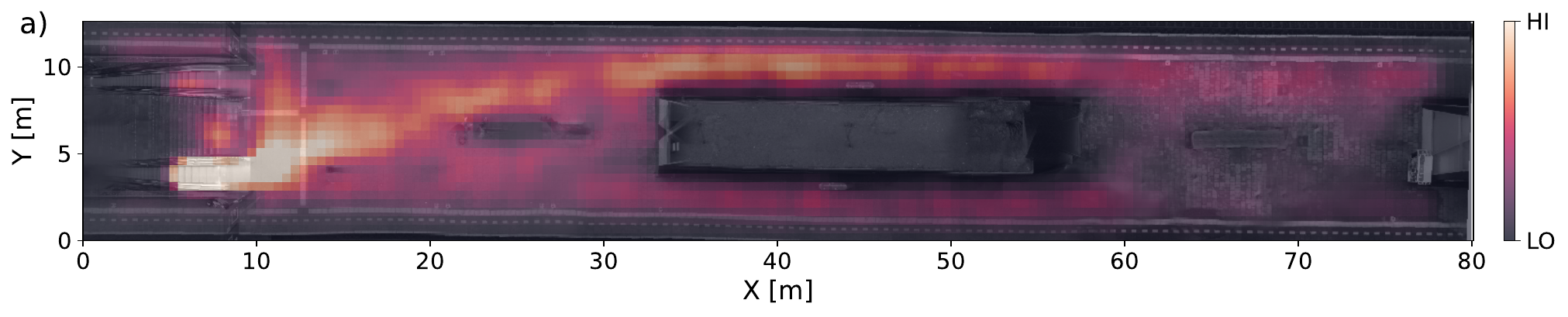}
    \end{subfigure}
    \begin{subfigure}[ht]{\textwidth}
        \centering
        \includegraphics[width=\textwidth]{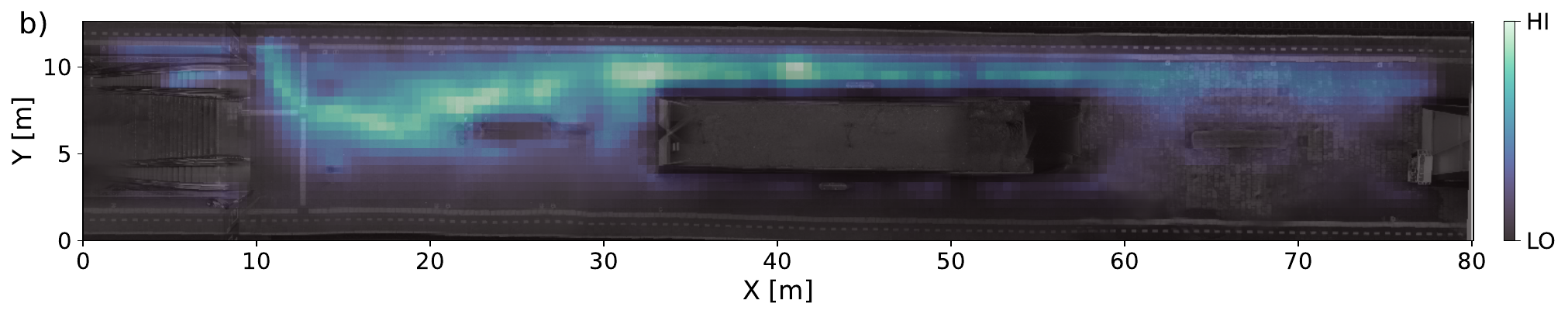}
    \end{subfigure}
    \begin{subfigure}[ht]{\textwidth}
        \centering
        \includegraphics[width=\textwidth]{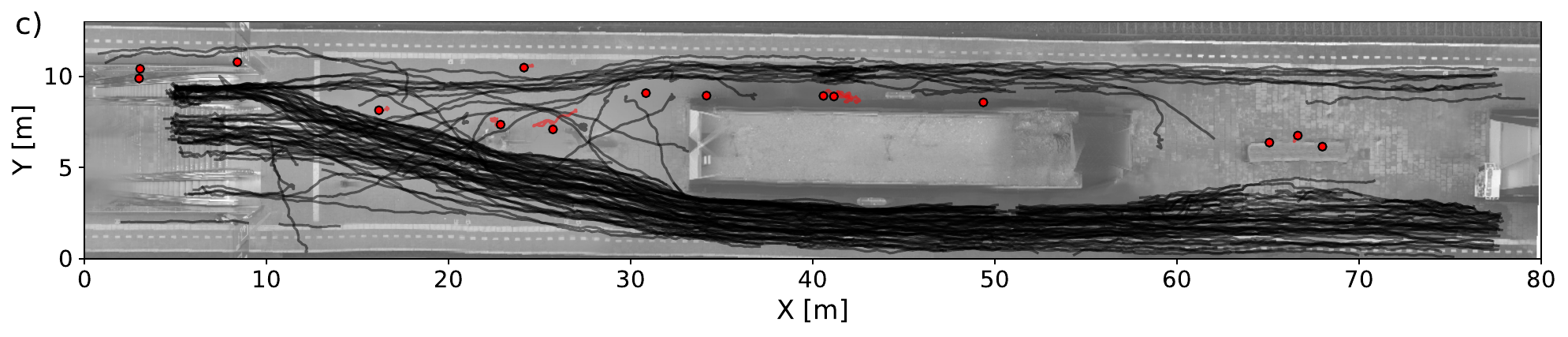}
    \end{subfigure}

    \caption{(a) Estimated one-particle marginal $p_1$ of positions of passing pedestrians at snapshot times, aggregated over all snapshots. Comparing with Figure~\ref{fig:agg_heatmap}(a), passing pedestrians are a completely different phenomenon, while some areas of activity overlap. (b) Heatmap showing the overlap between marginal densities of passing and waiting pedestrians (i.e. between Figure~\ref{fig:agg_heatmap}(a) and panel (a) of this figure, by taking the minimum of the two values at each position). For instance, the area between the entrance and first building show significant overlap. (c) Trajectories of passing (black) and waiting (red) pedestrians in $20\,$s leading up to an example snapshot being taken. While passing pedestrians move around waiting ones, waiting pedestrians barely change position by our definition.}
    \label{fig:passing_vs_waiting}
\end{figure}

For DPPs, one further processing step is needed. The DPP models we use are discrete (for reasons addressed in Section~\ref{subsec:dpp}), so snapshots are discretized onto a grid. A choice of grid requires a balance between computational cost and resolution. Inference requires the eigendecomposition of an $N \times N$ kernel matrix, where $N$ is the number of grid squares, at a cost of $\mathcal{O}(N^{3})$~\cite{Kulesza_2012}. We choose a grid spacing of $50\,$cm, close to the shoulder breadth of an adult pedestrian (commonly estimated at $40-50\,$cm, see e.g.~\cite{Weidmann_1992,Zanlungo_2020}). This results in $N^2 = 4160$ possible positions, keeping inference computationally feasible. The data is discretized by assigning each point to the center of the nearest cell, and then removing duplicate points. These are mathematically impossible in a DPP model, and are uncommon: $0.76$\% of points in the inference area (specified in Section~\ref{sec:application}) are removed.

\subsection{The influence of passing pedestrians on waiting behavior}\label{subsec:waiting_passing}

It is worth reiterating that our analysis focuses exclusively on pedestrians classified as waiting, and does not explicitly model the influence of passing pedestrians (i.e. those not considered waiting by the heuristic described in Appendix~\ref{app:data_prep}). Although for most situations, this assumption seems to be valid, there are plausible scenarios where passing pedestrians may cause waiting individuals to change positions. However, such events are comparatively infrequent in the present dataset, and, for the purpose of the phenomena we study, this is a reasonable approximation, as evidenced by the practically useful modeling results discussed in Section~\ref{sec:application}.

A large part of the influence of passing pedestrians is already reflected in the spatial intensity. Individuals rarely choose to wait in regions of high pedestrian traffic, such as near the staircase or the entrances of the central building, as can be seen in Figures~\ref{fig:agg_heatmap}(a) and~\ref{fig:passing_vs_waiting}(a-b). Moreover, a visual inspection of the footage indicates that waiting pedestrians typically remain stationary once they have settled on a position, while passing pedestrians tend to adjust their trajectories around them (Figure~\ref{fig:passing_vs_waiting}(c)). Finally, the behavior of the fitted models does not suggest systematic biases that would indicate a major role for interactions with passing pedestrians within the regime we consider, as is later discussed in Section~\ref{sec:application}. Although a unified treatment of both groups representing a more complete model would be of interest for future work, the above observations suggest that the present simplification is sufficient for our current aims.

\begin{figure}[t]
    \centering
    \setlength{\tabcolsep}{4pt}

    \begin{tabular}{ccc}
        \begin{subfigure}[t]{0.31\textwidth}
            \centering
            \includegraphics[width=\linewidth,keepaspectratio]{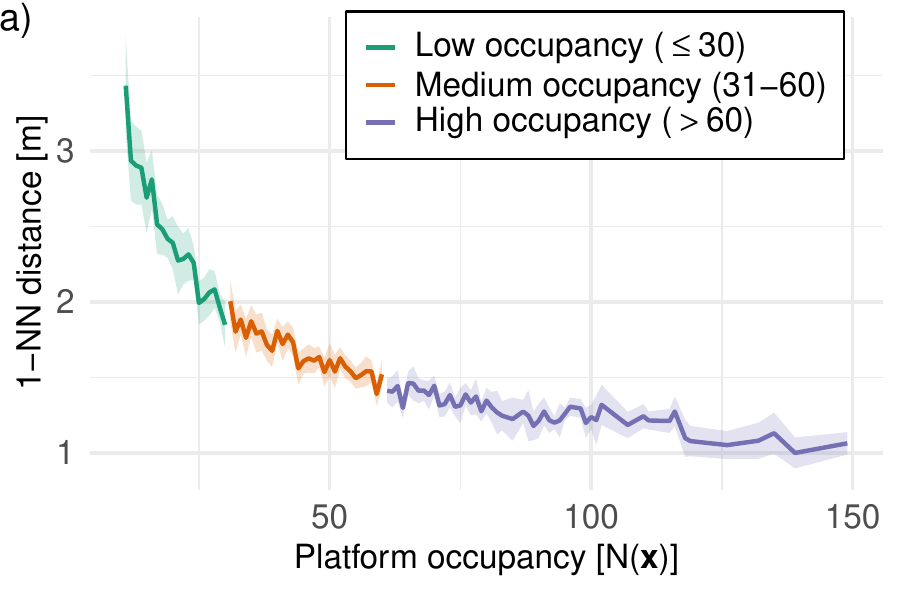}
        \end{subfigure}
        &
        \begin{subfigure}[t]{0.31\textwidth}
            \centering
            \includegraphics[width=\linewidth,keepaspectratio]{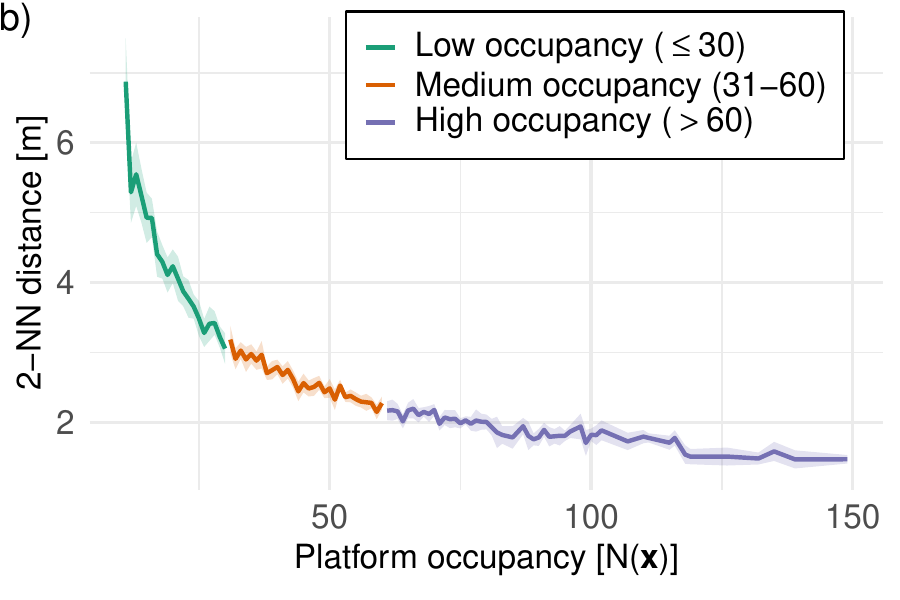}
        \end{subfigure}
        &
        \begin{minipage}[t]{0.31\textwidth}
            \centering
            \vspace{0pt}
        \end{minipage}
        \\[0.8em]

        \begin{subfigure}[t]{0.31\textwidth}
            \centering
            \includegraphics[width=\linewidth,keepaspectratio]{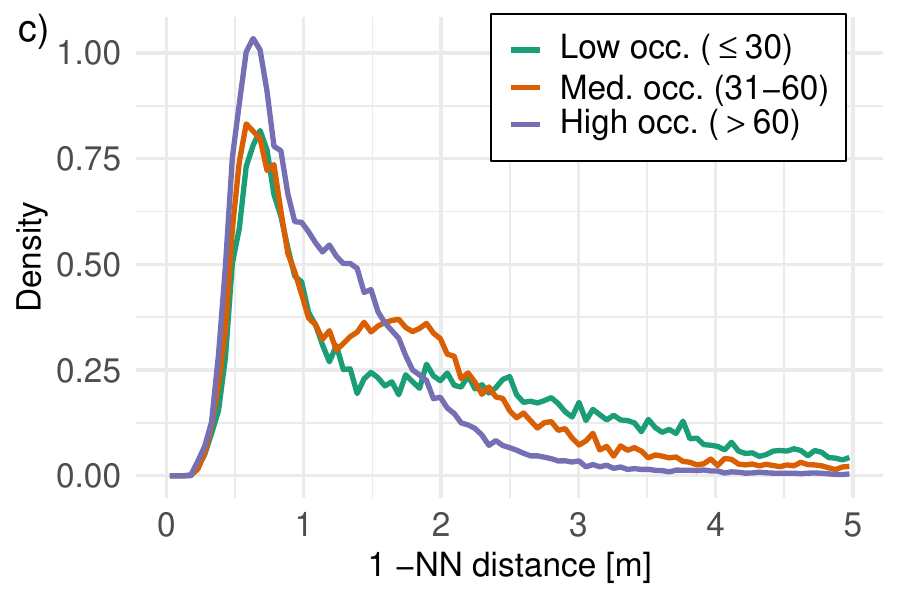}
        \end{subfigure}
        &
        \begin{subfigure}[t]{0.31\textwidth}
            \centering
            \includegraphics[width=\linewidth,keepaspectratio]{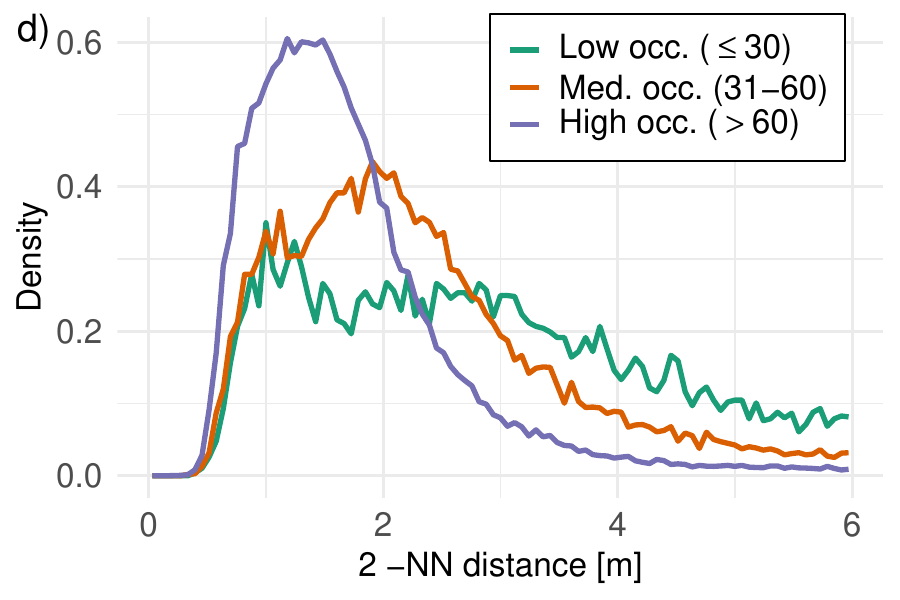}
        \end{subfigure}
        &
        \begin{subfigure}[t]{0.3\textwidth}
            \centering
            \includegraphics[width=\linewidth,keepaspectratio]{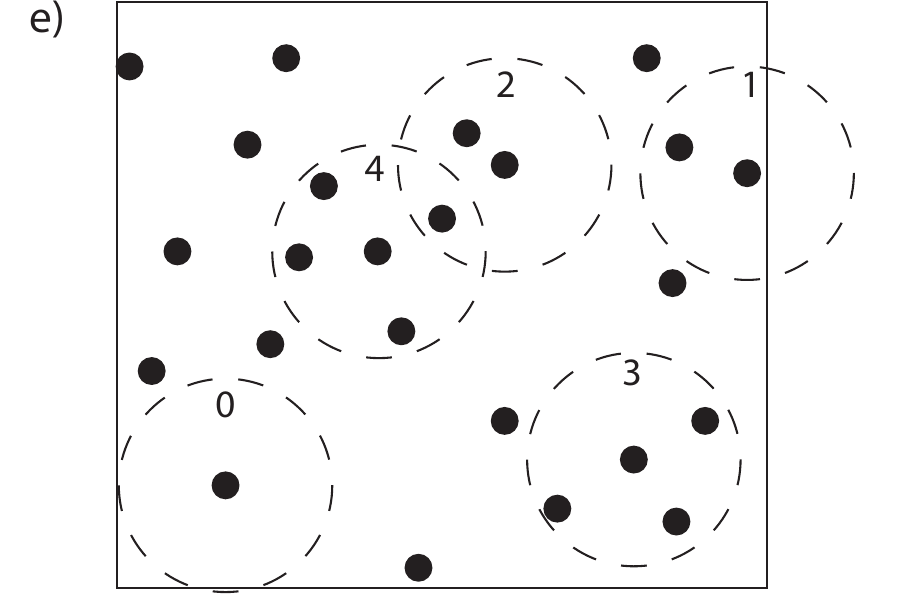}
        \end{subfigure}
        \\[0.8em]

        \begin{subfigure}[t]{0.31\textwidth}
            \centering
            \includegraphics[width=\linewidth,keepaspectratio]{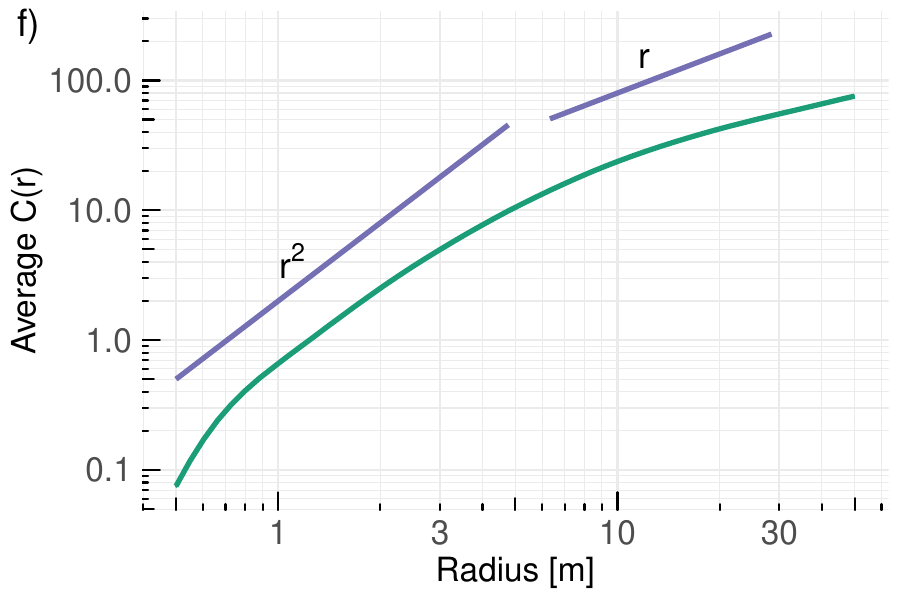}
        \end{subfigure}
        &
        \begin{subfigure}[t]{0.31\textwidth}
            \centering
            \includegraphics[width=\linewidth,keepaspectratio]{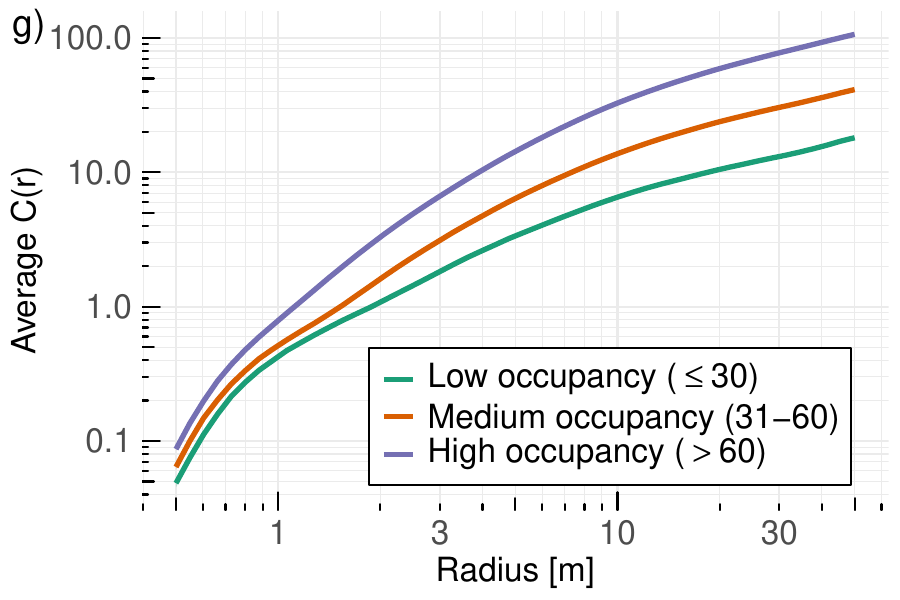}
        \end{subfigure}
        &
        \begin{subfigure}[t]{0.31\textwidth}
            \centering
            \includegraphics[width=\linewidth,keepaspectratio]{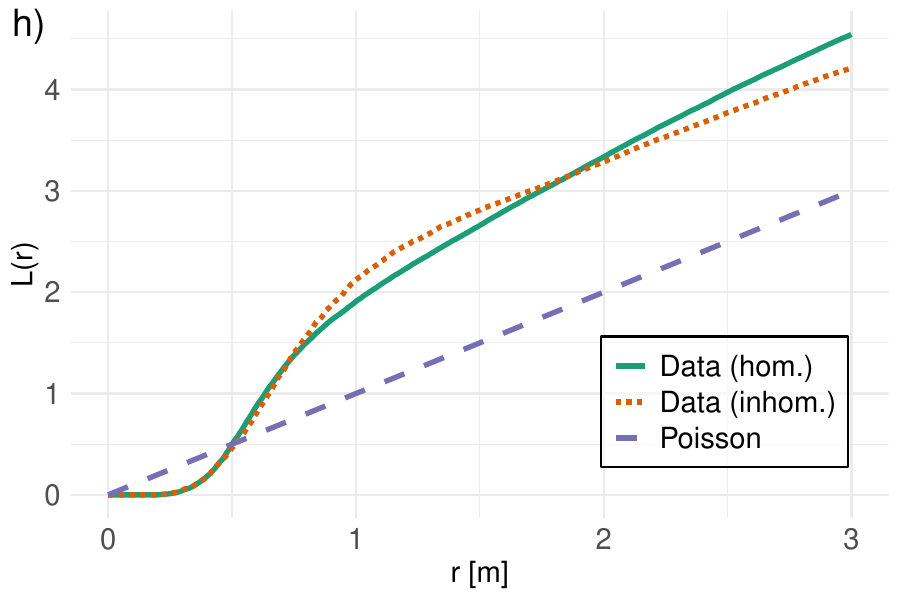}
        \end{subfigure}
    \end{tabular}
    \caption{Distance-based features~\ref{item:D1}-\ref{item:D4} of the dataset.
    (a) Average NN distance for each crowd size.
    (b) Average 2-NN distance for each crowd size.
    (c) Probability density function (PDF) of NN distances, separated by occupancy.
    (d) PDF of 2-NN distances.
    (e) Illustration of how points are counted for $\mathcal{C}(r)$.
    (f) Average $\mathcal{C}(r)$ over all snapshots.
    (g) Average $\mathcal{C}(r)$ separated by occupancy.
    (h) Average L-function for the dataset (solid and dotted representing homogeneous and inhomogeneous estimates, respectively) along with the theoretical L-function for a Poisson point process, a non-interactive process introduced in Section~\ref{sec:poisson} (dashed), which is linear with slope $1$.}
    \label{fig:data_features}
\end{figure}

\subsection{Key quantitative physical features of the data}\label{subsec:data_features}

There is a series of characteristics of the dataset we would like to replicate with our models, which will be generative in nature. Firstly, regarding the inhomogeneous intensities, we would like the points generated when simulating the model to match approximately the \emph{one-particle marginal probability density} of the real data, essentially matching the heatmap seen in Figure~\ref{fig:agg_heatmap}(a). Formally, conditional on $N(\x)=n$, we can define this as
\begin{equation}\label{eq:one-particle_marginal}
    p_1^{(n)}(x)
    =
    \int_{\X}\cdots\int_{\X}
    f(x,x_2,\dots,x_n \mid N(\x)=n)\,dx_2\cdots dx_n,
\end{equation}
where the integral is over $n-1$ folds and $f$ is the probability density function defined in \eqref{eq:spp_density}. In this article, we will typically consider the one-particle marginal density over the aggregated configuration consisting of the union of all snapshots. We denote this simply by $p_1$. In Section~\ref{subsec:poisson_inference} we discuss how to estimate this quantity, which we then use as an estimate of the intensity function $\beta$. For processes without interaction, this is unproblematic (see Section~\ref{sec:poisson}). However, when interaction is present, this comes with some caveats, discussed in Sections~\ref{sec:application} and~\ref{sec:discussion}. 

An effective model should replicate the interaction structure visible in the data. For both the estimation and validation done in Section~\ref{sec:application}, we restrict our attention to snapshots that contain at least $10$ waiting pedestrians (on the full platform). This is a heuristic choice to ensure that enough interaction is present in the dataset, leaving us with $2118$ snapshots and $105693$ pedestrian positions.\footnote{In fact, using the full dataset does not change results for the GPP models, and only slightly changes things for DPPs, see Appendix~\ref{app:generalization}.} It is instructive to consider multiple levels of platform occupancy to have a finer-grained view of the model's behavior in different situations. We partition the data into low occupancy scenarios (with $\leq 30$ pedestrians, corresponding to $825$ snapshots), medium occupancy (between $31$ and $60$ pedestrians; $654$ snapshots), and high occupancy scenarios ($>60$ pedestrians; $639$ snapshots). These limits partition the data into roughly even blocks, while all three on average sit on level of service A~\cite{Fruin_1971}. Since there is no unique way to quantitatively capture the complex interactions we observe, we choose to proceed by considering a series of nearest-neighbor metrics. These distance-based metrics are as follows (cf. also summary in Figure~\ref{fig:data_features}):

\begin{enumerate}[label=(D\arabic*), ref=(D\arabic*)]
    \item \label{item:D1} Average NN-distances: nearest-neighbor (NN) and second nearest-neighbor (2-NN) distances averaged over the configurations of a specific size (i.e. level of occupancy of the platform) should evolve similarly. See Figure~\ref{fig:data_features}(a-b).
    \item \label{item:D2} NN-distance distributions: the empirical probability distributions of the NN and 2-NN distances at different occupancy levels should have similar features (specifically peaks and tails in the corresponding estimated probability density functions should have matching location and behavior). See Figure~\ref{fig:data_features}(c-d).
    \item \label{item:D3} Average ball count $\mathcal{C}(r)$: this is a function of a radius $r$ which returns the number of points around any given point within this radius, averaged over all points (see Figure~\ref{fig:data_features}(e) for an illustration). Formally, for a configuration $\{x_1, \dots, x_n\}$ and a radius $r > 0$, this is given by
    \[ \mathcal{C}(r) = \frac{1}{n} \sum_{i=1}^n \sum_{\substack{j=1 \\ j \neq i}}^{n} \1\{ \|x_i - x_j \| \leq r \}, \]
    where $\|x_i - x_j \|$ denotes the (two-dimensional) Euclidean distance between points $i$ and $j$, and $\1$ denotes an indicator function. This snapshot-level metric can then be averaged across comparable configurations. See Figure~\ref{fig:data_features}(e-g).
    \item \label{item:D4} Average L-function: This metric requires a slightly more involved introduction. The L-function is a modified version of Ripley's K-function~\cite{Ripley_1977}, it is defined as $L(r) = \sqrt{K(r)/\pi}$, where $K$ is Ripley's K-function. This is not a function of the data, but rather of the distribution of random configurations $\x \in \X_n^*$, and is defined as 
    \begin{equation*}\label{eq:K_theo}
        K(r) = \frac{1}{|\X|} \E \left[ \sum_{i=1}^n \sum_{\substack{j=1 \\ j \neq i}}^n \frac{1}{\beta(x_i)\beta(x_j)} \1 \{\|x_i - x_j\| \leq r\} \right],
    \end{equation*}
    where $r \geq 0$. Note that since $K$ is a function of the unknown distribution of $\x$, it must be estimated. Empirical estimators are available, with slightly different definitions depending on whether we assume $\beta$ to be homogeneous. If we assume this to be the case, we can estimate $K$ by~\cite[Chapter 7.4.2]{Baddeley_2015}
    \begin{equation*}\label{eq:K_hat_hom}
        \hat{K}_{\text{hom}}(r) = \frac{|\X|}{n(n-1)} \sum_{i=1}^n \sum_{\substack{j=1 \\ j \neq i}}^n \1\{\|x_i - x_j \| \leq r \}. 
    \end{equation*}
    Note that this is very similar to $\mathcal{C}(r)$, differing primarily in its normalization. However, homogeneity of the intensity (i.e. stationarity) is a strong assumption, so we may consider an estimator which allows for some control of the effects of the inhomogeneity by plugging in an estimator of $\beta$, denoted by $\hat{\beta}$ (see Section~\ref{subsec:poisson_inference}):
    \begin{equation*}\label{eq:K_hat_inhom}
        \hat{K}_{\text{inhom}}(r) = \frac{1}{|\X|} \sum_{i=1}^n \sum_{\substack{j=1 \\ j \neq i}}^n \frac{1}{\hat{\beta}(x_i) \hat{\beta}(x_j)} \1\{\|x_i - x_j \| \leq r \}.
    \end{equation*}
    If we assume that $\beta$ is constant, we can take $\hat{\beta} = n/|\X|$ to recover the homogeneous estimator $\hat{K}_{\text{hom}}$. To get an empirical estimate of the L-function (i.e., $\hat{L}_{\text{hom}}$ or $\hat{L}_{\text{inhom}}$), we can now just plug in the corresponding K-function estimator. We then take an average value over all configurations. See Figure~\ref{fig:data_features}(h). 
\end{enumerate}
The metrics~\ref{item:D3} and~\ref{item:D4} are similar in spirit, with the former used extensively to characterize structure in e.g. (disordered) solids, liquids and complex matter~\cite{Goharshadi_2025} (also known here as the cumulative radial distribution function), and has also been employed in pedestrian dynamics literature~\cite{Pouw_2020}. The latter, by contrast, is a classical metric in spatial point process literature~\cite{Baddeley_2015}. The average ball count $\mathcal{C}(r)$ refrains from incorporating the inhomogeneity (which is unknown), and therefore does not require estimating any parameters beforehand. Plotting $\mathcal{C}(r)$ in log-log-scale, see Figure~\ref{fig:data_features}(f), we see that for medium and large radii, the curve agrees with the quadratic scaling law we would expect if there were no interaction present, transitioning to the linear scaling as soon as the radius reaches a size where boundary effects begin to have an effect. By contrast, at small radii, the growth is steeper than the quadratic line, indicating a strong repulsive effect at short ranges, which the proposed models should also replicate. An advantage of Ripley's L-function is that it allows for easy comparison with the no-interaction case. If the data consists of realizations of a Poisson process (homogeneous or inhomogeneous), the L-function is linear, allowing for the informal interpretation that an L-function below the ``Poisson line'' represents a more regular (i.e. repulsive) pattern, while values above the line represent more clustered (i.e. attractive) patterns. In addition, Ripley's K-function (and correspondingly the L-function, see~\ref{item:D4}), allows for some control of the intensity, which sets it apart from $\mathcal{C}(r)$. However, while $\hat{L}_{\text{inhom}}$ requires less stringent assumptions than the homogeneous version $\hat{L}_{\text{hom}}$, in~\cite[Chapter 7.10.2]{Baddeley_2015} the authors state that nonparametric estimates of the intensity (such as those we use in this article, see Section~\ref{sec:poisson}) result in more biased estimates of the K-function, and hence the L-function. As a result, this additional layer of complexity is of limited use in our case, as shall be clear from practical examples discussed in Section~\ref{sec:application}. Nevertheless, we consider both the homogeneous and inhomogeneous version of the L-function during evaluation, as it is unclear which is more meaningful in our context. When we estimate the L-function empirically for each snapshot and then take the mean over the entire dataset, the effect of the inhomogeneity is reduced by averaging over many replications, and the difference to the homogeneous version is less pronounced. A characteristic dent at short ranges is visible for both versions in Figure~\ref{fig:data_features}(h), indicating a strong repulsive interaction, which suitable models should replicate. Crucially, although the L-function here lies above that of the Poisson process for most of its trajectory, this does not necessarily indicate a clustering effect, as the inhomogeneity may falsely indicate such an effect, see e.g. discussion in~\cite[Chapter 7.3.5.1]{Baddeley_2015}.

\section{Poisson-type models as a baseline: limitations of disregarding interaction}\label{sec:poisson}

To acquire a fundamental understanding of the way we can model spatial inhomogeneity in the intensity, we turn to models without interaction in this section. We give a brief formal introduction to such models, discuss their limitations, and review the literature on nonparametric estimation of the intensity function. 

The most straightforward way to model a spatial point process is to assume there are no interactions between points. This leads to the inhomogeneous Poisson point process, which serves as a (rather standard) benchmark model. A Poisson point process is fully characterized by its spatial intensity function $\beta:\X\to[0,\infty)$. The total number of points in this model is random, and it has a Poisson distribution with mean equal to $\int_\X \beta(x) \d x$. When conditioning on the total number of points $N(\x) = n$, the locations of the points are jointly independent. In other words, the $n$ points in the configuration are an independent and identically distributed sample with density
\[ f(\x | N(\x) = n) \propto \prod_{x \in \x} \beta(x); \quad \x \in \X_n^*. \]
This conditional look at the Poisson model, also known as a \emph{binomial point process}, is  often preferable when simulating patterns for comparison, since it allows us to match observed occupancy levels of the train platform exactly. For inference, it does not make much of a difference whether we consider conditioning, as in the absence of interactions we can simply aggregate replicated point patterns into a single pattern without altering the inference. In both the Poisson case and the binomial case, statistical inference is reduced to a density estimation problem~\cite{Diggle_1988}. 

While the inference of the spatial intensity is relatively straightforward when assuming complete independence between points (i.e. no interaction) this is an unrealistic assumption. In reality, most natural phenomena that can be modeled as spatial point processes incorporate some aspect of interaction~\cite{Baddeley_2015,Moller_2004,Diggle_2013}. While the kernel-based methods we will discuss in the following can still be used, additional care needs to be taken when selecting the bandwidth. Several pragmatic approaches to bandwidth selection accommodating these scenarios exist, taking advantage of methods developed for non-interactive settings. Common examples include minimizing the mean squared error in state estimation while assuming the data comes from a stationary (i.e. homogeneous) Cox model~\cite{Diggle_1985}, or a non-model-based technique relying on a fundamental theorem in spatial processes~\cite{Cronie_2019}. The latter approach, while not assuming any underlying model class (unlike the former), seems to work best when the underlying process does not have repulsive interactions. For processes with repulsive interactions, none of the current methods seem to work reliably. However, having multiple samples may provide an avenue to develop new, principled approaches to bandwidth selection. In Section~\ref{subsec:cv}, we discuss a likelihood cross-validation approach leveraging this opportunity. Nevertheless, we begin by carefully introducing the kernel-based intensity estimation method in the next section, before investigating how well purely intensity-based Poisson-type models approximate the data in~\ref{subsec:poisson_inference}.

\subsection{Parametric and nonparametric estimation methods for Poisson models}
In some settings, whether the data is simpler or there is more information available than in our case, it is useful to model the intensity function parametrically. For example, in forest ecology settings, one possibility is to rewrite it as a loglinear function of different spatial covariates~\cite{Baddeley_2015,Ba_2023}. In this case, estimating the coefficients for these covariates is equivalent to logistic regression or maximum entropy techniques (see~\cite{Baddeley_2015} and references therein). However, for waiting pedestrians, this approach is less attractive, as there is no clear set of spatial covariates that can be measured (unlike in forestry or geological settings, where classical choices include soil composition, precipitation, or pH levels, among others~\cite{Ba_2023}). Nonparametric estimation techniques are therefore better suited here. In this regime, Bayesian approaches~\cite{Heikkinen_1998} or spline smoothing~\cite{Ogata_1998} have been used, but by far the most common approach is kernel density estimation, see e.g.~\cite{Baddeley_2015,Diggle_1985,vanLieshout_2024,Cronie_2019}. As mentioned above, estimating the intensity function of a Poisson or binomial point process is essentially a density estimation problem, for which statistical properties of kernel estimators are very well studied (see e.g.~\cite{Tsybakov_2009}). These tools can be applied to the spatial point process setting directly, with only slight modifications to account for boundary issues~\cite{Diggle_1985,Ramlau_1983,vanLieshout_2012}.

Formally, for a point configuration $\{x_1, \dots, x_n\}$ in $\X$, we get an intensity estimate of the form
\begin{equation}\label{eq:kde_intensity}
    \hat{\beta}_n(x; h) \propto \frac{1}{e(x)} \frac{1}{h^2} \sum_{i=1}^n \mathcal{K} \left(\frac{x - x_i}{h}\right),
\end{equation}
where $h > 0$ is a bandwidth and $\mathcal{K} \colon \R^2 \to [0, \infty)$ is a kernel function, i.e. a two-dimensional probability density function which is even in all its arguments and positive in a neighborhood of the origin~\cite{vanLieshout_2024}. The term $e(x)$ is an edge correction, given by
\begin{equation}\label{eq:global_correction}
    e(x; h) = \frac{1}{h^2} \int_{\X} \mathcal{K} \left( \frac{x - z}{h} \right) \d z.
\end{equation}
This is a global correction, in that it does not depend on the data points $x_i$. A well-known local alternative is Diggle's edge correction factor~\cite{Diggle_1985}, which essentially calculates the same factor as \eqref{eq:global_correction} for each $x_i$, and inserts it into the sum:
\begin{equation}\label{eq:kde_local}
    \tilde{\beta}_n(x; h) \propto \frac{1}{h^2} \sum_{i=1}^n \frac{1}{e(x_i)} \mathcal{K} \left(\frac{x - x_i}{h}\right).
\end{equation}
Note that \eqref{eq:kde_intensity} and \eqref{eq:kde_local} are defined up to a normalizing constant due to the edge correction terms. However, this normalizing constant is easily computable in practice.

Though the choice of smoothing kernel plays only a minimal role in practice, the same cannot be said for the choice of smoothing bandwidth, as it very crucially affects the accuracy of the model~\cite{Diggle_1985}. In density estimation settings, and when performance is assessed by a mean-squared error criterion, an (asymptotically) optimal bandwidth selection can be attained by using a least-squares cross validation approach~\cite{Tsybakov_2009}. In the context of spatial point processes, assuming the data actually comes from a Poisson process, the most established approach is a likelihood cross-validation (LCV) approach, which uses the Poisson likelihood as a validation criterion~\cite[Chapter 5]{Loader_1999}. Beyond kernel estimators with global bandwidths, adaptive methods also exist, which modify the bandwidth based on local differences in the data~\cite{vanLieshout_2024}. In our context, we stick to the global LCV method, opting for a simple approach for we can use cross-validation methods to later incorporate information about the interaction structure of the data (see Section~\ref{subsec:cv}).

\begin{figure}[t]
    \centering
    \begin{subfigure}[ht]{0.32\textwidth}
        \centering
        \includegraphics[width=\textwidth]{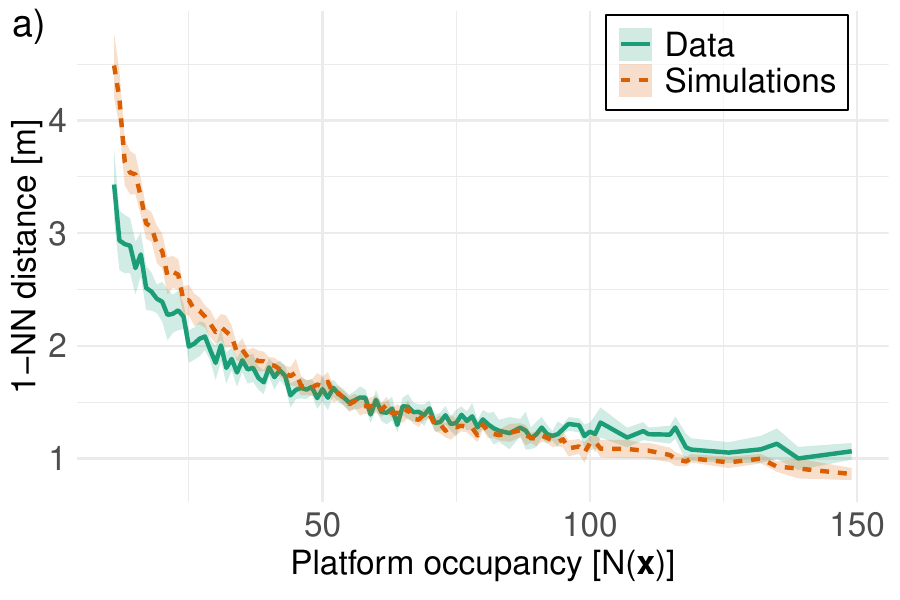}
    \end{subfigure}
    \begin{subfigure}[ht]{0.32\textwidth}
        \centering
        \includegraphics[width=\textwidth]{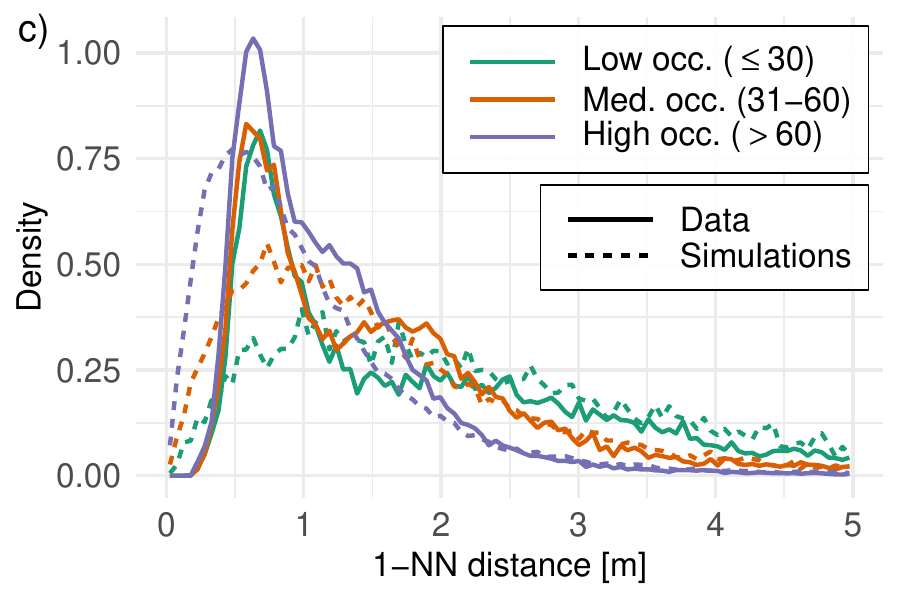}
    \end{subfigure}
    \begin{subfigure}[ht]{0.32\textwidth}
        \centering
        \includegraphics[width=\textwidth]{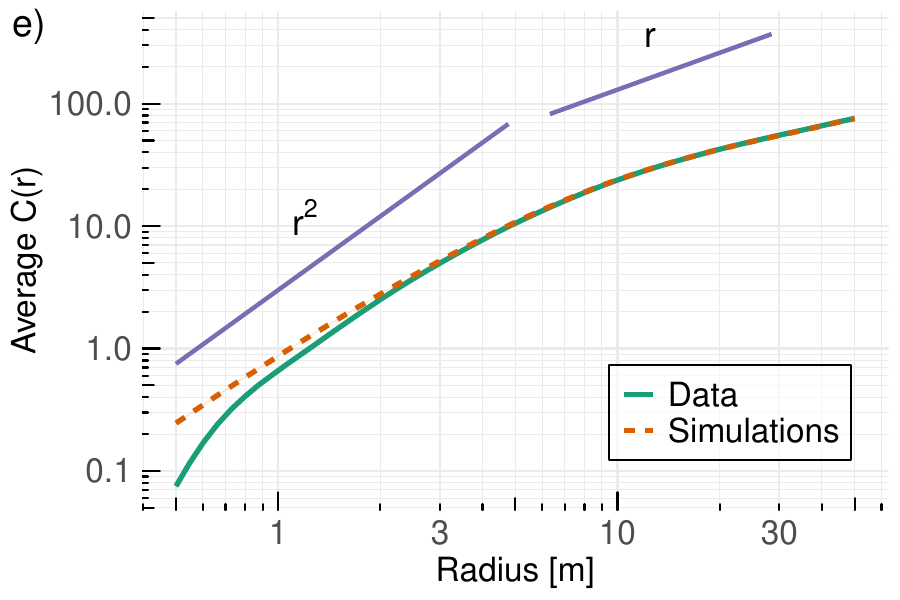}
    \end{subfigure}

    \begin{subfigure}[ht]{0.32\textwidth}
        \centering
        \includegraphics[width=\textwidth]{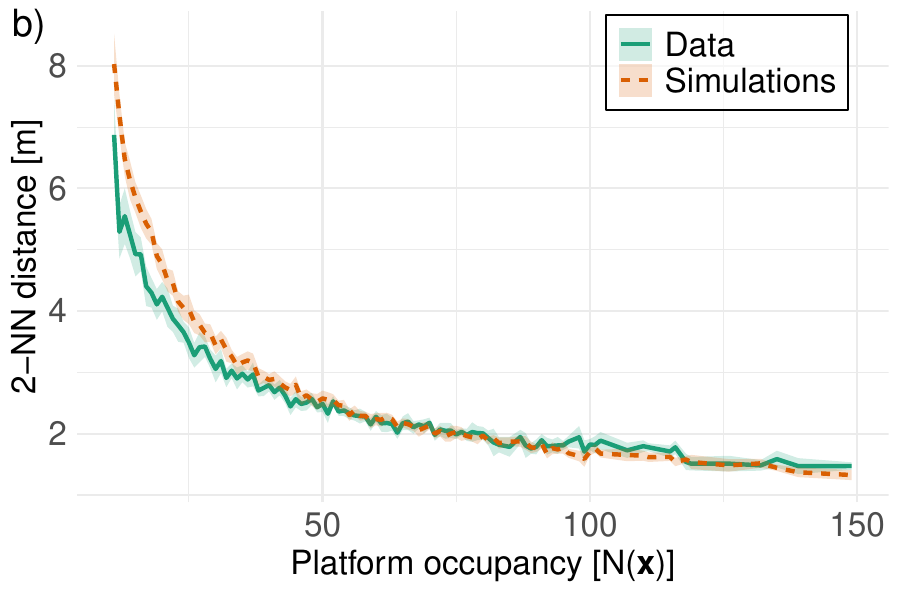}
    \end{subfigure}
    \begin{subfigure}[ht]{0.32\textwidth}
        \centering
        \includegraphics[width=\textwidth]{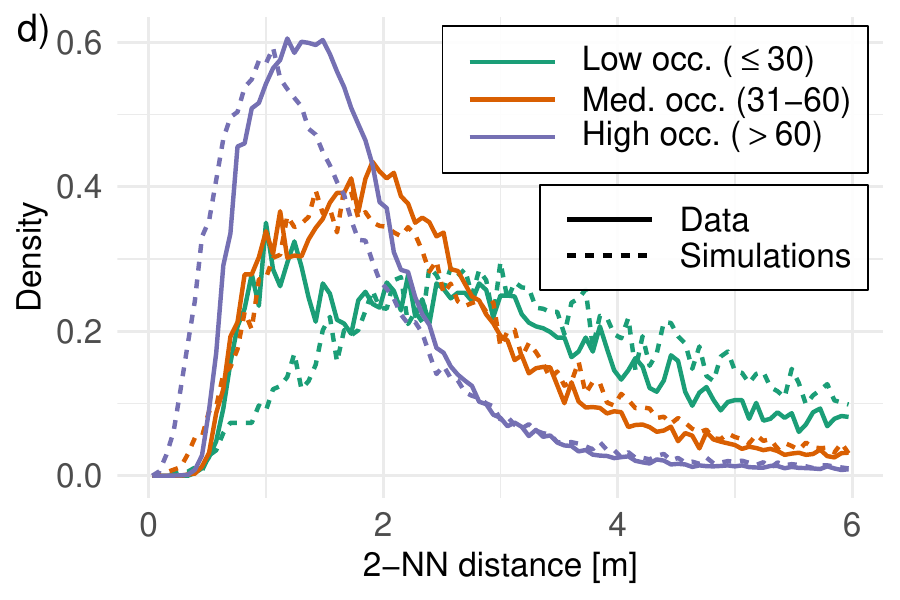}
    \end{subfigure}
    \begin{subfigure}[ht]{0.32\textwidth}
        \centering
        \includegraphics[width=\textwidth]{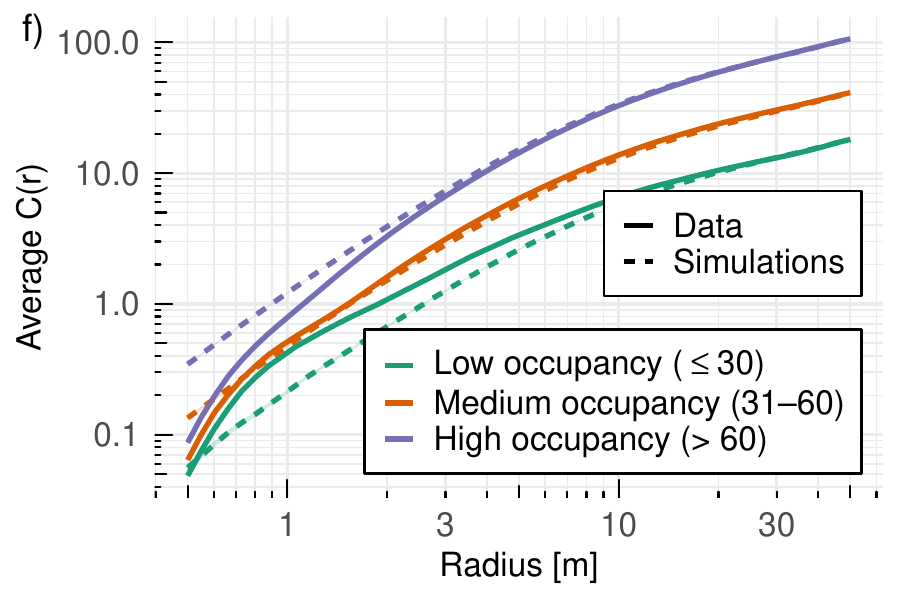}
    \end{subfigure}
    \caption{Comparison of distance-based statistics~\ref{item:D1}-\ref{item:D3} for Poisson simulations versus the dataset. (a) Average NN distance. (b) Average 2-NN distance. (c) PDF of nearest neighbor distances, separated by occupancy.  (d) PDF of 2-NN distances. (e) Average $\mathcal{C}(r)$ over all snapshots in log-log scale. (f) Average $\mathcal{C}(r)$ separated by occupancy. Clearly, Poisson processes are unable to capture repulsive interactions at short ranges.}
    \label{fig:poisson_features}
\end{figure}

\subsection{Nonparametric estimation of the inhomogeneous intensity function based on train station data}\label{subsec:poisson_inference}
In order to estimate the intensity function for this model, we use kernel density estimation with the commonly used Gaussian kernel, i.e. 
\[ \mathcal{K}(x) = \frac{1}{2\pi} \exp\left( \frac{-x^T x}{2 h} \right) \]
for $x \in \R^2$. We select the bandwidth $h$ using Poisson likelihood cross-validation, and apply Diggle's local edge correction as in \eqref{eq:kde_local}. The resulting two-dimensional density can then be used to simulate realizations of a binomial process. For each configuration in the dataset, we simulate a corresponding configuration from the model which has the same number of points to get an accurate comparison. We compare the key features introduced in Section~\ref{subsec:data_features} to the simulations in Figure~\ref{fig:poisson_features}. In Figure~\ref{fig:poisson_features}(a) and (b), we can see a comparison of the average NN and 2-NN distances at several occupancy levels. At higher densities, the model performs well, while at low ranges the simulations actually overestimate the distance between points. This phenomenon may be attributed to the fact that pedestrians may appear as pairs or groups on a train platform, which can skew these statistics significantly when comparing low-density scenarios. A comparison of the distributions of NN and 2-NN distances in Figure~\ref{fig:poisson_features}(c) and (d) is more insightful. Here, we see that for the Poisson model the peak in the distribution of NN distances is less pronounced, with heavier tails, at all occupancy levels. 2-NN distances are captured more accurately, but again miss the mark significantly at low densities. Comparing the average $\mathcal{C}(r)$ over configurations in Figure~\ref{fig:poisson_features}(e) and (f), we see that, as expected, the Poisson model gives us the expected quadratic slope until we reach radii that exceed the platform. This is mathematically justified, since for independent observations, the expected ball count should be proportional to the volume of the balls, and therefore proportional to the square of the radius. This in turn means that we do not capture the repulsive effect we see in the data at shorter ranges. Overall, this comparison exemplifies that we are losing accuracy in modeling by ignoring the repulsive interaction between points at close ranges, leading us to explore more complex models.

\section{Repulsive-interaction point process models for waiting pedestrians} \label{sec:models}

In the context of spatial point processes there are two classes that are well suited to model scenarios with repulsive interactions, namely Determinantal Point Processes (DPPs for short) and Gibbs Point Processes (GPPs for short). In this section, we introduce both processes by giving a short review of the literature concerning their statistical inference before introducing the mathematical formalism with which we can define them, as well as the foundations of the inference strategy we will use for each model. As a brief overview, DPPs, as their name suggests, encode their likelihood structure through a determinant of an operator, the \emph{DPP-kernel}, henceforth naturally enabling the modeling of repulsive interactions. Due to their nature, DPPs allow an analytical form of the likelihood, which is highly desirable for statistical inference. Furthermore, in discrete settings, it is easy and computationally effective to sample from these processes. The main pitfall of DPPs is the restrictive nature of interactions they can capture. In contrast, GPPs are a much more general class of models, relying on an intensity function and an interaction function which are multiplicatively combined. Their flexible mathematical structure allows for a wide range of continuous interaction regimes. However, this flexibility comes at a cost. An intractable normalizing constant in the likelihood makes exact sampling impossible in all but a few special cases, necessitating the use of approximate methods.

\subsection{Determinantal point processes} \label{subsec:dpp}

DPPs were originally introduced in the field of random matrix theory~\cite{Mehta_1960}, but were first described as a general class of models in statistical physics, where they were used to describe the distribution of Fermion systems at thermal equilibrium~\cite{Macchi_1975}. More recently, they have found a diverse range of application in machine learning, primarily in the area of recommender systems~\cite{Chen_2018,Gillenwater_2018}. Spatial modeling applications exist, but are less common~\cite{Lavancier_2015}. Statistical estimation of DPPs continues to be an active research area, with a significant amount of work done primarily on parametric models, including some contributions involving inhomogeneous settings~\cite{Affandi_2014,Lavancier_2021}. However, these works generally assume a parametric form for both the spatial intensity and the interaction. There are different approaches, including maximum likelihood methods~\cite{Lavancier_2015}, composite likelihood and Palm likelihood methods~\cite{Moller_2007}, as well as minimum contrast estimation~\cite{Christophe_2017}, which is a fast method based on matching Ripley's K-function or the pair correlation function (informally, a normalized derivative of the K-function with respect to the radius $r$, see~\cite[Chapter 7.6]{Baddeley_2015}) to the data. Statistical guarantees are generally only possible in stationary settings, though these or similar methods can be used in non-stationary settings as well~\cite{Baddeley_2015}. Nonparametric inference has been less extensively studied. In the discrete case, both expectation-maximization (EM)~\cite{Gillenwater_2014} and fixed-point methods~\cite{Mariet_2015} have been proposed to nonparametrically estimate the entire DPP-kernel at once. In continuous settings, a fixed-point algorithm to solve a maximum likelihood-based problem has also been proposed, although this seems to work best in stationary settings~\cite{Fanuel_2021}. Estimating the entire DPP-kernel at once chooses to disregard any separation between the spatial intensity and the interaction between points. However, separating the two is a desirable feature from the viewpoint of interpretability, as having a parametric model for the repulsion may lead to a better understanding of the underlying physical mechanisms governing pedestrian behavior. 

To introduce DPPs formally, let $L \colon \X \times \X \to \R$ be a positive semidefinite kernel operator with nonnegative eigenvalues, guaranteeing proper normalization. This operator is often referred to as a \emph{likelihood kernel}, but in this article we will call it the \emph{DPP-kernel} to differentiate it from the kernels used in the nonparametric estimation of intensity functions (as in Section~\ref{sec:poisson}). Then, if a configuration $\x \in \X^*$ is considered to be distributed according to a DPP\footnote{
    Technically, the operator $L$ defines a so-called \emph{L-ensemble}, a special case of a DPP. However, the restrictions on the general class of DPPs necessary for this are usually satisfied in practice, see the discussion in~\cite[Section~2.2]{Kulesza_2012}.
}, its probability density function is given by
\begin{equation}\label{eq:dpp_density}
    f(\x) \propto \det (L_\x), 
\end{equation}
where $L_\x$ is an $|\x| \times |\x|$ matrix with entries defined by $L(x, y)$ for each $x, y \in \x$. When $\X$ is discrete and finite, the DPP-kernel $L = L_{\X}$ is a matrix, and the normalizing constant in \eqref{eq:dpp_density} is given by $\det(L + I)$, where $I$ is the $|\X| \times |\X|$ identity matrix. When $\X$ is continuous, the normalizing constant is given by $\prod_{n=1}^\infty (\lambda_n + 1)$, where $(\lambda_n)_{n \geq 1}$ are the eigenvalues of the operator $L$.

Conveniently, DPPs naturally allow for conditioning on the number of points in the configuration without computationally complicating sampling~\cite[Chapter 5.2.2]{Kulesza_2012}. Conditional DPPs are known as \emph{k-DPPs}, where $k$ refers to the number of points in the configuration. The density of a configuration $\x_k = \{x_1, \dots, x_k\} \in \X_k^*$ is given by 
\begin{equation}\label{eq:k_dpp_density}
    f(\x_k) = \frac{\det (L_{\x_k})}{e_k(\Lambda)},
\end{equation}
where $e_k(\Lambda)$ is the $k^{\text{th}}$ elementary symmetric polynomial for $\Lambda \coloneqq \{ \lambda \colon \lambda \text{ eigenvalue of } L \}$, which can be computed efficiently~\cite[Algorithm 7]{Kulesza_2011}. The density is the same regardless of whether we are in the discrete or continuous regime, only the dimension of $\Lambda$ changes.

The operator $L$ can be decomposed as
\begin{equation}\label{eq:dpp_decomposition}
    L(x, y) = \sqrt{\beta(x)} \Gamma(x, y) \sqrt{\beta(y)}
\end{equation}
for $x, y \in \X$, where $\beta \colon \X \to R_{> 0}$ is a measure of intensity for each possible position in $\X$, and $\Gamma \colon \X \times \X \to [-1, 1]$ is a symmetric positive semidefinite kernel operator satisfying $\Gamma(x, x)=1$ for all $x \in \X$. This factorization effectively decouples the spatial inhomogeneity from the interaction behavior. Specifically, the operator $\Gamma$ can be interpreted as a measure of interaction, which we call \emph{DPP interaction kernel}. Note that the diagonal of $L$ is given precisely by the terms $\beta(x)$ for $x \in \X$. This decomposition, due to~\cite{Kulesza_2012}, gives an intuitive way to model the influence of the environment and the interaction between points separately. In a discrete setting, say $\X = \{1, \dots N\}$, this becomes more intuitive, as the operator $L$ is simply a matrix, for which the diagonal entries encode likelihoods of certain positions in the discrete space being chosen, while the off-diagonals can be interpreted as encoding the strength of repulsion between position $i$ and position $j$ for $i, j \in \X$.

When $\X$ is discrete and finite, sampling is simple and fast in both the unconditional case and the $k$-DPP case. Sampling algorithms are based on spectral decompositions of the DPP-kernel, due to~\cite{Hough_2006}, and their main computational complexity is the calculation of an eigendecomposition for the kernel matrix $L$. For a more extensive treatment, see~\cite[Chapter 2.4.4]{Kulesza_2012} for DPPs and~\cite[Chapter 5.2.2]{Kulesza_2012} for $k$-DPPs. In the continuous case, exact sampling is oftentimes infeasible except in a limited number of special cases, which leads to approximate sampling methods such as truncation-based methods~\cite{Lavancier_2015} or low-rank approximations of the kernel~\cite{Affandi_2013}. Unfortunately, these methods do not have easily available implementations in the case of $k$-DPPs, which makes continuous DPPs less attractive in our setting.

\subsection{Likelihoods for discrete DPPs}

As discussed above, due to simulation methods for continuous DPPs being rather complicated and their availability being limited, we restrict ourselves to the discrete case $\X = \{1, \dots, N\}$ for $N \in \N$.  To model patterns of waiting pedestrians on an appropriately spaced discrete grid $\X$, we want to employ the decomposition property of DPPs to estimate separately the underlying intensity $\beta$ (i.e. the scalar intensity values at each grid position) and the DPP interaction kernel $\Gamma$. As is briefly mentioned in Section~\ref{sec:intro}, the spatial intensity is hard to parametrize meaningfully. Although important from a viewpoint of understanding the attractiveness of different spatial regions, it can also be considered a nuisance parameter distorting the estimation of the interaction. We choose to model it nonparametrically using kernel density estimation as in Section~\ref{sec:poisson}, enabling us to capture heterogeneous patterns while not being concerned with restrictive parametrizations. On the other hand, we choose to model the interaction parametrically, as we are more interested in its shape and physical interpretation.

We can choose to model the interaction kernel as a parametric matrix $\Gamma_\theta$ for a parameter vector $\theta \in \R^p,\ p \in \N$. The main advantage of DPPs is their tractable probability density, motivating the use of likelihood-based estimation methods. Considering the intensity $\beta$ as \emph{fixed}, the log-likelihood of $\theta$ given a sequence of observed configurations $\x_1, \dots, \x_T$ is
\begin{equation*}
    \ell (\theta | \x_1, \dots, \x_T) = \left( \sum_{t=1}^T \log \det L_{\x_t} (\theta) \right) - T \log \det \left( L(\theta) + I \right),
\end{equation*}
where $I$ is the $N \times N$ identity matrix and $L(\theta)$ is the DPP-kernel of the model with interaction kernel $\Gamma_\theta$ (and given intensity terms). For $k$-DPPs, only the normalizing term changes, and the log-likelihood becomes
\begin{equation*}
    \ell (\theta | \x_1, \dots, \x_T) = \left( \sum_{t=1}^T \log \det L_{\x_t} (\theta) \right) - \sum_{t=1}^T \log e_{k_t}(\Lambda(\theta)),
\end{equation*}
where $k_t = |\x_t|$, and $\Lambda(\theta)$ is the set of eigenvalues of $L(\theta)$. Recall that $e_k(\Lambda)$ denotes the $k$-th elementary symmetric polynomial for $\Lambda$. While in general these log-likelihoods are not guaranteed to be concave even in discrete settings, likelihood-based methods do show promise, particularly when $\Gamma_\theta$ is not too complex~\cite{Affandi_2014}. In our setting, a grid search based on direct likelihood computations is sufficient to find an optimal value, as will be seen later in Sections~\ref{subsec:numerics}~and~\ref{sec:application}.

\subsection{Gibbs Point Processes} \label{subsec:gpp}

Gibbs point processes are much more versatile than DPPs. Virtually any kind of interpoint interaction can feasibly be modeled, including repulsive and attractive interactions, as well as their combinations. This flexibility comes at a price, however. Simulation and inference are greatly complicated by intractable normalizing constants in the likelihoods of these processes. Due to this, many inference approaches are currently based on heuristics or rely on restrictive assumptions. In particular, processes with underlying spatial inhomogeneity without any clear parametric structure are difficult to deal with, and as a result not extensively studied. As in the previous section, a brief literature review is followed by a short formal introduction of the modeling framework and an overview of the estimation methods we intend to use in the waiting pedestrian setting.

Gibbs point processes were first used in statistical physics to model interacting particles in the continuum~\cite{Ruelle_1969,Preston_1976}, and later used in a variety of different contexts, including forestry~\cite{Stoyan_1998}, anatomy~\cite{Diggle_2006}, and materials science~\cite{Ohser_2000}. Their mathematical theory is well-developed, with the subclass of Markov point processes, Gibbs point processes with finite ranges of interaction, being extensively studied in the context of spatial statistics~\cite{vanLieshout_2000,Moller_2004,Dereudre_2019,Daley_2008,Daley_2003}. Likelihood-based statistical inference relies greatly on Markov Chain Monte Carlo (MCMC) methods~\cite{Geyer_1994} due to intractable normalizing constants in the formulation of the likelihood. In stationary settings, i.e. settings with constant intensity, some asymptotic properties are understood for certain models, even if a central limit theorem is not available~\cite{Dereudre_2017}. By contrast, as far as the authors are aware, likelihood-based inference in inhomogeneous settings has been an outstanding open problem (see e.g.~\cite{Ogata_1986}) and remains largely unexplored. To circumvent the issues that crop up due to the intractability of the likelihood, inference for Gibbs processes often relies on a pseudolikelihood construction due to~\cite{Besag_1975}, which uses a conditional property of Markov point processes to avoid the calculation of the normalizing constant~\cite{Baddeley_2000a,Jensen_1991,Jensen_1994}. Under strict structural assumptions on the intensity, this has been shown to work well in inhomogeneous settings for certain models~\cite{Ba_2023, Coeurjolly_2025}. However, statistical estimation for models with inhomogeneous intensities in more general non-parametric forms remains an open problem.

For GPPs, we consider a continuous area of observation, as discretizing the space does not lead to an obvious simplification of the problem as for DPPs. As in the case of DPPs, GPPs can be described through the combination of an intensity function and an interaction function. In the following, we will restrict our attention to pairwise interactions, although an extension to higher-order interactions is possible, see e.g.~\cite[Chapter 4.3]{vanLieshout_2000}. The probability density function for a configuration $\x \in \X^*$ is given by 
\begin{equation}\label{eq:gpp_density}
    f(\x) = Z^{-1} \prod_{x \in \x} \beta(x) \prod_{\{x, y\} \subset \x} \gamma(x, y),
\end{equation}
where $\beta \colon \X \rightarrow \R_{\geq 0}$ is the \emph{intensity function} and $\gamma \colon \X \times \X \rightarrow  \R_{\geq 0}$ is the \emph{interaction function}. A value $\gamma \equiv 1$ corresponds to no interaction, while values $\gamma < 1$ capture repulsive interactions, and $\gamma > 1$ capture attractive interactions. In the above $Z = Z(\beta, \gamma) > 0$ is a normalizing constant, which is in general intractable. This density function is defined with respect to the distribution of a unit-rate Poisson process, which results in the normalizing constant having the form
\begin{equation*}
    Z = \int_{\X^*} \prod_{x \in \mathbf{y}} \beta(x) \prod_{\{x, y\} \subset \mathbf{y}} \gamma(x, y) \d \mathbf{y} = \sum_{n=0}^\infty \frac{e^{-|\X|}}{n!} \int_\X \cdots \int_\X \prod_{i=1}^n \beta(u_i) \prod_{1 \leq i < j \leq n} \gamma(u_i, u_j) \d u_1 \cdots \d u_n.
\end{equation*}
The above statement (and in particular establishing the second equality) is not trivial, and requires a rigorous treatment as in~\cite[Chapters 5 and 7]{Daley_2003}~\cite[Chapters 9, 10 and 15]{Daley_2008}, as well as in~\cite{Dereudre_2019}. For our purposes, it is more relevant to focus on a conditional setting. Specifically, if we condition on the number of points in the configuration say $N(\x) = n$, the probability density function (with respect to the usual Lebesgue measure) becomes
\begin{equation}\label{eq:gpp_density_conditional}
    f(\x | N(\x) = n) = Z_n^{-1} \prod_{i=1}^n \beta(x_i) \prod_{1 \leq i < j \leq n} \gamma(x_i, x_j),
\end{equation}
where the (still intractable in general) normalizing constant is given by 
\begin{equation*}
    Z_n = \int_\X \cdots \int_\X \prod_{i=1}^n \beta(u_i) \prod_{1 \leq i < j \leq n} \gamma(u_i, u_j) \d u_1 \cdots \d u_n.
\end{equation*}
This is a very flexible modeling framework, allowing for essentially any combination of attractive and repulsive interactions. However, this flexibility comes at the cost of rather complicated sampling and inference procedures. The intractability of the density makes exact sampling computationally very expensive, in the cases where it is possible, and in many cases it remains an open question~\cite{Moller_2004,Berthelsen_2002}. For most models, one has to resort to approximate sampling methods, primarily MCMC procedures, see~\cite[Chapter 3]{vanLieshout_2000} or~\cite[Chapter 7]{Moller_2004}.

\subsection{Pseudolikelihoods for GPPs}

We approach the inference of the density of a GPP in a semiparametric manner, similar to what is done for DPPs. We again estimate the intensity function $\beta$ nonparametrically, and we parametrize the interaction $\gamma$ by a parameter vector $\theta \in \R^p$ for $p \in \N$. While for most standard choice of interaction functions the log-likelihood is concave, there is no closed form for the partition function. MCMC methods have been used to approximate the normalizing constant, but this approach is computationally quite expensive~\cite{Geyer_1994}. Therefore, a more common approach relies on using an approximation of the likelihood, Besag's pseudolikelihood~\cite{Baddeley_2000a,Besag_1975,Jensen_1991}. This approach gives a computationally attractive alternative to maximum likelihood, while retaining good statistical properties~\cite{Ba_2023,Jensen_1991}. In our setting, we choose to consider the process conditionally, given the number of points in each realization. This complicates things, as the standard form of the pseudolikelihood relies on the use of the \emph{Papangelou conditional intensity}~\cite[Chapter 10.4]{Daley_2008}, which is (informally) defined as the ratio 
\begin{equation*}
    \rho(y \vert \x) = \frac{f(\x \cup \{y\})}{f(\x)}
\end{equation*}
for a configuration $\x \in \X^*$ and a point $y \in \X \backslash \x$. This ratio does not exist for the conditional density defined in \eqref{eq:gpp_density_conditional}, as the configurations $\x \cup \{y\}$ and $\x$ have different sizes. Following the approach in~\cite[Appendix D]{Moller_2022} we can nevertheless use the original construction of~\cite{Besag_1975} to define a meaningful conditional pseudolikelihood in the following way.

Consider point configurations of size $n$. Given a random configuration $X\in\X_n^*$ of size $n$ define $X_{-i} = (X_1, \dots, X_{i-1}, X_{i+1}, \dots, X_n)$ for $i = 1, \dots, n$. Effectively this is almost the configuration $X$, but where point $X_i\in\X$ was removed. Analogously, define $\x_{-i} = (x_1, \dots, x_{i-1}, x_{i+1}, \dots, x_n)$ for a deterministic configuration. Note crucially that the random variables $X_{-1}, \dots, X_{-n}$ are identically distributed. Additionally, for any $i = 1, \dots, n$, conditional on $N(X) = n$ and $X_{-i} = x_{-i}$ the random variable $X_i$ depends only on $\x \backslash \{x_i\}$. Also, $X_i$ has a (conditional) probability density given by
\begin{equation*}
    f(x | \x \backslash \{x_i\}) = \tilde Z (\x \backslash \{x_i\})^{-1} \beta(x) \prod_{\substack{y \in \x \backslash \{x_i\} \\ y \neq x}} \gamma(x, y),
\end{equation*}
with
\begin{equation*}
    \tilde Z (\x \backslash \{x_i\}) = \int_{\X} \beta(u) \prod_{\substack{t \in \x \backslash \{x_i\} \\ t \neq u}} \gamma(u, t) \d u.
\end{equation*}
Notably, computing $\tilde Z (\x \backslash \{x_i\})$ requires only an integral over $\X$, which can be evaluated much more effectively than the normalizing constant in \eqref{eq:gpp_density_conditional}. Now consider a parametrization $\gamma_\theta$ for the interaction function $\gamma$ and take $\beta$ as fixed. The conditional log-pseudolikelihood function for the parameter vector $\theta$ given a single configuration $\x$ is then a product of the conditional densities $f(x | \x \backslash \{x_i\})$ for $i = 1, \dots, n$, given by 
\begin{equation*}
    \text{CPL}_n(\theta | \x) = \sum_{i=1}^n \left( \log \beta(x_i) + \sum_{{\substack{y \in \x \backslash \{x_i\}}}} \log \gamma_\theta (x_i, y) - \log \tilde Z(\x \backslash \{x_i\}; \theta) \right).
\end{equation*}
We can omit the intensity term $\sum_{i=1}^n \log \beta(x_i)$ in the optimization, since it does not depend on the parametrization $\theta$. However, note that the terms involving $\tilde Z$ still depend both on $\theta$ and $\beta$. Nevertheless, this expression can still be evaluated effectively, as it does not require the computation of high-dimensional integrals. Finally, given $T$ independent configurations $\x_1, \dots, \x_T$, the conditional log-pseudolikelihood is the sum over the individual log-pseudolikelihoods.

\section{Semiparametric inference of spatial point processes with replicated spatial patterns: a novel methodology}\label{sec:inference}

In this section, we begin in~\ref{subsec:cv} by proposing a statistical inference approach to estimate the intensity and interaction function of a spatial point process in the presence of replicated spatial patterns. Using these independent realizations from the same process, we can leverage cross-validation methods to select the bandwidth for the nonparametric kernel estimate of the intensity function, introduced in Section~\ref{sec:poisson}. Then, we use a likelihood-based approach to estimate the parameters of a parametric interaction function. In~\ref{subsec:numerics}-\ref{subsec:numerics_results} we then test this procedure in a controlled setting, generating synthetic data to assess stability and consistency of the estimation method. This sets up the application to the dataset of waiting pedestrians in Section~\ref{sec:application}.

\subsection{A unified cross-validation approach to intensity and interaction inference}\label{subsec:cv}

In this subsection, we present an approach to iteratively estimate the intensity and interaction, for both DPPs and GPPs. As previously discussed in Section~\ref{sec:poisson}, we would like to estimate the background intensity nonparametrically. To this end, we use kernel smoothing methods with likelihood-based bandwidth selection, just as we did for processes without interaction in~\ref{subsec:poisson_inference}. However, we would like to incorporate some knowledge of the interaction into our choice of the kernel bandwidth. This leads to a $k$-fold cross validation approach, in which we alternate between estimating the parameter (or parameters) of interaction and the bandwidth for the intensity estimate. The approach can be summarized as follows for general spatial point processes with density as in \eqref{eq:spp_density}:
\begin{algorithm}\label{alg:cv}
    Input: observations $\{\x_1, \dots, \x_T\}$, parametric interaction model $\Gamma_\theta$
    \begin{enumerate}
        \item Split the observations into $k$ folds of training data and test data. For each fold:
        \begin{enumerate}[label=(\roman*)]
            \item Estimate an initial bandwidth $h_0$ for the KDE under the assumption that there is no interaction, and generate an intensity estimate $\hat\beta_{h_0}$
            \item Plug-in the estimated intensity and maximize the log-likelihood (or log-pseudolikelihood) under $f_{\theta, h_0}$ for $\theta$, obtaining an estimate $\hat\theta_0$
            \item Let $\{c \cdot h_0 \colon c \in C\}$ be a set of candidate bandwidths, where $C$ is a set of coefficients. From this, generate $|C|$ candidate models using the intensities obtained from these bandwidths and the interaction from $\hat\theta_0$
            \item Evaluate each candidate model by calculating the log-likelihood (log-pseudolikelihood) of the test set under each model
        \end{enumerate}
        \item Sum the log-likelihoods (log-pseudolikelihoods) for each coefficient across folds and choose the coefficient with the maximum value, $\hat{c}$
        \item Re-estimate a bandwidth $h_0^{all}$ using Poisson likelihood cross-validation from all observations $\{\x_1, \dots, \x_T\}$, and get a final choice of bandwidth $\hat{h} = \hat{c} \cdot h_0^{all}$
        \item Using the intensity estimate $\beta_{\hat{h}}$, estimate interaction parameters $\hat{\theta}$ by maximizing the log-likelihood (log-pseudolikelihood) over the observations
    \end{enumerate}
    Return $(\hat{h}, \hat{\theta})$.
\end{algorithm}

For DPPs, since we consider a discrete ground set, the initial estimate of the bandwidth is done by assuming the kernel matrix is diagonal (i.e. no interaction), and then running a grid search over candidate bandwidths, again employing a cross-validation with an arbitrary number of folds $k < T$. When the ground set is continuous, as is the case for GPPs, we obtain an initial bandwidth estimate using leave-one-out cross validation under Poisson likelihood, as in Section~\ref{subsec:poisson_inference}. It is worth noting that choosing an initial bandwidth is not strictly necessary, as we could just as well do a grid search from the start. However, it gives us a reasonable starting point that is also based on maximizing likelihood, ensuring comparability. Algorithm~\ref{alg:cv} can be iterated more than once by using the estimated $\hat{h}$ and $\hat{\theta}$ of the first iteration as starting bandwidths and interaction values within folds to further refine the choice. For DPPs and GPPs, we test this procedure on synthetic data in Section~\ref{subsec:numerics} and demonstrate its performance on the train station data in Section~\ref{sec:application}. See Appendix~\ref{app:stability} for a brief discussion of the numerical stability of Algorithm~\ref{alg:cv} across iterations.

\subsection{Performance analysis in a controlled setting}\label{subsec:numerics}

The inference approach discussed in Section~\ref{subsec:cv} is a novel way to separate the influences of intensity and interaction. To evaluate the performance of this semiparametric method, we run a series of experiments in a controlled setting. Synthetic data is generated from pre-specified models, after which the stability of the estimation procedure described in Algorithm~\ref{alg:cv} is tested by running it on batches of configurations and comparing results to gauge the variance of the estimator. Then, the consistency of the estimation is tested by running the procedure on batches of increasing size, i.e. at each step the previous batch is augmented by a fixed number of samples. 

For both the discrete (DPP) and the continuous (GPP) setting, we select a simple inhomogeneous intensity function to be a mixture of Gaussian bumps. Specifically, we choose a mixture of four two-dimensional Gaussian probability density functions, each of which has a different location (mean) and variance. Each component is multiplied by a different mixture weight, modeling different scales of intensity across the observation area. This choice of a smooth, multiscale intensity function covers many physical settings, in which locations attractiveness varies smoothly and has different strengths in different areas~\cite{Baddeley_2015,Ba_2023,Cronie_2019}. In particular, it mirrors what we see in the train station data (see Section~\ref{sec:data}), at least when we restrict our view to certain sections of the platform. When boundaries are known, such as when considering only a predetermined rectangle within the platform area, we expect a certain degree of smoothness in the spatial inhomogeneity, which is reflected by this choice of underlying intensity.

We simulate two synthetic datasets for each process. First, we simulate from a process without interaction. For discrete DPPs, this corresponds to simulating from a DPP with a diagonal DPP-kernel (equivalent to sampling the points of the lattice without replacement according to discrete probability distribution given by the intensity). For GPPs, this corresponds to simulating from an inhomogeneous Poisson process. Testing the estimation on these datasets constitutes a kind of sanity check, where the goal is to ensure that the inference is able to detect that no interaction is present, or at least interpret this as very weak interaction. Second, we simulate data from each interactive model with specified parameters. The goal of this experiment is to verify whether the procedure we propose is able to recover the parameters when there is no model mismatch.

\subsection{Formal setup of the experiments}

\begin{figure}[t]
    \centering
    \includegraphics[width=\textwidth]{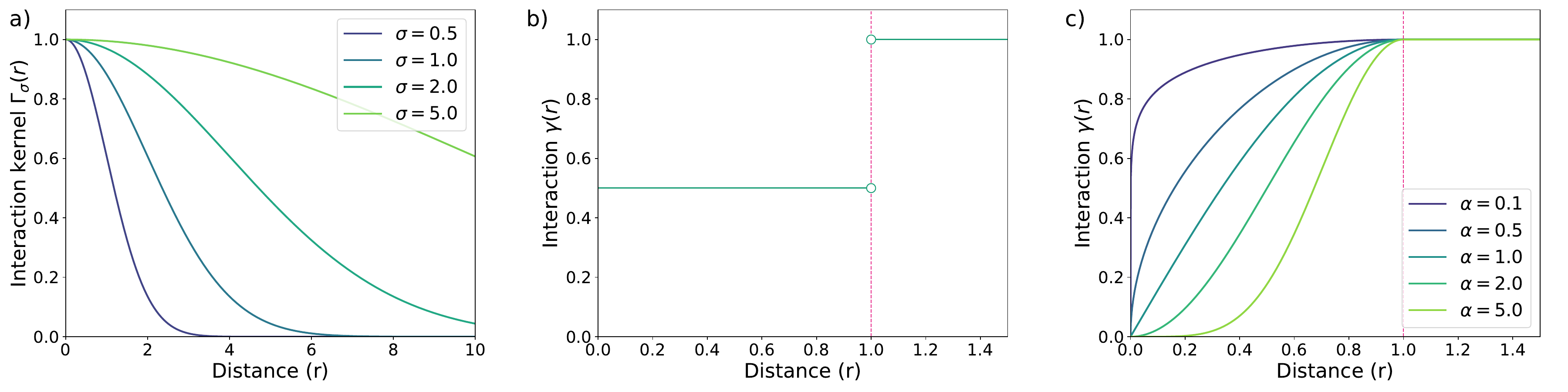}
    \caption{Depiction of the interaction components of the DPP (a) and GPP (b-c) models. Note that, due to different parametrizations, for the DPP model values close to $0$ correspond to no interaction, and for the GPP models no interaction corresponds to values close to $1$. Panel (a) shows a Gaussian DPP interaction kernel with multiple values of $\sigma$. At a fixed distance $r$, a higher value represents stronger repulsion, which decays with distance. A value of $1$ (achieved only at $r=0$) indicates full exclusion. (b) Strauss interaction function with $R=1$ and $\vartheta=0.5$. As illustrated, $\vartheta$ controls the strength of interaction. (c) DGS interaction function with $R=1$ and multiple values of $\alpha$, illustrating the role of $\alpha$ as a shape parameter. As for DPPs, interaction strength decays with distance.}
    \label{fig:interactions}
\end{figure}

We begin with the discrete setting, investigating the performance of our estimation procedure for DPPs on a regular lattice $G = \{1,...,N\} \times \{1,...,M\}$ for $N, M \in \N$. We model the interaction by using a Gaussian kernel, which for $\sigma > 0$ and $x, y \in G$ is given by 
\begin{equation}\label{eq:gaussian_kernel}
    \Gamma_{\sigma} (x, y) = \exp \left( -\frac{\|x - y \|_2^2}{2\sigma} \right),
\end{equation}
where $\| \cdot \|_2$ denotes the Euclidean norm. This results in a one-parameter model for the interaction. The interaction kernel is visualized for several values of $\sigma$ in Figure~\ref{fig:interactions}(a). Unlike Gibbs interaction functions, a higher value of $\Gamma_\sigma$ indicates a stronger repulsion, which decays as the distance between points increases.

For this experiment, we choose $G$ to be a two-dimensional $25 \times 25$ point lattice. We define the ground-truth intensity in two coordinates as \(\tilde{\beta}_0 : \mathbb{R}^2 \to \mathbb{R}\), given by
\[ \tilde{\beta}_0(x,y)
= \sum_{k=1}^4 
w_k \exp\!\left(
    -\frac{(x-\mu_{k,1})^2 + (y-\mu_{k,2})^2}{2v_k^2}
\right)
+ b,
\]
where
\[
\mu = \big((30,30),\, (70,70),\, (30,70),\, (70,30)\big);
\ 
v = (7.5,\, 10,\, 5,\, 5);
\ 
w = (0.8,\, 1.2,\, 0.6,\, 1.0);
\  
b = 0.01.
\]
Finally, let $\beta_0: \X \to \mathbb{R}$ be the function assigning the intensities to each point of the lattice $\X$ by evaluating $\tilde{\beta}_0$ and then transforming to one dimension by ordering lattice positions lexicographically from top-left to bottom-right. When we disregard interaction, the DPP-kernel $L$ is then given simply by $\text{diag}(\beta_0(1), \dots, \beta_0(625))$. When we include interaction, we get 
\[L_{ij} = \sqrt{\beta_0(i)} \Gamma_{\sigma}(i, j) \sqrt{\beta_0(j)};\quad i, j \in \{1, \dots, 625\}.\]
Note that $\beta_0$ is not a probability density, and its scale is chosen so that the expected number of points in a sample from this DPP is comparable with the scenarios stemming from the pedestrian data we will model. In particular, the expected cardinality can be explicitly calculated (see e.g.~\cite[Section 2.3]{Kulesza_2012}), and is approximately $60.35$ when simulating from the process without interaction, and $21.15$ when simulating with interaction, setting $\sigma = 2.5$ in \eqref{eq:gaussian_kernel}. We simulate 10 000 samples from both processes for testing. 

For the Gibbs processes, we consider two different models based on different interaction functions. First, we consider a \emph{Strauss interaction function}, which takes a constant value between $0$ and $1$ up to a fixed radius of interaction $R>0$. Formally, for a configuration of points $\x \in \X^*$ where $\X = \R^2$, this leads to a density function for the model given by
\begin{equation*}
    f(\x) \propto \beta(\x) \vartheta^{s(\x)}, \quad \text{where } s(\x) = \sum_{\{x, y\} \subset \x} \1\{\|x - y\| < R\}.
\end{equation*}
Here, $s(\x)$ is the number of pairs in $\x$ at distance less than $R$ and $\vartheta \in (0, 1)$ is a constant. This model has two parameters, $\vartheta$ and $R$, see Figure~\ref{fig:interactions}(b) for reference. As described in~\cite{Baddeley_2000a}, estimating $\vartheta$ is easier than estimating $R$, since the family of densities is only exponential with respect to $\vartheta$. However, we can use a profile pseudolikelihood approach to estimate both parameters, running a grid search over the $\vartheta$ values for each choice of $R$ and choosing the combination with the highest pseudolikelihood.

The second model we consider is a modified version of the \emph{Diggle-Gates-Stibbard (DGS) interaction}~\cite{Diggle_1987}, which interpolates smoothly between $0$ and $1$ in a range of values specified by a finite interaction radius $R>0$, as before. We add a parameter $\alpha \geq 0$ which controls the shape of this curve, providing more flexibility. Formally, the modified DGS interaction function has the form
\begin{equation*}
    \gamma(x, y) = 
    \begin{cases}
    \Biggl( \sin \Bigl( \dfrac{\pi \, \|x - y\|}{2 R} \Bigr) \Biggr)^{2\alpha}, & \text{if } \|x - y\| < R, \\[2mm]
    1, & \text{if } \|x - y\| \ge R.
    \end{cases}
\end{equation*}
See Figure~\ref{fig:interactions}(c) for a visualization. Estimating $\alpha, R$ comes with the same caveat as before, in that we need to resort to profile pseudolikelihood estimation to get a combined estimate. In the following we will refer to the modified DGS model simply as the DGS model.

For the continuous models, consider an observation window given by $\X = [0, 100] \times [0, 100]$. Define the ground truth intensity as $\beta_0 : \X \to \R$, given by
\[ \beta_0(x,y)
= \sum_{k=1}^4 
w_k \exp\!\left(
    -\frac{(x-\mu_{k,1})^2 + (y-\mu_{k,2})^2}{2v_k^2}
\right)
+ b
\]
where
\[
\mu = \big((30,30),\, (70,70),\, (30,70),\, (70,30)\big);
\ 
v = (7.5,\, 10,\, 5,\, 5);
\ 
w = (0.04,\, 0.06,\, 0.03,\, 0.05);
\ 
b = 0.001.
\]
For GPPs, we can immediately take $\beta_0$ to be the intensity function. Note that this is the same ground truth as for DPPs, except for the weights and baseline value being reduced by a factor of $10$, so that the expected number of points is again reasonable. We simulate $10 000$ samples from an inhomogeneous Poisson process on $\X$ with intensity $\beta_0$, as well as from Strauss/DGS models with interaction radii of $R=5$ and parameters $\vartheta = 0.5, \alpha = 2$. Empirically, we get an average number of points per sample of $74.46$ for the process without interaction. Simulating from the Strauss process results in an average number of points per sample of $67.54$, and the DGS model has an average of $69.53$.

\subsection{Results: stability and consistency}\label{subsec:numerics_results}

\begin{figure}[p]
    \begin{minipage}{\textwidth}
    \centering
    \begin{subfigure}[ht]{0.49\textwidth}
        \centering
        \includegraphics[width=\textwidth]{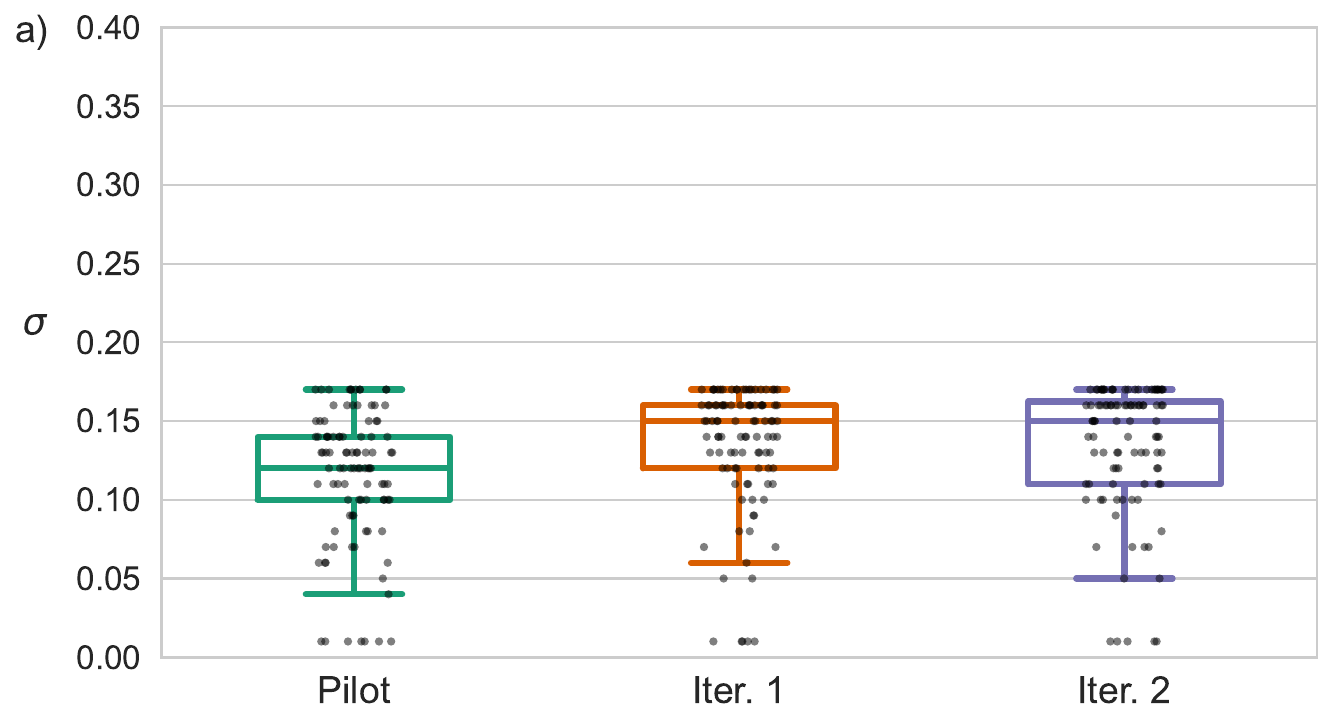}
    \end{subfigure}
    \begin{subfigure}[ht]{0.49\textwidth}
        \centering
        \includegraphics[width=\textwidth]{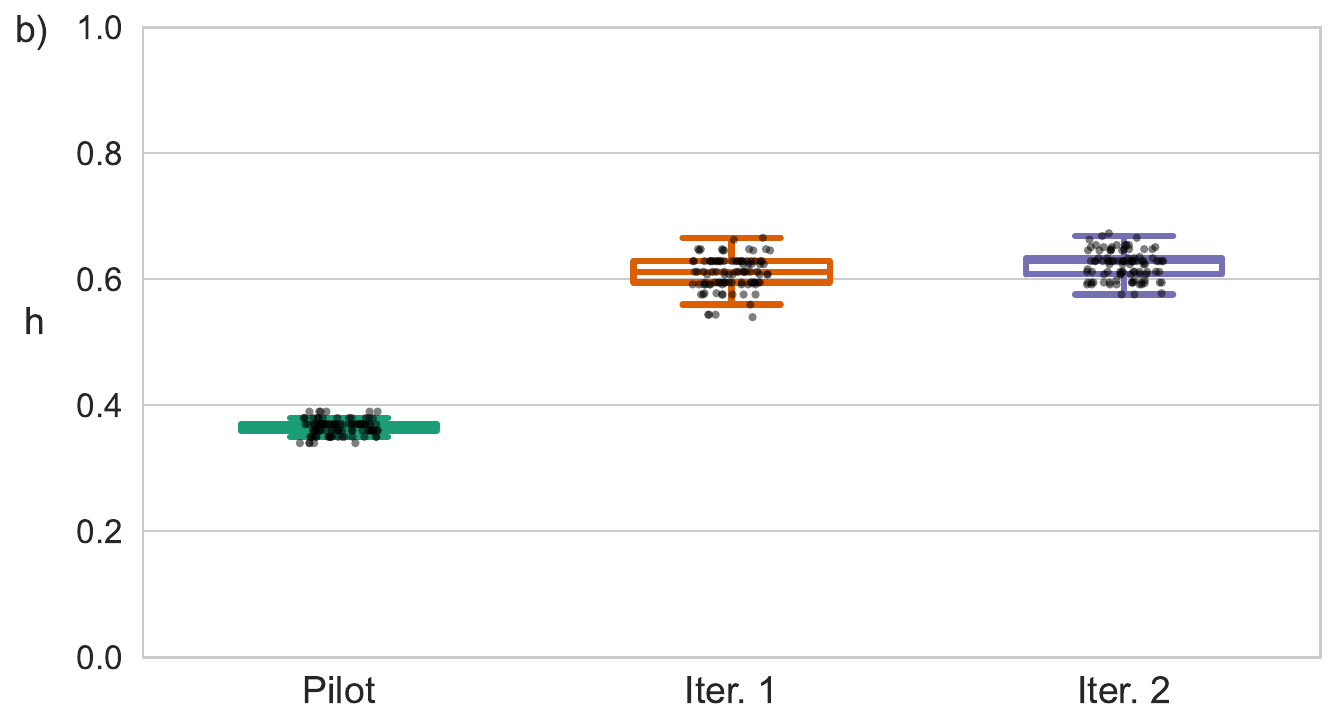}
    \end{subfigure}
    \begin{subfigure}[ht]{0.49\textwidth}
        \centering
        \includegraphics[width=\textwidth]{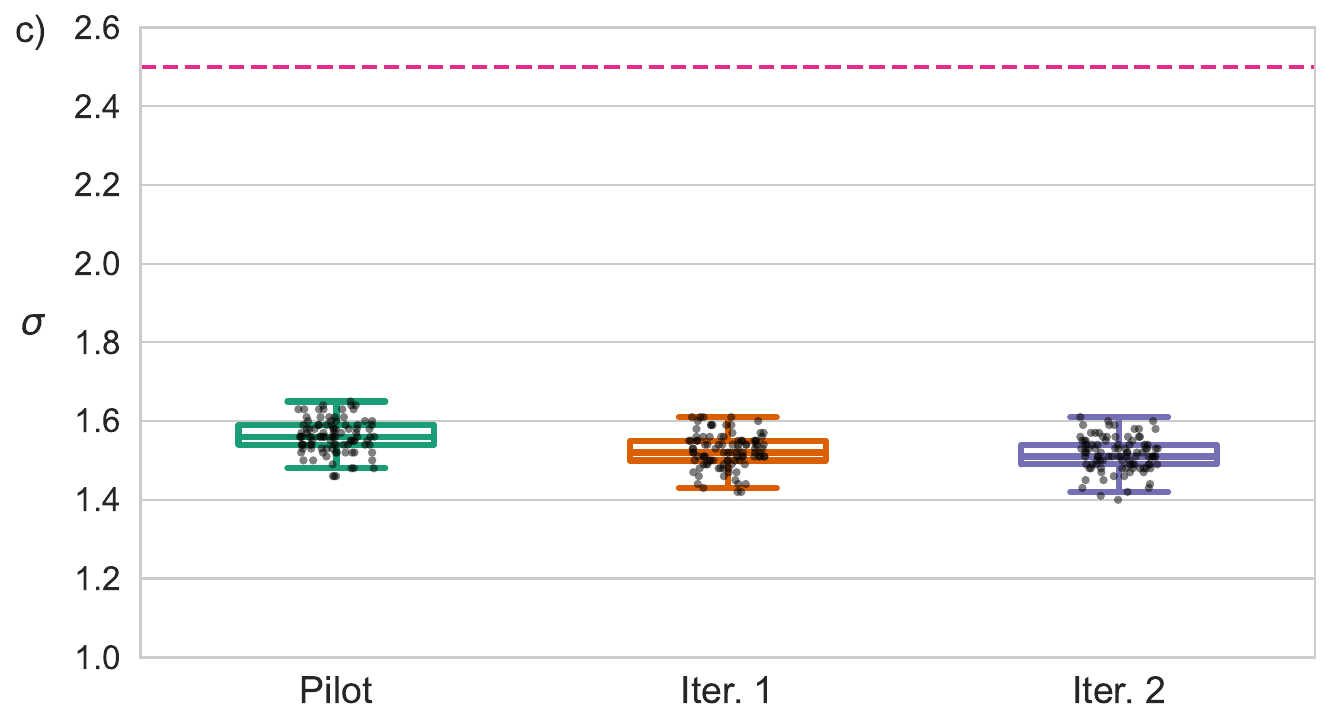}
    \end{subfigure}
    \begin{subfigure}[ht]{0.49\textwidth}
        \centering
        \includegraphics[width=\textwidth]{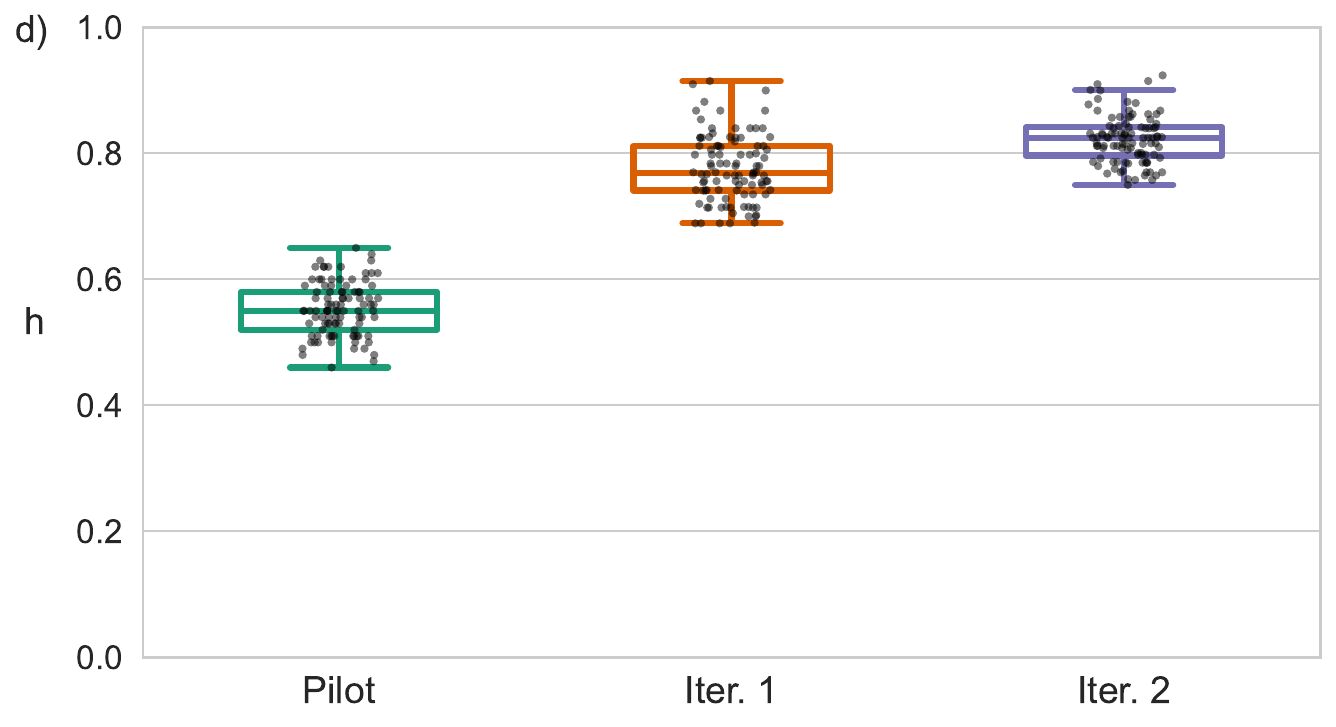}
    \end{subfigure}
    \caption{Distribution of parameter estimates for batched DPP experiments without interaction ((a) and (b)) and with true interaction parameter $\sigma = 2.5$ ((c) and (d)). (a) and (c) show estimates for $\sigma$, by iteration of the estimation procedure in Algorithm~\ref{alg:cv}. For experiment with interaction (c), true parameter as dashed line. (b) and (d) show estimates of the bandwidth $h$ for the inferred intensity. While estimation of $h$ is quite stable within each iteration, the estimator of $\sigma$ has higher variability. Nevertheless, estimated values are generally close to $0$ for the case without interaction, which indicates little interaction is being detected. For the interactive case, the range of $\sigma$ is relatively small, yet the estimate is rather biased.}
    \label{fig:dpp_boxplots}
    \end{minipage}

    \begin{minipage}{\textwidth}
    \centering
    \begin{subfigure}[ht]{0.49\textwidth}
        \centering
        \includegraphics[width=\textwidth]{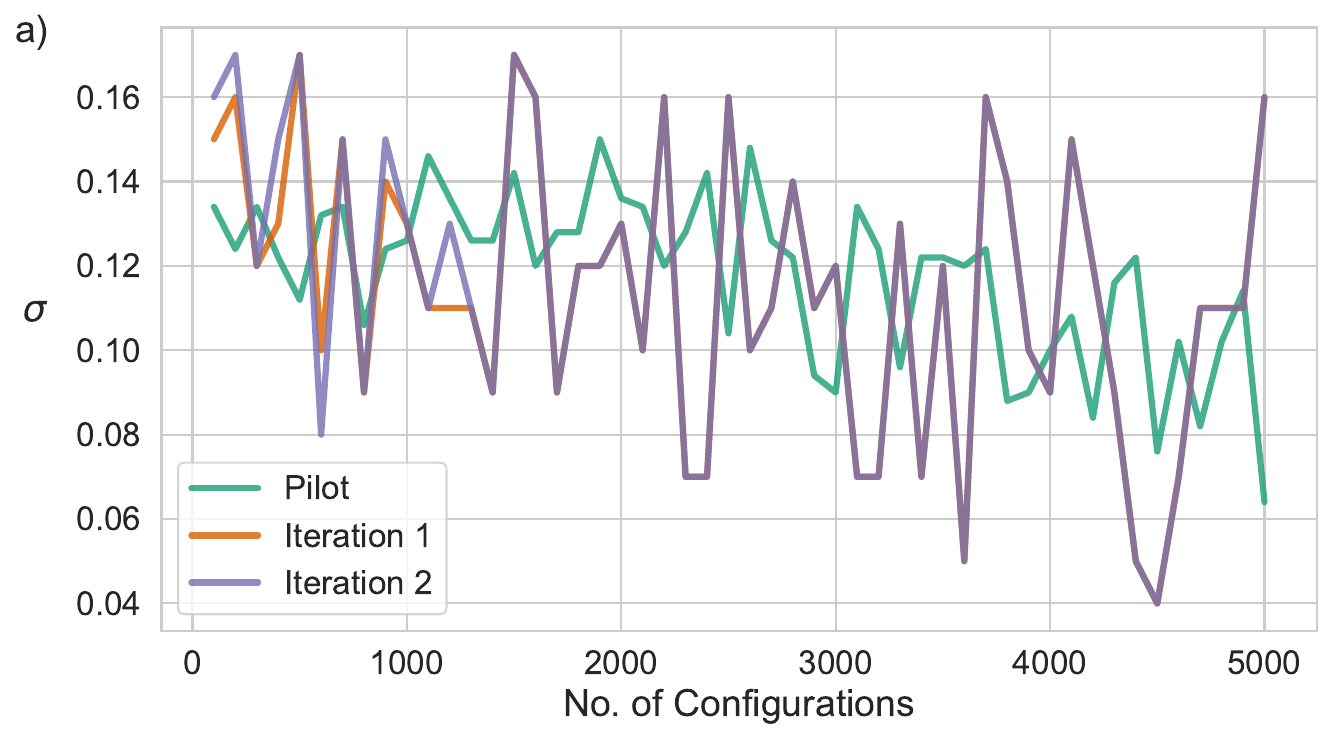}
    \end{subfigure}
    \begin{subfigure}[ht]{0.49\textwidth}
        \centering
        \includegraphics[width=\textwidth]{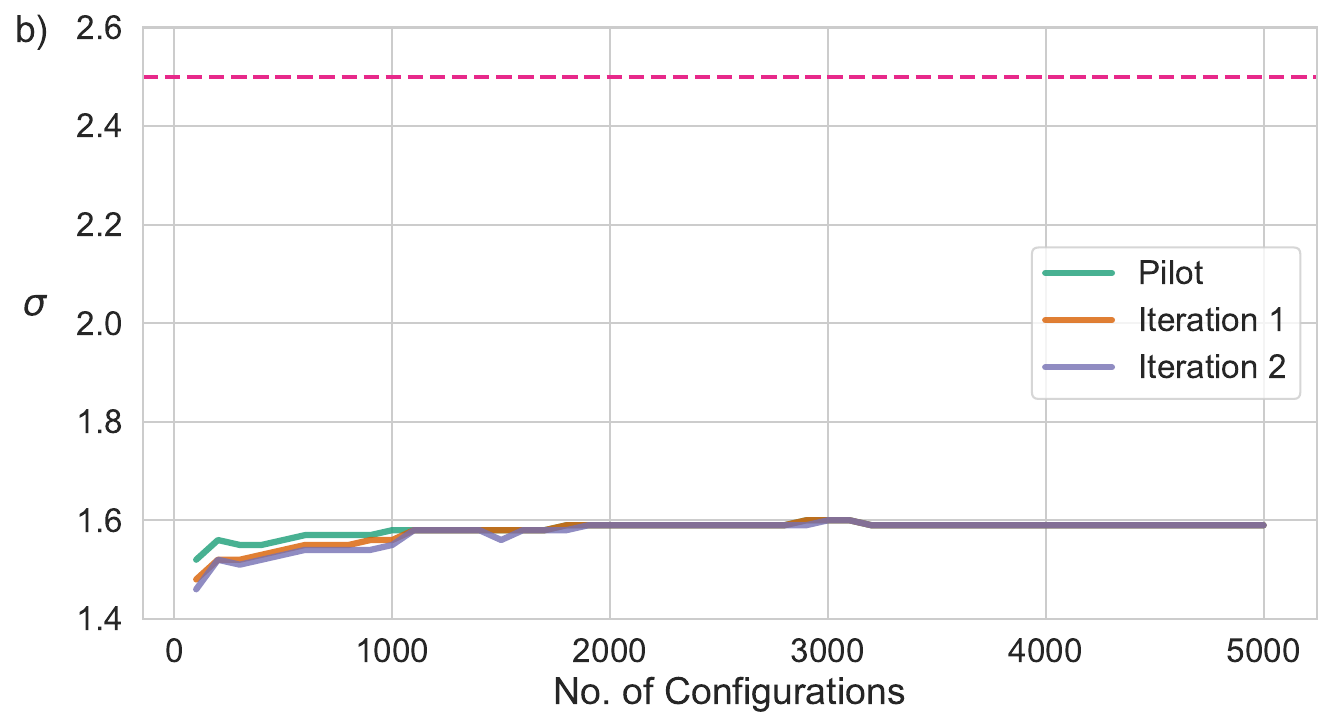}
    \end{subfigure}
    
    \caption{Convergence of $\sigma$ during consistency experiments. (a) convergence for no-interaction data. Observe slow trend towards $0$. (b) convergence for data generated from a DPP with true $\sigma$ of $2.5$ (dashed line). While convergence can be clearly observed, the estimate is biased. In both cases, when the number of configurations is large, the estimates for iteration 1 and 2 coincide, as the bandwidth is no longer altered between iterations.}
    \label{fig:dpp_consistency}
    \end{minipage}
\end{figure}

\begin{figure}[p]
\centering

\begin{minipage}{\textwidth}
\centering
\begin{subfigure}{0.32\textwidth}
    \includegraphics[width=\textwidth]{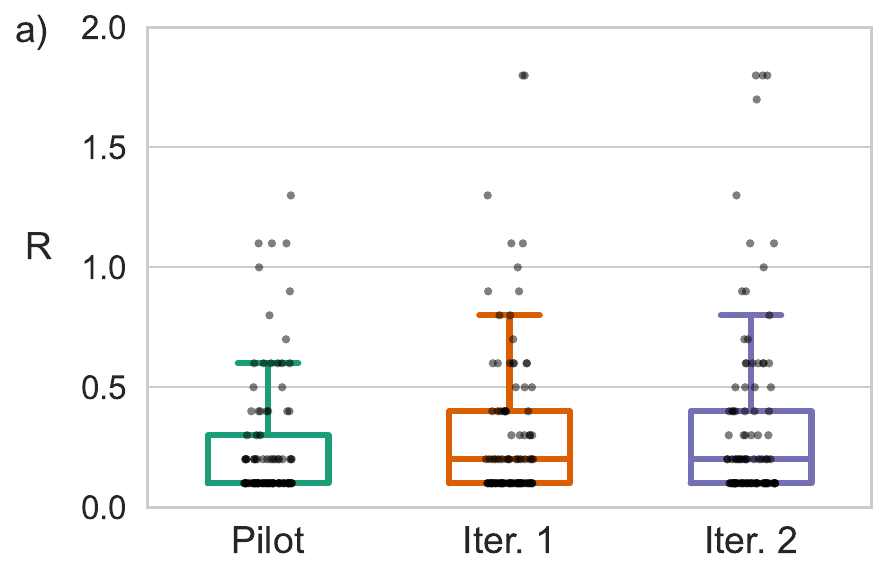}
\end{subfigure}
\begin{subfigure}{0.32\textwidth}
    \includegraphics[width=\textwidth]{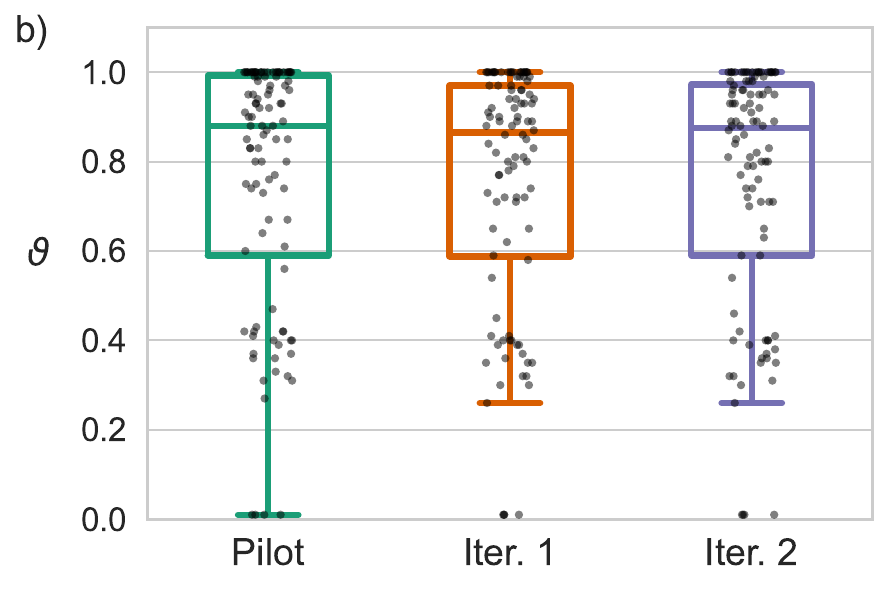}
\end{subfigure}
\begin{subfigure}{0.32\textwidth}
    \includegraphics[width=\textwidth]{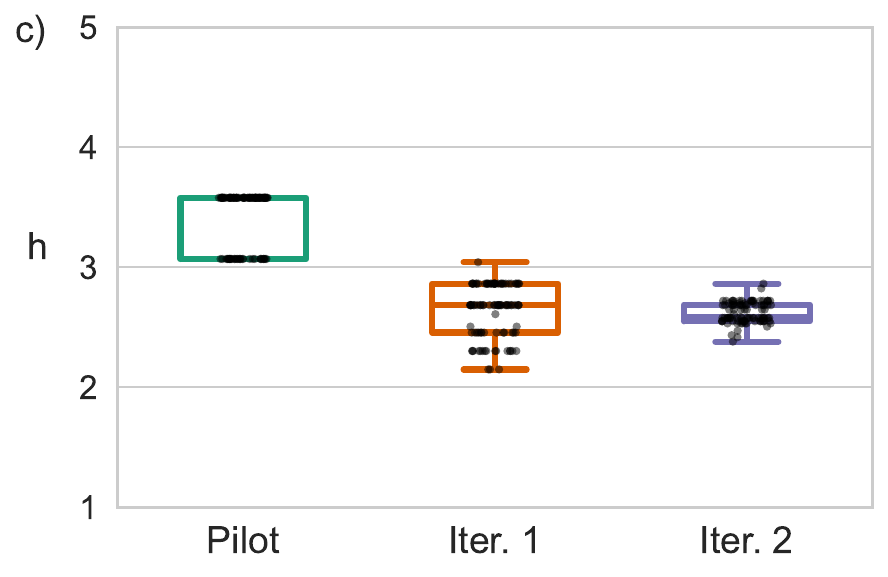}
\end{subfigure}

\begin{subfigure}{0.32\textwidth}
    \includegraphics[width=\textwidth]{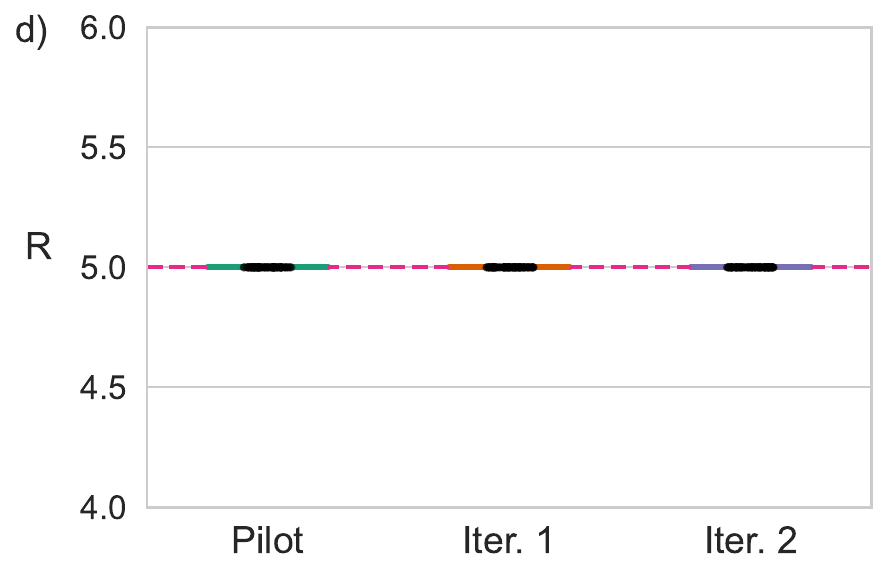}
\end{subfigure}
\begin{subfigure}{0.32\textwidth}
    \includegraphics[width=\textwidth]{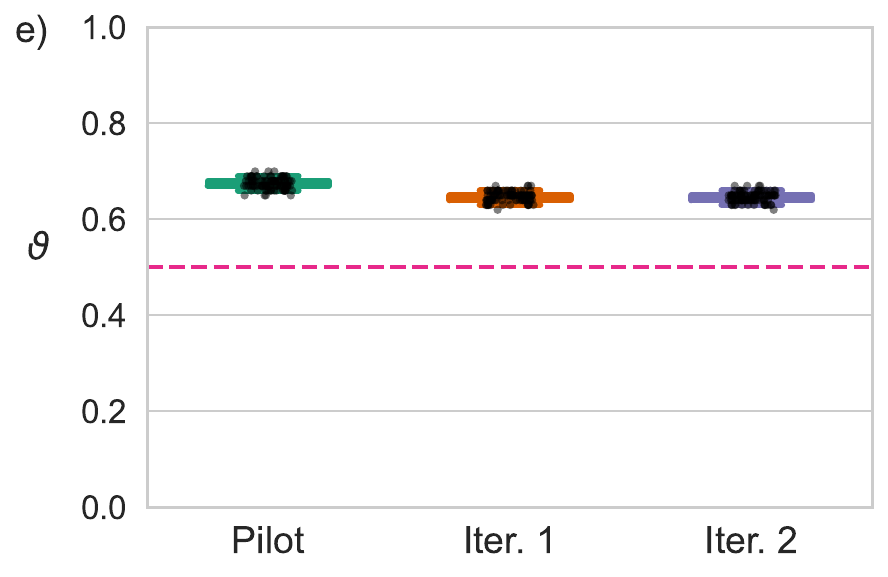}
\end{subfigure}
\begin{subfigure}{0.32\textwidth}
    \includegraphics[width=\textwidth]{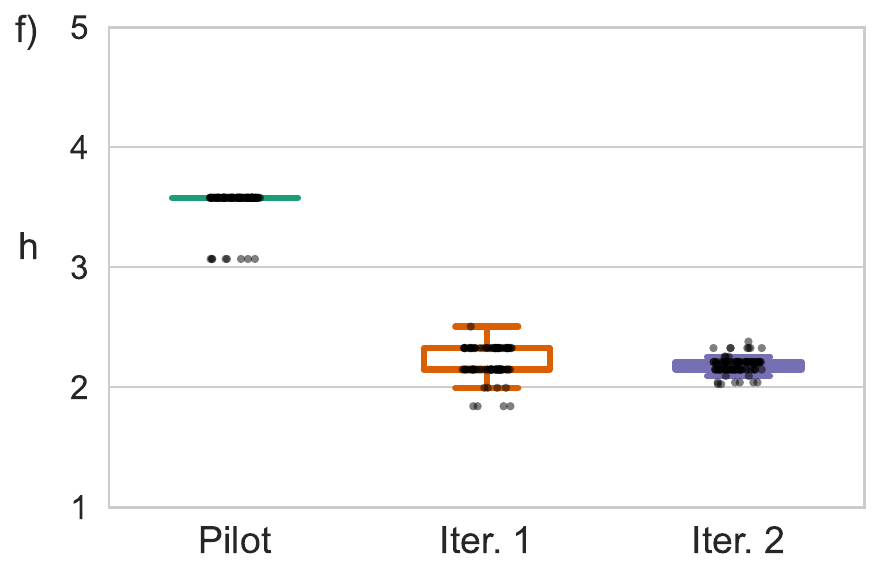}
\end{subfigure}

\captionof{figure}{Distribution of $R, \vartheta$ estimates and bandwidth $h$ for batched Strauss experiments without interaction (a-c) and with true parameters $R=5, \vartheta=0.5$ (d-f), with each boxplot corresponding to an iteration of Algorithm~\ref{alg:cv}. In the no interaction case, either a small $R$ or $\vartheta$ close to $1$ is, on average, correctly estimated. When interaction is present, estimates remain quite stable, but there is some bias in the estimation of $\vartheta$.}
\label{fig:strauss_boxplots}
\end{minipage}

\vspace{1em}

\begin{minipage}{\textwidth}
\centering
\begin{subfigure}{0.32\textwidth}
    \includegraphics[width=\textwidth]{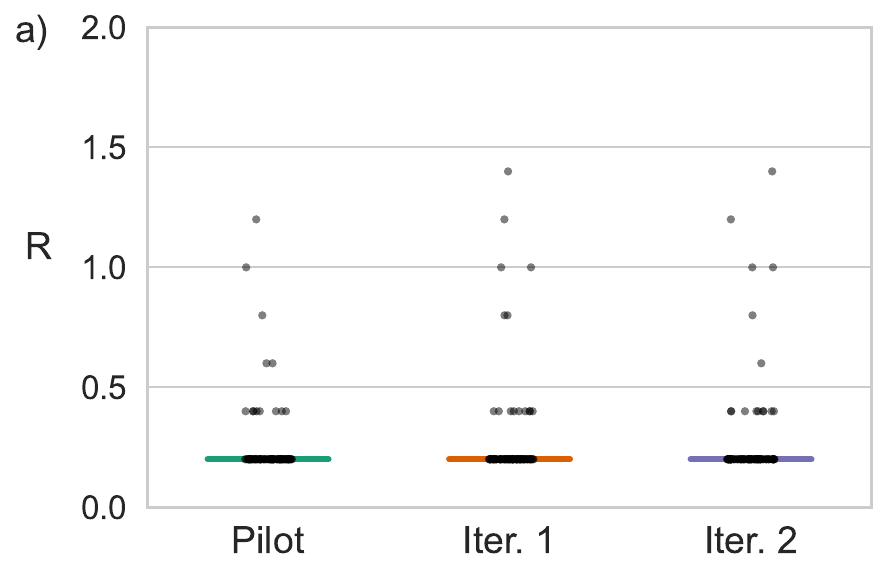}
\end{subfigure}
\begin{subfigure}{0.32\textwidth}
    \includegraphics[width=\textwidth]{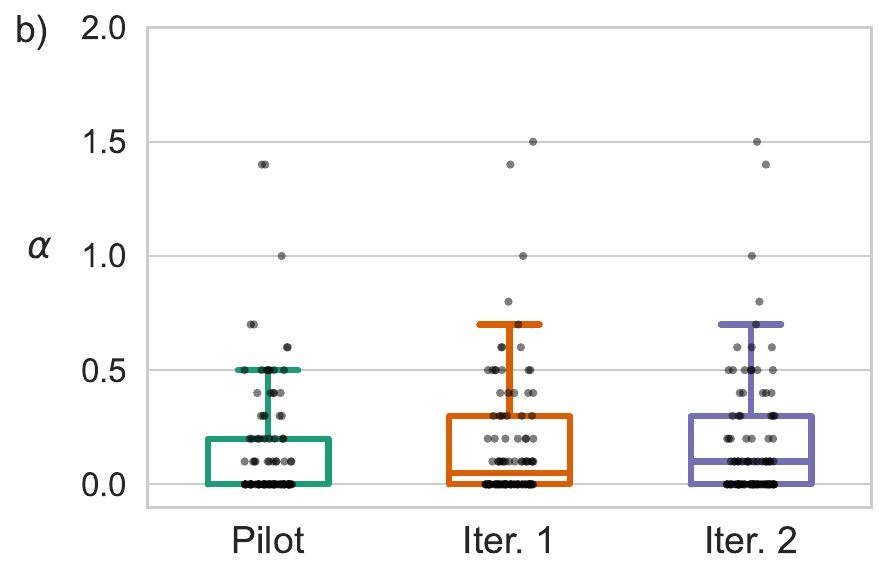}
\end{subfigure}
\begin{subfigure}{0.32\textwidth}
    \includegraphics[width=\textwidth]{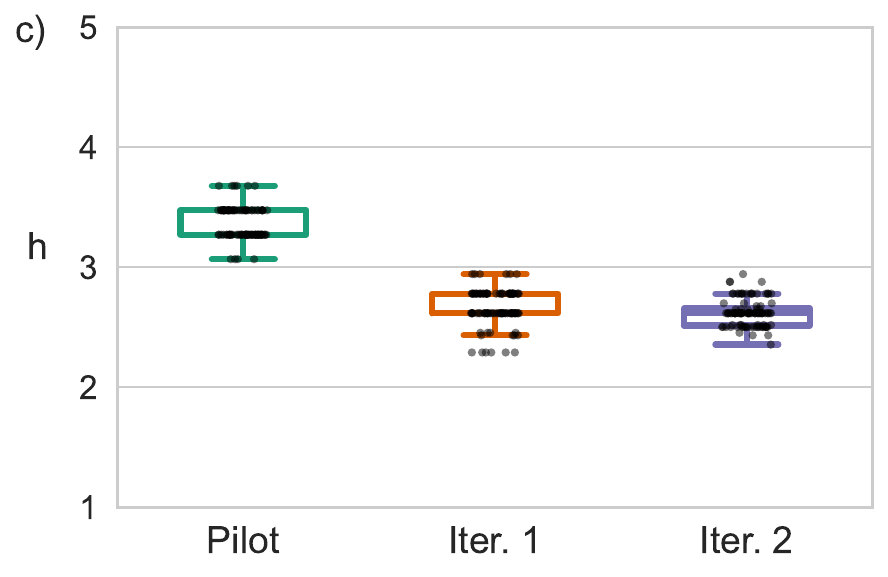}
\end{subfigure}

\begin{subfigure}{0.32\textwidth}
    \includegraphics[width=\textwidth]{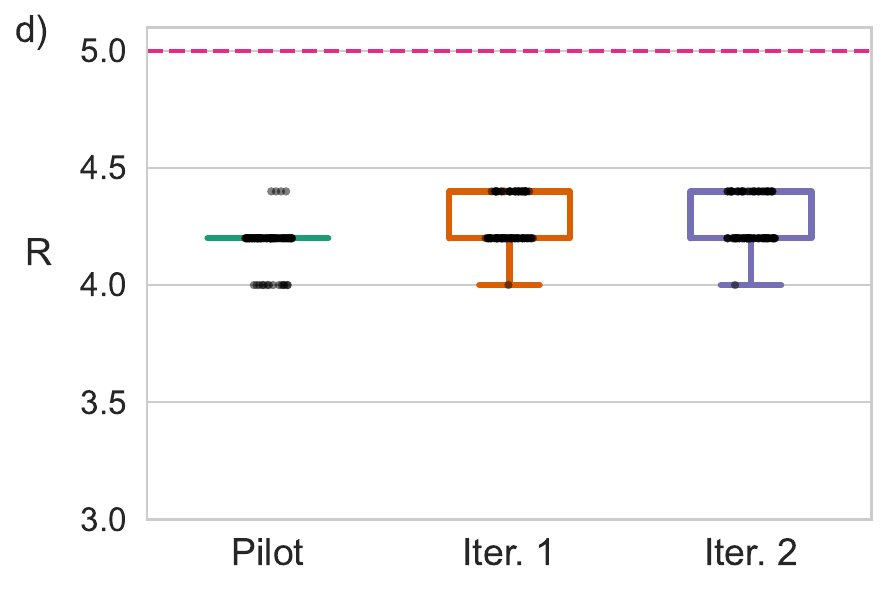}
\end{subfigure}
\begin{subfigure}{0.32\textwidth}
    \includegraphics[width=\textwidth]{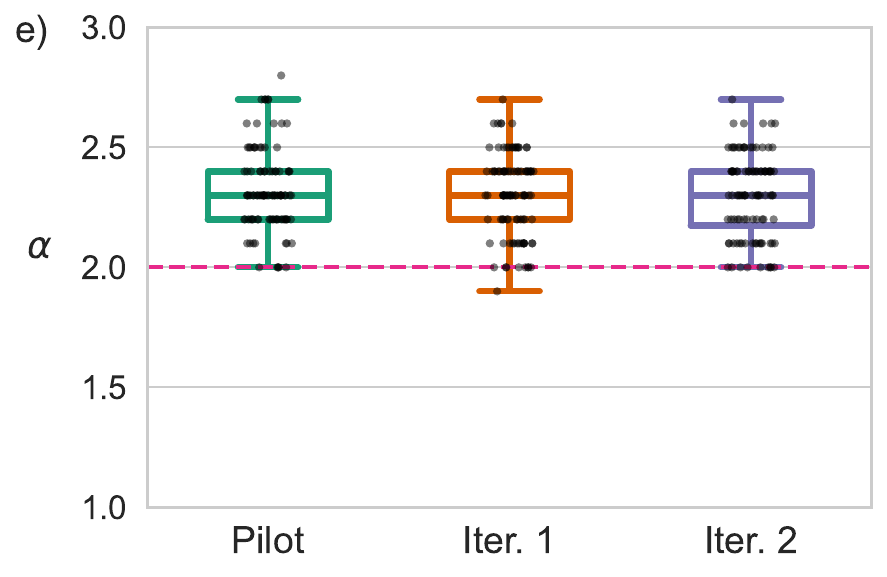}
\end{subfigure}
\begin{subfigure}{0.32\textwidth}
    \includegraphics[width=\textwidth]{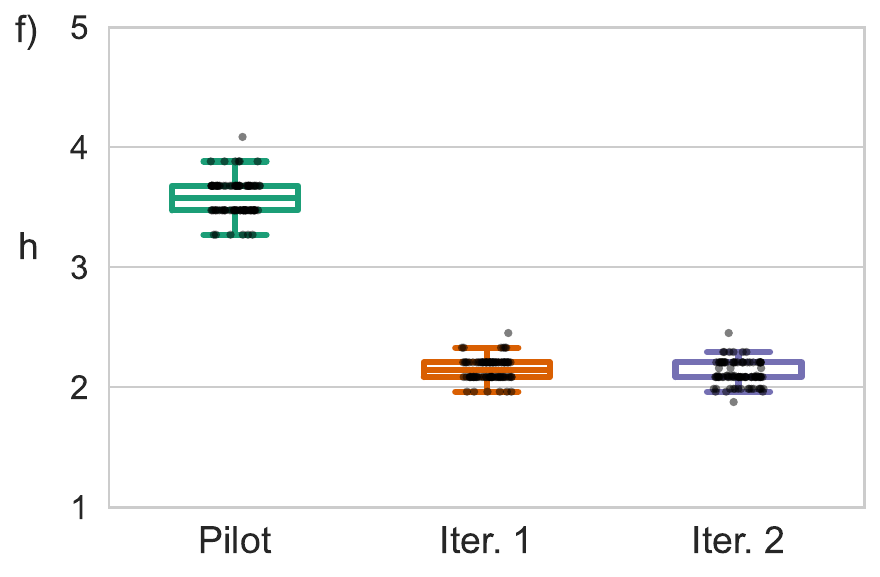}
\end{subfigure}

\captionof{figure}{Distribution of $R, \alpha$ estimates and bandwidth $h$ for batched DGS experiments without interaction (a-c) and with true parameters $R=5, \alpha=2$ (d-f). Each boxplot corresponds to an iteration of Algorithm~\ref{alg:cv}. When no interaction is present, the procedure detects this by estimating $R$ (a) close to $0$, leading to $\alpha$ (b) no longer being identifiable. With interaction, both $R$ and $\alpha$ estimates are quite stable, but an underestimation of $R$ (d) leads to $\alpha$ being slightly overestimated (e).}
\label{fig:dgs_boxplots}
\end{minipage}

\vspace{1em}

\begin{minipage}{\textwidth}
\centering
\includegraphics[width=\textwidth]{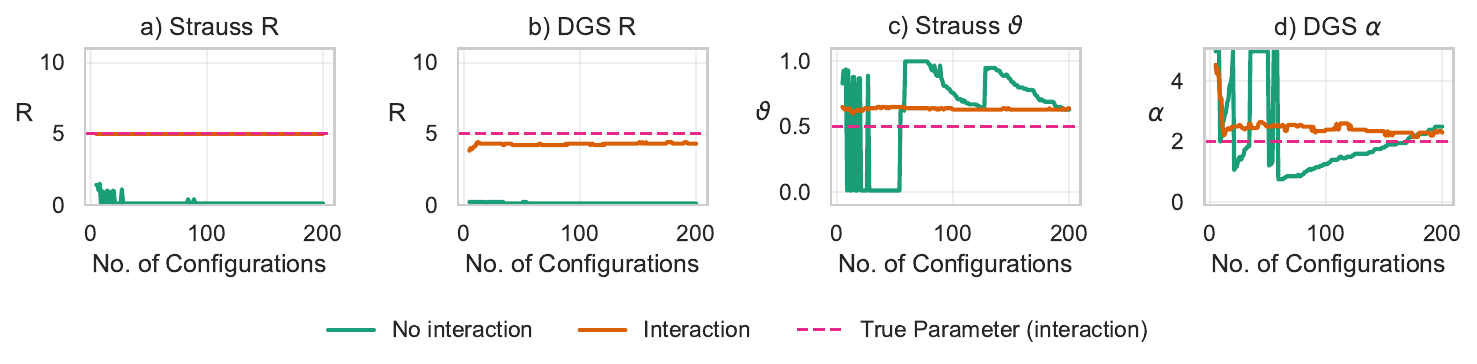}
\captionof{figure}{Convergence of parameter estimates for Strauss ((a) and (c)) and DGS models ((b) and (d)). For interactive case, true parameter marked with dashed line. Strauss model estimates $R$ well, yet shows some bias for $\vartheta$ in interactive settings. DGS model shows some bias in both the estimation of $R$ and of $\alpha$ in the interactive case. In the non-interactive case, both models are able to identify a lack of repulsion either through a small $R$ or corresponding $\vartheta$ / $\alpha$.}
\label{fig:gpp_consistency}
\end{minipage}

\end{figure}

For both DPPs and GPPs, we run the same stability and consistency experiments on two different datasets, one generated from a process without interaction, and one generated from the corresponding interactive model. For the stability experiments, we generate $10000$ samples from each model, and then run the estimation procedure on batches of $100$ samples each. Results are summarized in Table~\ref{table:dpp_numerics_results} (for DPPs) and Table~\ref{table:gpp_numerics_results} (for GPPs). In addition, we visualize results in Figures~\ref{fig:dpp_boxplots},~\ref{fig:strauss_boxplots} and~\ref{fig:dgs_boxplots}. To investigate the consistency of the estimation, the approach differs slightly between DPPs and GPPs. For DPPs, we increase the size of the dataset from which we estimate by $100$ samples each time, from $100$ to $5000$ samples. For GPPs, the procedure converges much more quickly, and we therefore increase the size of the dataset by one configuration at each step, from $5$ to $200$ samples. The evolution of the estimators is visualized in Figures~\ref{fig:dpp_consistency} and~\ref{fig:gpp_consistency}.

Results from the batched experiments without interaction (see Figure~\ref{fig:dpp_boxplots} (a) and (b)) show that we stably estimate small values of $\sigma$, between the minimal grid value of $0.01$ and $0.17$. As a case in which $\sigma = 0$ is degenerate, the best we can hope for is such small values, indicating that DPPs are able to generally detect a lack of, or at least scarcity, of interaction. When interaction is present, the parameter estimate is quite biased. This is due to the estimation of the intensity being imperfect, as is further elaborated in Appendix~\ref{app:dpp_just}. As an overview, the intensity is estimated directly from the one-particle marginal density $p_1$ over the aggregated snapshots, which, in the likelihood during parameter estimation, causes much of the spacing between points to already be caused by the spatial inhomogeneity, leading to lower estimates for $\sigma$. 

A similar story is told by the consistency experiment, the results of which can be seen in Figure~\ref{fig:dpp_consistency}. While when no interaction is present no clear convergence can be seen, there is a general downwards trend, indicating that as more data is available, the procedure tends to choose smaller values of $\sigma$, in line with the generating process. In contrast, when interaction is present in the data, we clearly see that the procedure converges to a choice of parameter of about $1.6$, as compared to the true value of $\sigma = 2.5$. As could be seen in the batched experiments, this represents a clear underestimation of the repulsion, indicative of the strong influence an imperfect estimate of the intensity can have on the procedure.

\begin{table}[t]
    \centering
    \begin{tabular}{ |l||c|c| }
        \hline
        \multicolumn{3}{|c|}{DPPs - average est. interaction parameter and bandwidth} \\
        \hline
        Parameter           & No interaction ($\sigma=0$)   & Interaction ($\sigma=2.5$) \\
        \hline
        $\hat{\sigma}_0$    & $0.1208 \ (\pm 0.0230)$ & $1.5570 \ (\pm 0.0418)$ \\
        $\hat{\sigma}_1$    & $0.1332 \ (\pm 0.0385)$ & $1.5238 \ (\pm 0.0422)$ \\
        $\hat{\sigma}_2$    & $0.1257 \ (\pm 0.0468)$ & $1.5142 \ (\pm 0.0431)$ \\
        \hline
        $h_0$               & $0.3647 \ (\pm 0.0120)$ & $0.5512 \ (\pm 0.0431)$ \\
        $h_1$               & $0.6096 \ (\pm 0.0260)$ & $0.7739 \ (\pm 0.0501)$ \\
        $h_2$               & $0.6236 \ (\pm 0.0236)$ & $0.8238 \ (\pm 0.0403)$ \\
        \hline
    \end{tabular}
    \caption{Results for batched estimation for DPPs, assessing stability. Averages of parameter estimates ($\hat{\sigma}_i$) and bandwidth estimates ($h_i$) across batches of $100$ configurations each, for three iterations; standard deviations in parentheses. True parameters for no interaction and interaction case denoted by $\sigma$.}
    \label{table:dpp_numerics_results}
\end{table}

\begin{table}[ht]
    \centering
    \begin{tabular}{|l||c|c||c|c|}
        \hline
        \multicolumn{1}{|c||}{} &
        \multicolumn{2}{c||}{Strauss} &
        \multicolumn{2}{c|}{DGS} \\
        \cline{2-5}
        Parameter & No int. ($\vartheta = 1$) & Int. ($\vartheta = 0.5, R=5$) & No int. ($\alpha = 0$) & Int. ($\alpha = 2, R=5$) \\
        \hline
        $\hat{\vartheta}_0$ / $\hat{\alpha}_0$ & $0.7615 \ (\pm 0.2781)$ & $0.6747 \ (\pm 0.0115)$ & $0.1640 \ (\pm 0.2722)$ & $2.3255 \ (\pm 0.1596)$ \\
        $\hat{\vartheta}_1$ / $\hat{\alpha}_1$ & $0.7496 \ (\pm 0.2746)$ & $0.6473 \ (\pm 0.0108)$ & $0.1760 \ (\pm 0.2804)$ & $2.2875 \ (\pm 0.1497)$ \\
        $\hat{\vartheta}_2$ / $\hat{\alpha}_2$ & $0.7618 \ (\pm 0.2776)$ & $0.6739 \ (\pm 0.0106)$ & $0.1630 \ (\pm 0.2708)$ & $2.3200 \ (\pm 0.1578)$ \\
        \hline
        $R_0$            & $0.2560 \ (\pm 0.2658)$ & $5.0000 \ (\pm 0.0000)$ & $0.2480 \ (\pm 0.1552)$ & $4.1750 \ (\pm 0.0805)$ \\
        $R_1$            & $0.3100 \ (\pm 0.3404)$ & $5.0000 \ (\pm 0.0000)$ & $0.2680 \ (\pm 0.2083)$ & $4.2800 \ (\pm 0.0787)$ \\
        $R_2$            & $0.2500 \ (\pm 0.2524)$ & $5.0000 \ (\pm 0.0000)$ & $0.2440 \ (\pm 0.1512)$ & $4.1820 \ (\pm 0.0767)$ \\
        \hline
        $h_0$            & $3.4168 \ (\pm 0.2379)$ & $3.5392 \ (\pm 0.1384)$ & $3.3719 \ (\pm 0.1368)$ & $3.5851 \ (\pm 0.1139)$ \\
        $h_1$            & $2.6401 \ (\pm 0.2189)$ & $2.2156 \ (\pm 0.1270)$ & $2.6640 \ (\pm 0.1514)$ & $2.1385 \ (\pm 0.1012)$ \\
        $h_2$            & $3.3997 \ (\pm 0.1389)$ & $3.4911 \ (\pm 0.1100)$ & $3.3229 \ (\pm 0.1688)$ & $3.5345 \ (\pm 0.1155)$ \\
        \hline
    \end{tabular}
    \caption{Results for batched estimation for GPPs, assessing stability. Averages of interaction parameter estimates ($\hat{\vartheta}_i$ and $\hat{\alpha}_i$ for Strauss and DGS, respectively), interaction radius estimates ($R_i$), and bandwidth estimates ($h_i$) across batches of 100 configurations each given for three iterations; standard deviations in parentheses. True parameters for no interaction and interaction case denoted by $\vartheta, \alpha$, and $R$.}
    \label{table:gpp_numerics_results}
\end{table}

For DPPs, the iterative procedure we propose leads to larger choices of bandwidths, taking into account the effect the interaction has on the likelihood. In parallel, Figure~\ref{fig:dpp_boxplots} shows that when interaction is present, the estimate of the interaction parameter decreases slightly as the bandwidth increases. This is a natural effect of the adjusted intensity estimate correcting for a more regular spatial pattern, leading to less of the repulsion being ``left'' for the interaction parameter to cover.

For GPPs, the batched experiment (see Figures~\ref{fig:strauss_boxplots} and~\ref{fig:dgs_boxplots}) indicates that both the Strauss and DGS models are able to identify the lack of interaction in the no-interaction case, and both have some bias in the estimation in the interactive case. Note that there is no unique set of parameters describing no-interaction, and therefore the best-case scenario is that the inference method consistently infers a parameter combination corresponding to lack of interaction. This is indeed the case in the synthetic data, as in both models we tend towards estimating a very small radius of interaction $R$. In this case the interaction parameter ($\vartheta$ or $\alpha$, respectively) plays no role, and it is therefore no longer be identifiable, which results in greater variation in the estimates here (with no further consequences). Nevertheless, a lack of interaction can be identified, as $R$ tending towards $0$ indicates. Testing also showed that if we fix a larger interaction radius (e.g. $R=5$), both models estimate interaction parameters corresponding to no interaction, i.e. $\vartheta \approx 1$ and $\alpha \approx 0$. When interaction is present, the Strauss model is able to identify $R$ very well, see Figure~\ref{fig:strauss_boxplots}(d). However, even when the radius is correctly estimated, there is some bias in the estimation of $\vartheta$ (see Figure~\ref{fig:strauss_boxplots}(e)). This parameter is overestimated, corresponding to an underestimation of the repulsion strength. For the DGS model, we see that $R$ is slightly underestimated (cf. Figure~\ref{fig:dgs_boxplots}(d)), leading to $\alpha$ being slightly overestimated (an overestimation of the repulsion strength, see Figure~\ref{fig:dgs_boxplots}(e)). When the radius is fixed to the true value, we see similar behavior as for the Strauss process, as the DGS model also then tends to estimate values of $\alpha$ slightly below the true parameter value of $2$. This aligns with the behavior we see for DPPs, and confirms the distorting influence of the intensity. As for DPPs, if we run only the pseudolikelihood-based interaction parameter inference, providing the model with the true intensity, we are able to recover the true parameters.

Consistency testing (see Figure~\ref{fig:gpp_consistency}) again reveals similar behavior for the two GPP models. When no interaction is present, both the Strauss model and the DGS model resort to choosing the smallest possible radius of interaction $R$. As became apparent in the batched tests, this leads to the second parameter $\vartheta$ / $\alpha$ no longer being identifiable, as multiple combinations of parameters give rise to the same model without interaction. In the interactive case, both the Strauss and DGS interaction models settle on a choice of interaction radius quite quickly. While in the Strauss case this is the true radius, in the DGS case the radius is slightly underestimated. For the Strauss process, the interaction parameter $\vartheta$ seems to converge quickly around a value of about $0.65$, representing the same slight overestimation we saw in the batched experiments. The DGS process also converges quickly, but the underestimated interaction radius causes an overestimation of the interaction parameter $\alpha$. Supposing the true interaction radius were picked, the interaction parameter actually converges to a value of $1.7$, an underestimation of the interaction strength, matching results for the Strauss process and the DPPs.

\subsection{Conclusion: GPPs outperform DPPs}

In conclusion, the estimation procedure proposed in Algorithm~\ref{alg:cv} is more reliable for GPPs than for DPPs. When no interaction present, DPP models need a large amount of data to converge towards small values of $\sigma$, while for GPPs one of the two interaction parameters is quickly and reliably chosen to correspond to little or no interaction. While this means the individual parameters are not identifiable in this case, as multiple combinations of parameters lead to a no-interaction model, the model correctly captures this lack of interaction. When interaction is present, GPPs again perform better in recovering the true parameters than DPPs. They are less sensitive to propagating errors in the inference of the intensity. While biases in parameter estimates exist in both model classes, the existence of a second parameter in the GPP models allows for some balancing effects. In addition, the bias is more pronounced for DPPs, as the interaction parameter is further off from the true parameter.

\begin{figure}[t]
    \centering
    \includegraphics[width=\textwidth]{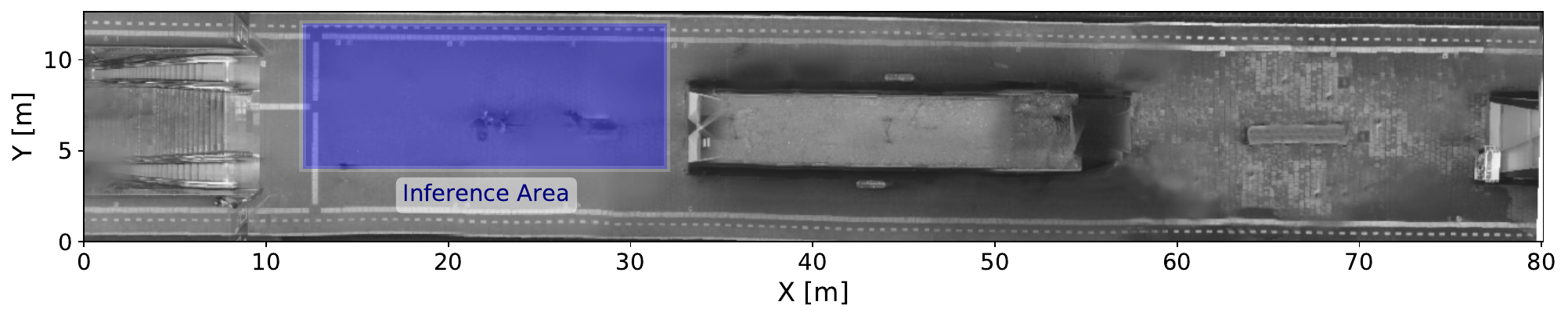}
    \caption{Observation area for inference. For simplicity, statistical inference (but not simulation) is restricted to these areas to ensure regular boundaries and less spatial variation in intensity.}
    \label{fig:rectangles}
\end{figure}

\section{Application to real-world replicated waiting pedestrian configurations}\label{sec:application}

In this section, we apply the inference method described and tested in Section~\ref{sec:inference} to the train station dataset of snapshots of waiting pedestrians introduced in Section~\ref{sec:data}. We begin with the DPP model with Gaussian interaction kernel, as this is the most straightforward. After discussing the results of the inference for this model in discrete space, we move on to continuous GPP models. Results are discussed for both the Strauss and DGS interaction functions, highlighting both limitations and advantages of the GPP approach.

\begin{figure}[p]
\centering
\begin{minipage}{\textwidth}
    \centering
    \begin{subfigure}[b]{\textwidth}
        \includegraphics[width=\textwidth]{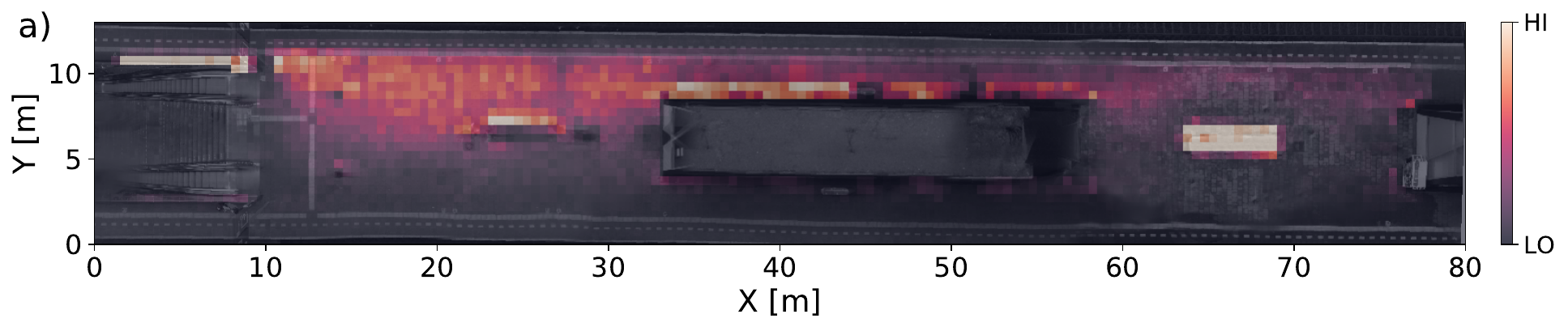}
    \end{subfigure}
    \begin{subfigure}[b]{\textwidth}
        \includegraphics[width=\textwidth]{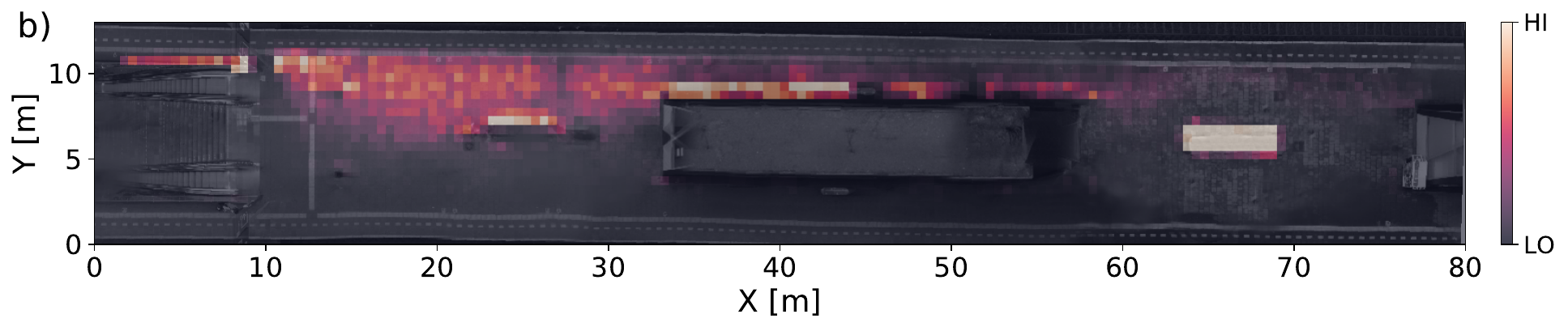}
    \end{subfigure}
    \caption{Histogram of (a) discretized data point configurations and (b) simulated point configurations from DPP, binning corresponds to the grid used for the discrete point process. We see high agreement in terms of points of interest, but DPP estimated one-particle marginal density is less spread out than can be seen in the data.}
    \label{fig:dpp_marginal}
\end{minipage}

\vspace{1em}

\begin{minipage}{\textwidth}
    \centering
    \begin{subfigure}[ht]{0.32\textwidth}
        \centering
        \includegraphics[width=\textwidth]{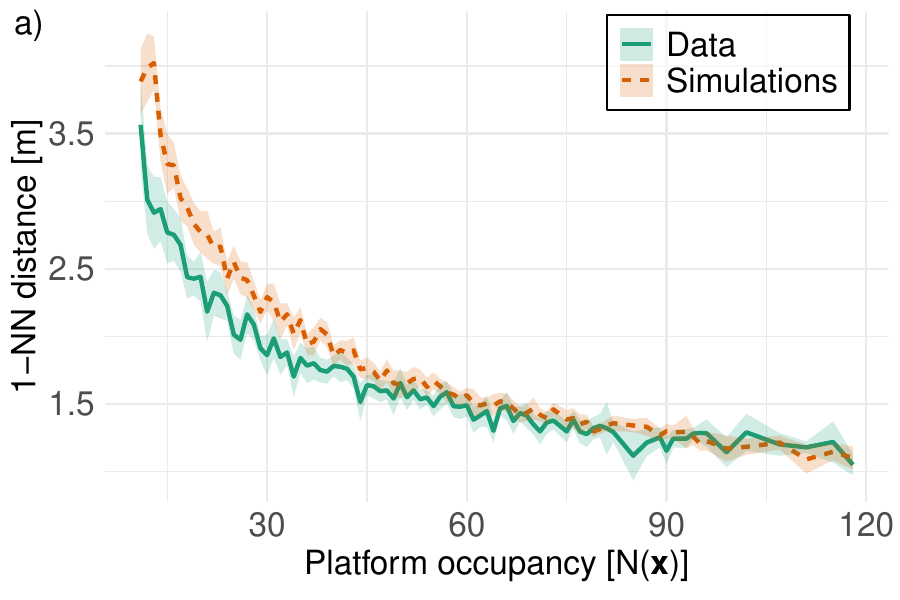}
    \end{subfigure}
    \begin{subfigure}[ht]{0.32\textwidth}
        \centering
        \includegraphics[width=\textwidth]{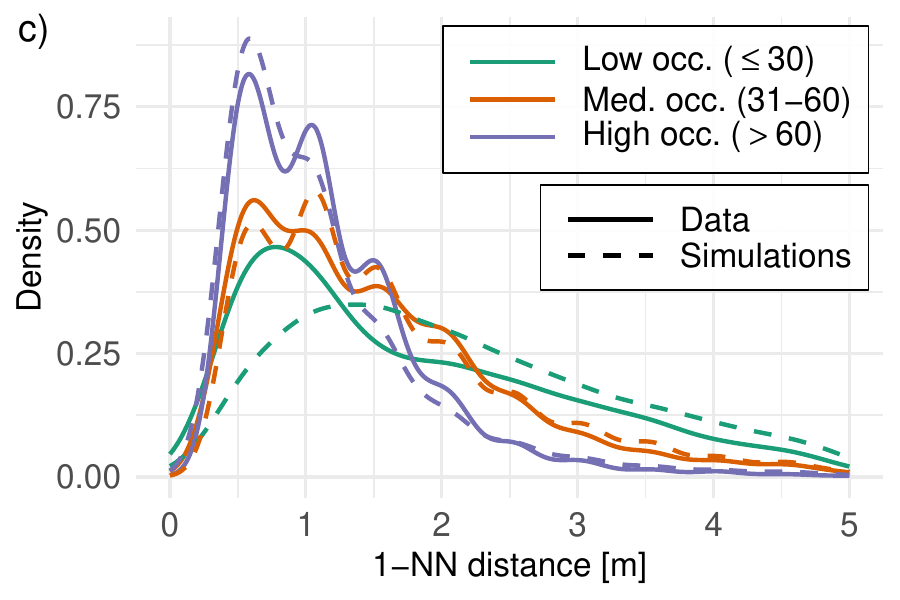}
    \end{subfigure}
    \begin{subfigure}[ht]{0.32\textwidth}
        \centering
        \includegraphics[width=\textwidth]{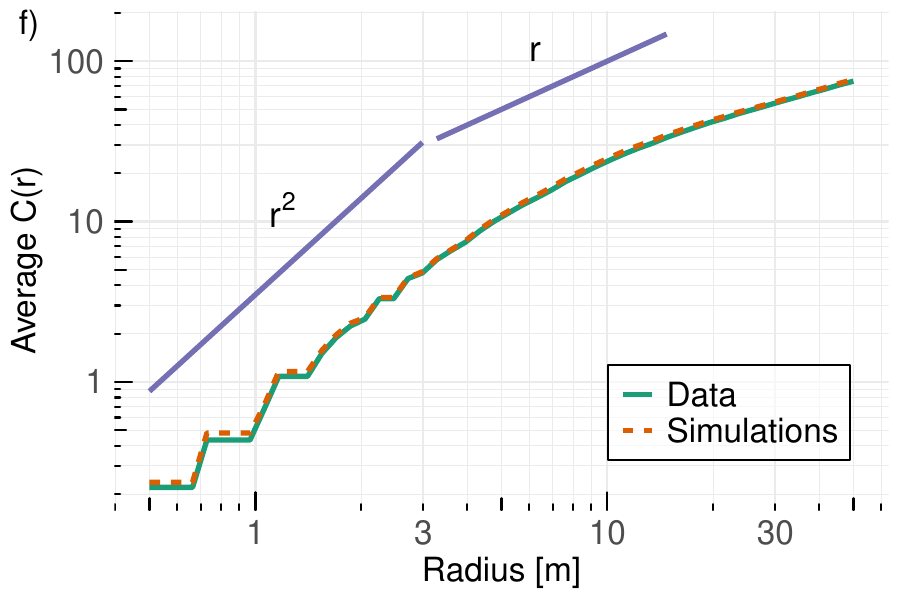}
    \end{subfigure}

    \begin{subfigure}[ht]{0.32\textwidth}
        \centering
        \includegraphics[width=\textwidth]{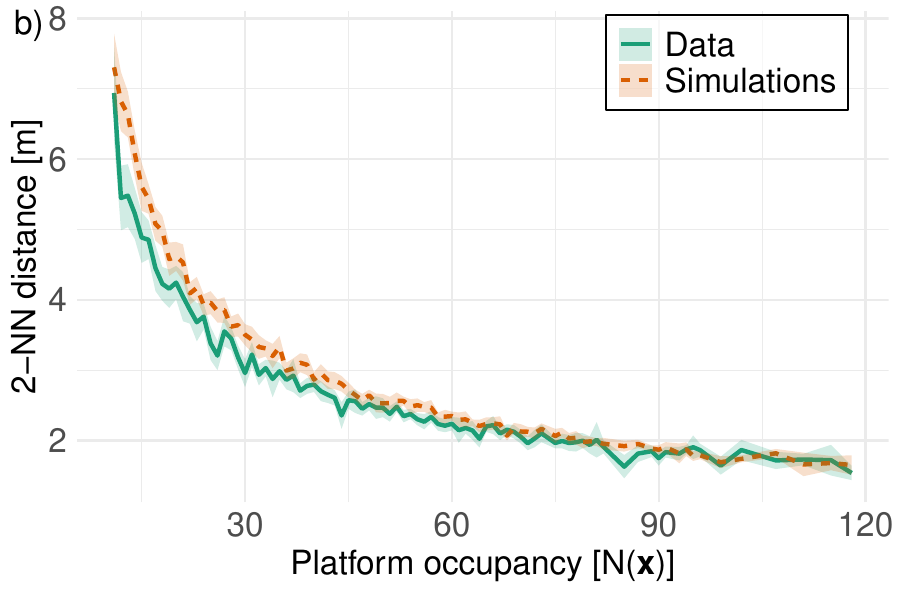}
    \end{subfigure}
    \begin{subfigure}[ht]{0.32\textwidth}
        \centering
        \includegraphics[width=\textwidth]{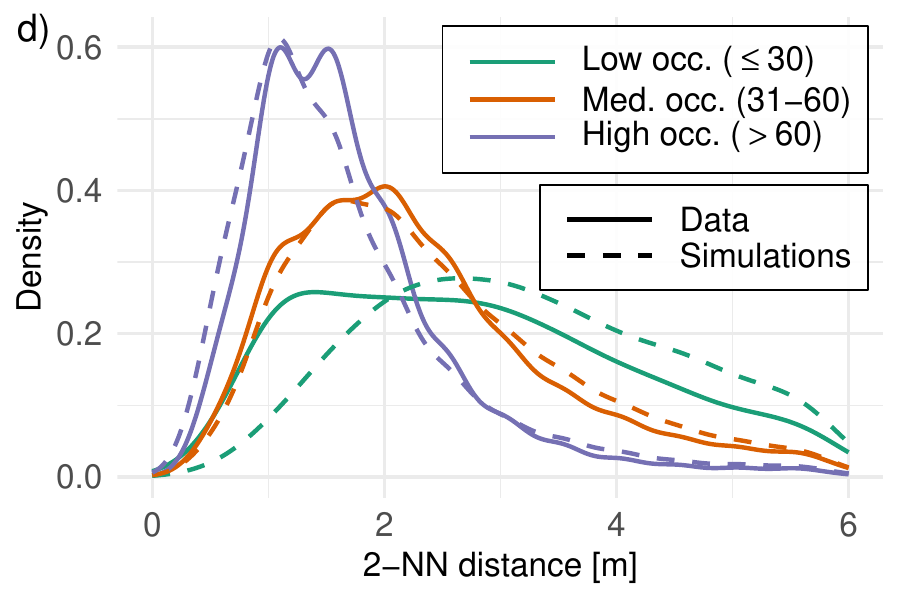}
    \end{subfigure}
    \begin{subfigure}[ht]{0.32\textwidth}
        \centering
        \includegraphics[width=\textwidth]{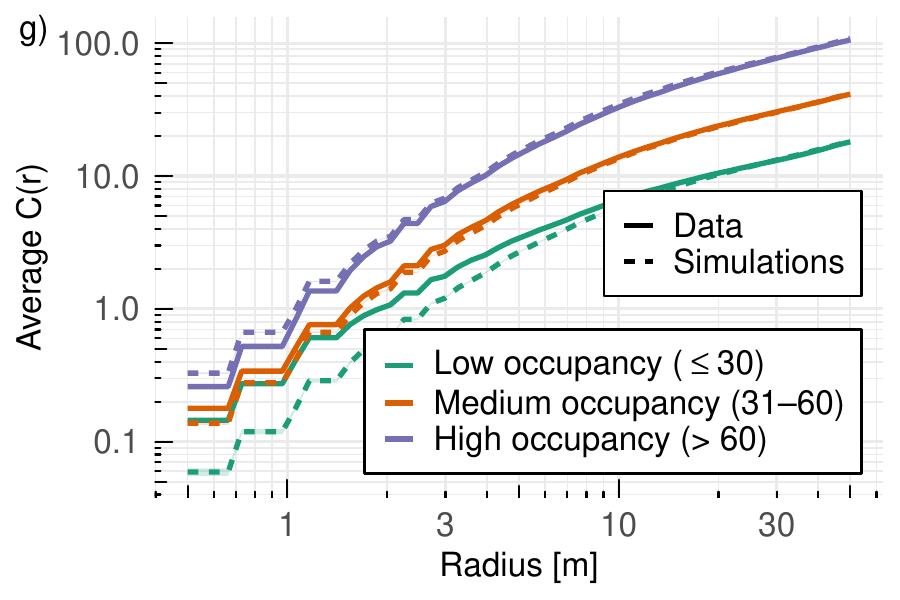}
    \end{subfigure}

    \caption{Comparison of distance-based statistics \ref{item:D1}-\ref{item:D3} (introduced in Section~\ref{subsec:data_features}) of DPP simulations with estimated $\sigma = 0.714$ (using Algorithm~\ref{alg:cv}) versus the dataset. (a) Average NN distance. (b) Average 2-NN distance. (c) PDF of nearest neighbor distances, separated by occupancy. A simple kernel smoothing is applied for visualization purposes, using Scott's rule of thumb for bandwidth selection~\cite{Scott_1992}. (d) PDF of 2-NN distances, same smoothing rule as for previous panel. (e) L-functions for DPP simulations vs. data (including correction for inhomogeneity). When inhomogeneity is ignored, curves show high agreement. When correction is introduced, repulsion is stronger for simulations than for the data. (f) Average $\mathcal{C}(r)$ over all snapshots in log-log scale. (g) Average $\mathcal{C}(r)$ separated by occupancy. DPPs underestimate repulsion significantly, yet capture interactions more faithfully than Poisson processes, especially at lower occupancy levels.}
    \label{fig:dpp_features}
\end{minipage}
\end{figure}

\begin{figure}[p]
\centering
\begin{minipage}{\textwidth}
    \centering
    \begin{subfigure}[ht]{0.32\textwidth}
        \centering
        \includegraphics[width=\textwidth]{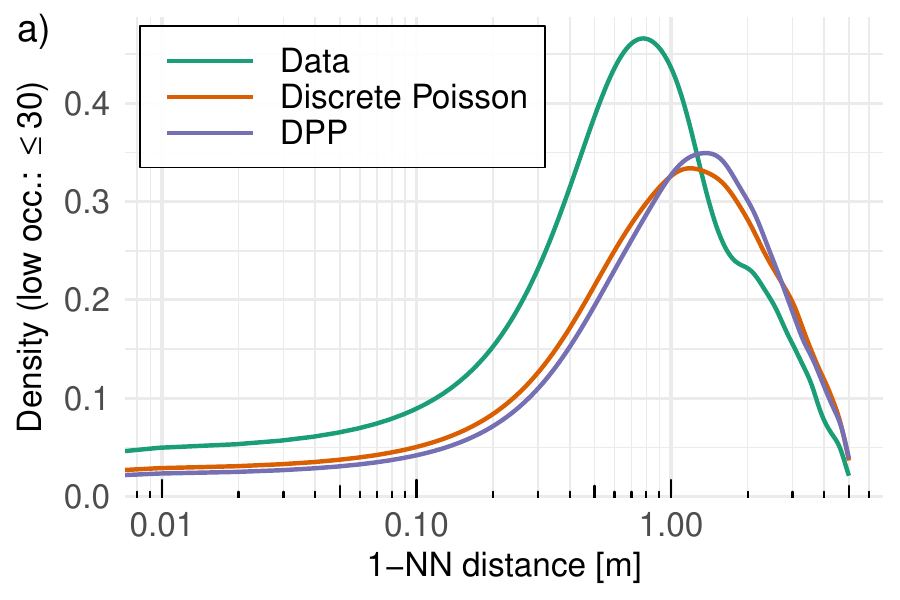}
    \end{subfigure}
    \begin{subfigure}[ht]{0.32\textwidth}
        \centering
        \includegraphics[width=\textwidth]{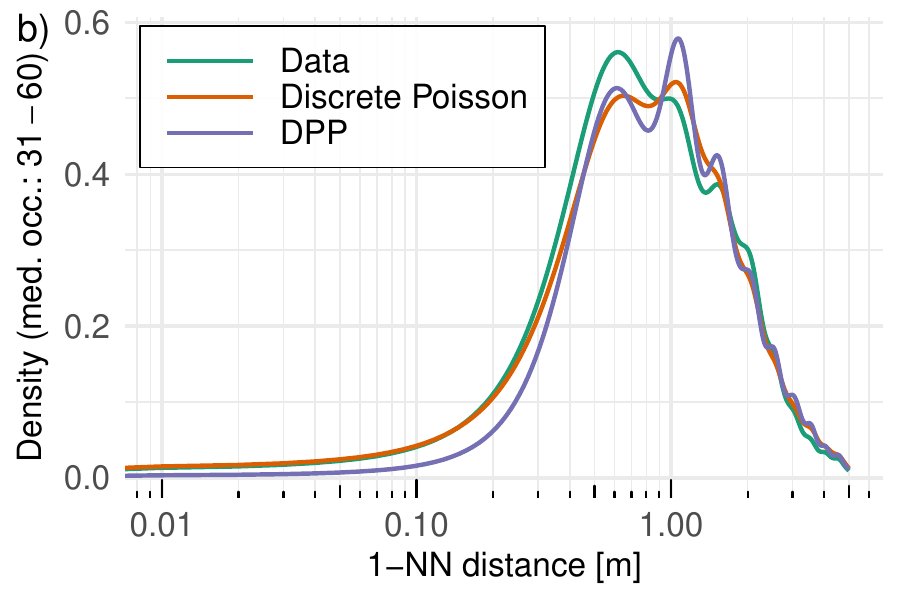}
    \end{subfigure}
    \begin{subfigure}[ht]{0.32\textwidth}
        \centering
        \includegraphics[width=\textwidth]{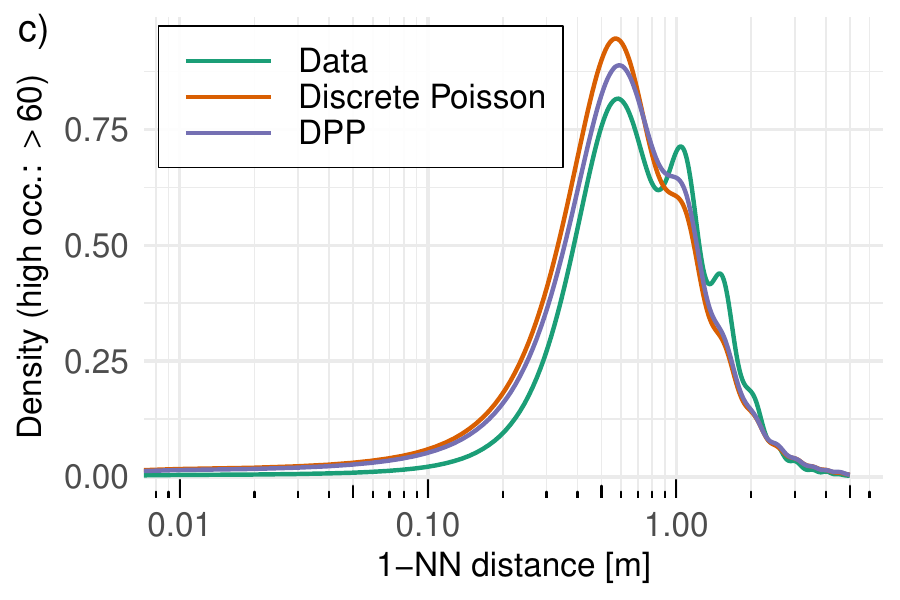}
    \end{subfigure}

    \begin{subfigure}[ht]{0.32\textwidth}
        \centering
        \includegraphics[width=\textwidth]{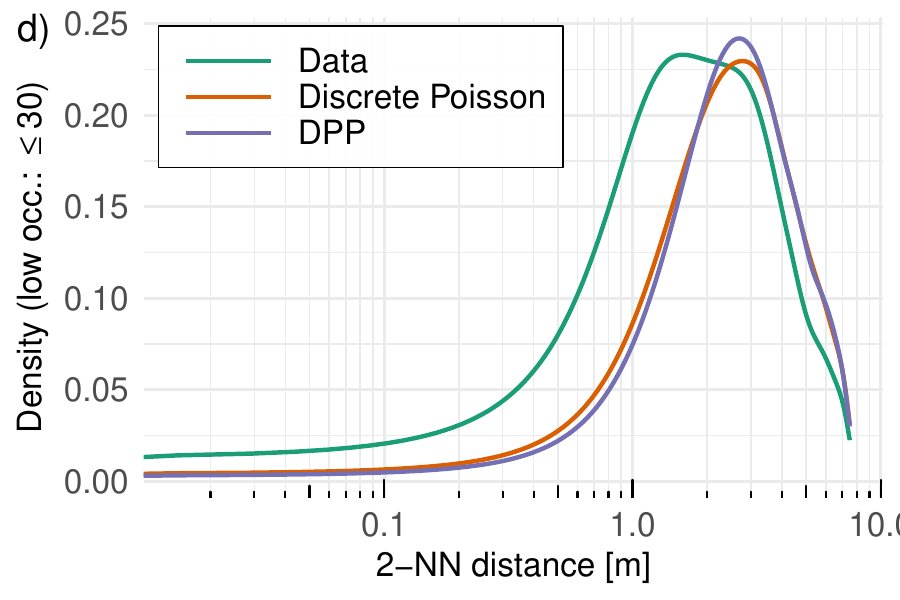}
    \end{subfigure}
    \begin{subfigure}[ht]{0.32\textwidth}
        \centering
        \includegraphics[width=\textwidth]{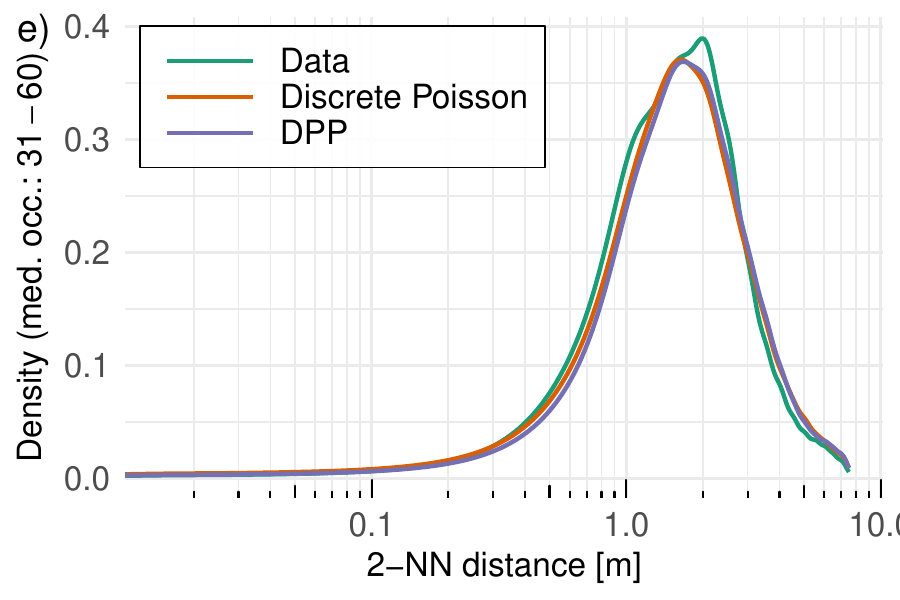}
    \end{subfigure}
    \begin{subfigure}[ht]{0.32\textwidth}
        \centering
        \includegraphics[width=\textwidth]{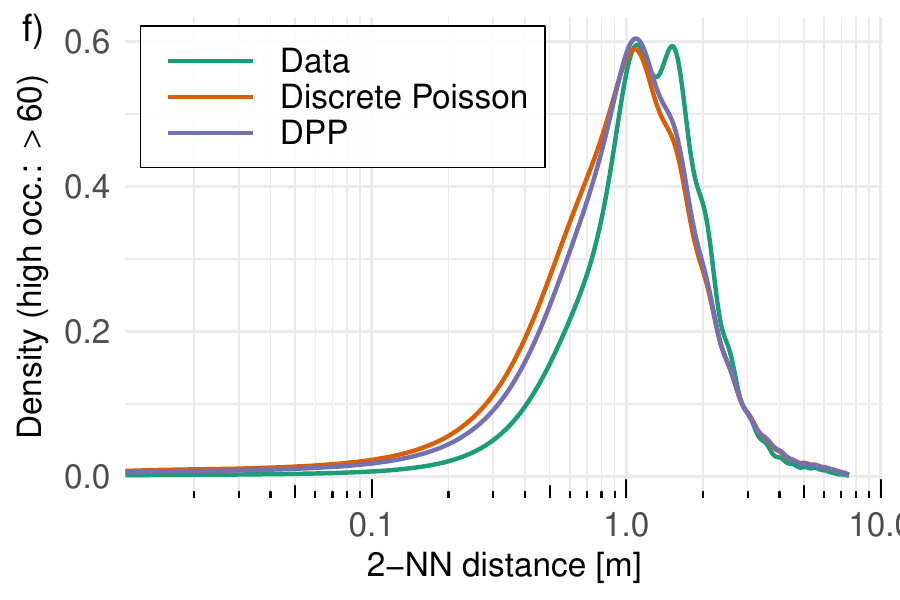}
    \end{subfigure}

    \begin{subfigure}[ht]{0.32\textwidth}
        \centering
        \includegraphics[width=\textwidth]{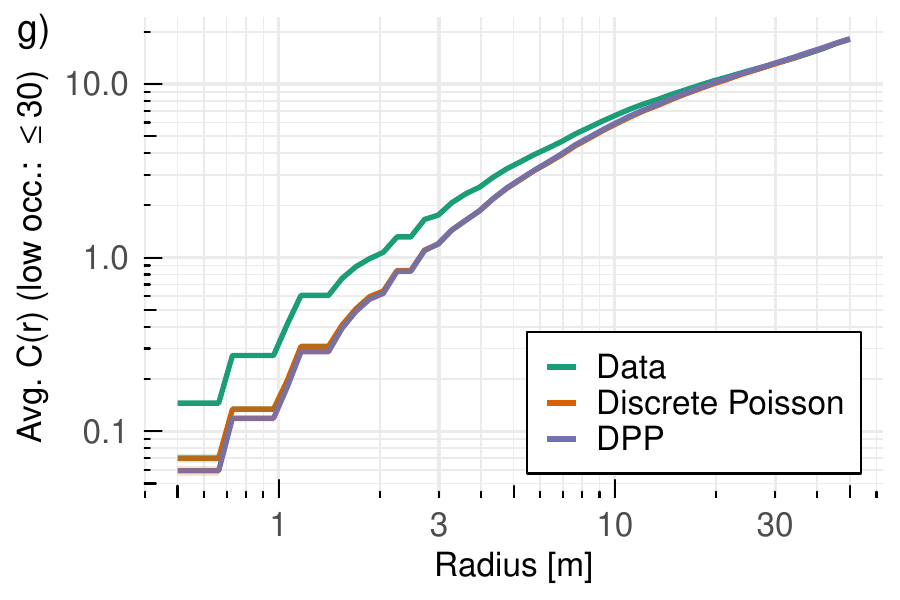}
    \end{subfigure}
    \begin{subfigure}[ht]{0.32\textwidth}
        \centering
        \includegraphics[width=\textwidth]{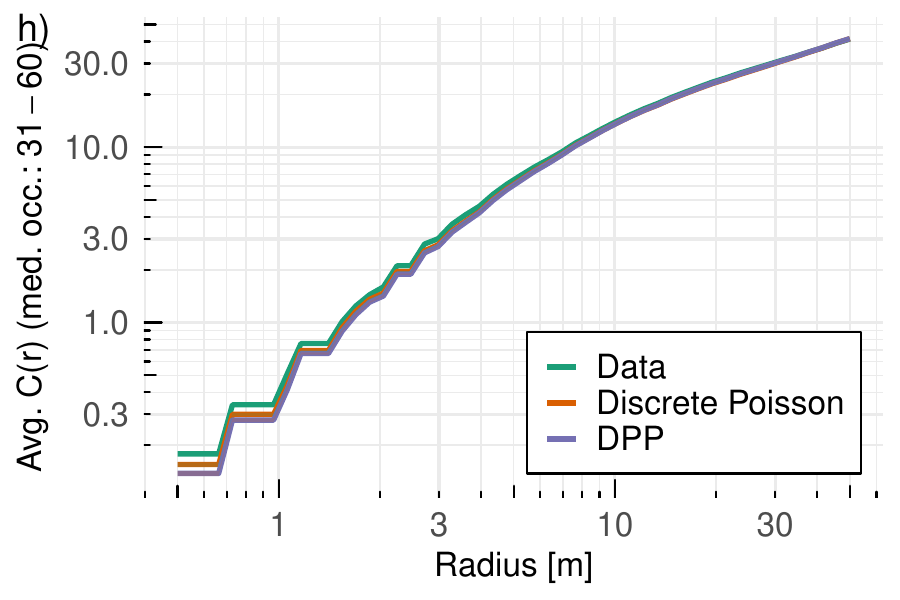}
    \end{subfigure}
    \begin{subfigure}[ht]{0.32\textwidth}
        \centering
        \includegraphics[width=\textwidth]{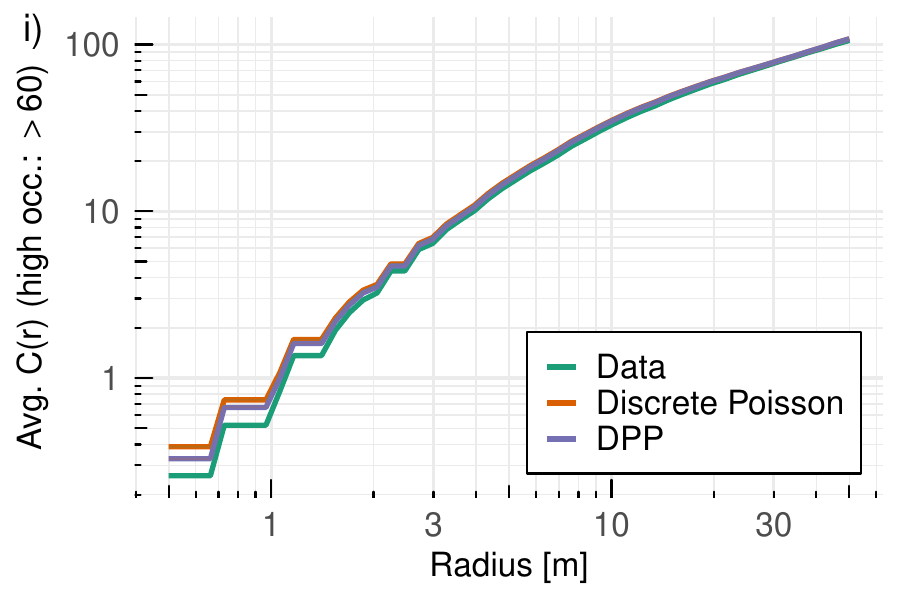}
    \end{subfigure}

    \begin{subfigure}[ht]{0.32\textwidth}
        \centering
        \includegraphics[width=\textwidth]{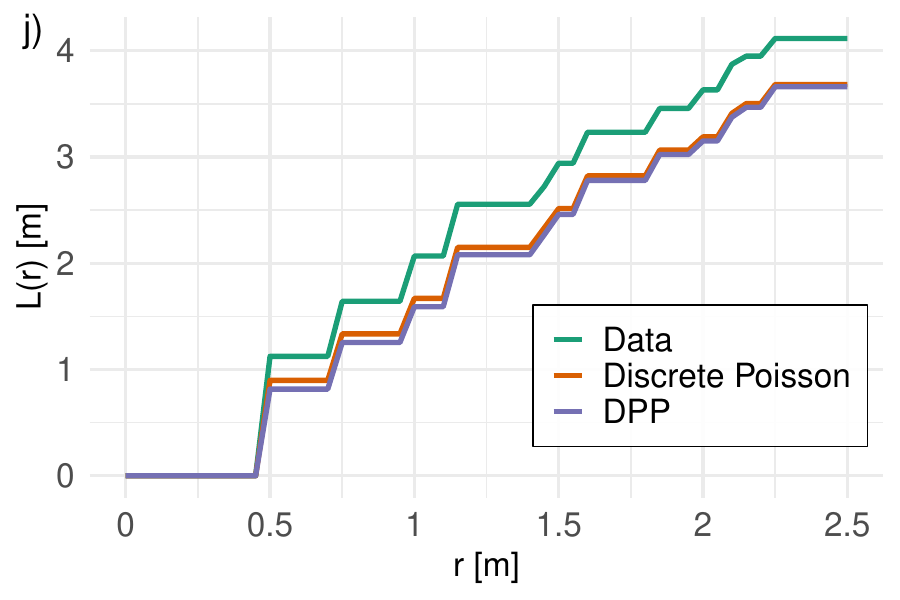}
    \end{subfigure}
    \begin{subfigure}[ht]{0.32\textwidth}
        \centering
        \includegraphics[width=\textwidth]{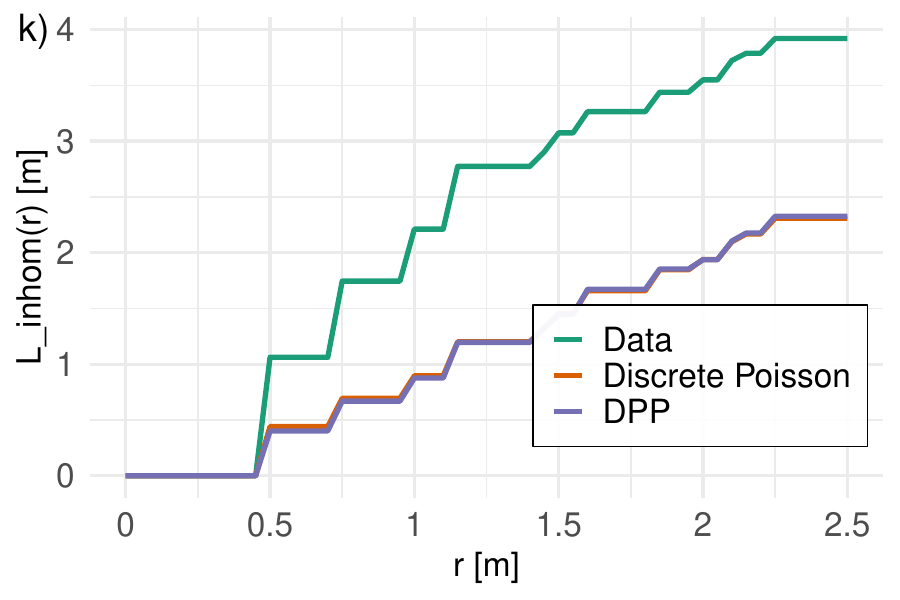}
    \end{subfigure}

    \caption{Side by side comparison of \ref{item:D2}-\ref{item:D4} for the Data, discrete Poisson, and DPP. Panels separated by occupancy, with NN and 2NN PDFs in logarithmic scale in $x$. PDFs undergo the same smoothing as in Figure~\ref{fig:dpp_features} (a) PDFs of NN distances at low occupancy. (b) PDFs of NN distances at medium occupancy. (c) PDFs of NN distances at high occupancy. (d) PDFs of 2-NN distances at low occupancy. (e) PDFs of 2-NN distances at medium occupancy. (f) PDFs of 2-NN distances at high occupancy. (g) Avg. $\mathcal{C}(r)$ at low occupancy. (h) Avg. $\mathcal{C}(r)$ at medium occupancy. (i) Avg. $\mathcal{C}(r)$ at high occupancy. (j) Homogeneous L-functions. (k) Inhomogeneous L-functions. Overall, DPPs offer only very slight benefits over inhomogeneous Poisson processes.}
    \label{fig:dpp_comparison}
\end{minipage}
\end{figure}

For the estimation, we restrict our attention to a subregion of the platform. We estimate both the interaction parameters and the bandwidth of the kernel estimate of the intensity on this reduced dataset to avoid irregular boundaries and to ensure the inference is done on a ``well-behaved'' part of the data, in the sense that large fluctuations of the intensity are rare, and a reasonably high number of points is ensured. See Figure~\ref{fig:rectangles} for an illustration of the location of this area, and Appendix~\ref{app:generalization} for a brief discussion on the effect of this choice. In order to simulate configurations of points for comparison with the real-world data, we use the full platform. We first generate an intensity for the full platform by using the bandwidth estimated in the inference procedure and producing a kernel estimate using the data on the full platform. Then, we use the estimated interaction parameters to simulate from the corresponding models in the full area of observation.

\subsection{Semiparametric likelihood estimation for discrete determinantal point processes}
\label{subsec:dpp_estimation}

As discussed in Section~\ref{subsec:dpp}, we estimate the DPP-kernel of the DPP model by separating the inference of the intensity from that of the interaction kernel. As described in Section~\ref{subsec:peaks}, the data is discretized to a $50\,$cm grid. Using Algorithm~\ref{alg:cv}, we select a bandwidth for the KDE of the intensity, and we estimate an interaction parameter $\sigma$ for a Gaussian interaction kernel, as described in Section~\ref{subsec:numerics}. Specifically, $\sigma$ is chosen by maximizing the log-likelihood over a grid of values between $0.5$ and $3$, separated by steps of $0.01$. We obtain a final bandwidth estimate of $210\,$mm for the intensity and an estimate of the interaction parameter of $\hat{\sigma} = 0.714$ (note that this is in grid units).

Using the estimated intensity and interaction kernel, we simulate configurations of points for comparison. In particular, we group the dataset into the number of points per sample $N(\x)$, and then simulate the exact same number of configurations per configuration size as in the dataset. To reiterate, note that for the simulations we use the entire platform, estimating an intensity function by using the bandwidth estimated on the rectangle in the previous step. By comparing the features described in Section~\ref{subsec:data_features}, we see that the estimate of $p_1$, the one-particle marginal density, is quite similar to the data, see Figure~\ref{fig:dpp_marginal}, which is not unexpected seeing as the intensity is directly estimated from a similar estimate of $p_1$, and the estimated interaction parameter is rather low. In terms of distance statistics, shown in Figure~\ref{fig:dpp_features}, we see that the determinantal model falls short in several categories. An apparent underestimation of the interaction parameter is indicated both by the underestimation of NN and 2-NN distances (Figure~\ref{fig:dpp_features}(a-d)), as well as in the average ball count $\mathcal{C}(r)$ (Figure~\ref{fig:dpp_features}(f-g)). Generally, the model seems to perform better at low occupancy, which is not unexpected, as the intensity should dominate interactive effects in this regime.

The shortcomings of the DPP model in practice are underlined in Figure~\ref{fig:dpp_comparison}, in which we compare it to simulations from a discrete analogue of a Poisson process. This is a DPP with a diagonal matrix as its DPP-kernel, where the diagonal entries are a discrete KDE intensity estimate as described in Section~\ref{sec:inference}. In panels (a)-(f), the PDFs of NN and 2-NN distances show that while DPPs can model the distribution well at low occupancy, so can a discrete Poisson process. Only at higher occupancy, where both fall short, does the DPP model perform slightly better. A similar story is told by panels (g)-(i), as the Average $\mathcal{C}(r)$ of the data is matched slightly better by the DPP at medium and high occupancy. Finally, considering L-functions shown in panels (j) and (k), while the repulsion seen in the data is captured well by the DPP, the Poisson model also captures this, indicating that this is mostly captured by the discretization itself.

These results indicate that, as already suspected in Section~\ref{subsec:numerics}, the inference procedure we use leads to significantly biased estimates of the interaction. Real-world experiments indicate an underestimation of the interaction, matching the results in a controlled setting. In part, this mismatch may be due to the rigid structure of determinantal models. The parametrization of the process via a positive semidefinite symmetric matrix enforces a strong propagation of repulsion over different ranges of interaction. In particular, if we want the process to have a strong repulsion at close range, as is indicated by the data, this repulsion will necessarily continue to be strong at longer ranges as well. As strong long-range repulsion is not visible in the data, this may cause the inference procedure to discard the possibility of strong interactions altogether. In addition, the computational advantages of DPPs require a discretization of the space, which introduces further bias to the model. Some information in the data is lost by discretizing, and computational advantages are lost if the grid is chosen to be too fine. These drawbacks notwithstanding, DPPs remain a computationally attractive approach that is able to incorporate some repulsive interaction, even if the current approach underestimates the corresponding parameters.

\begin{figure}[t]
    \centering
    \begin{subfigure}[b]{\textwidth}
        \centering
        \includegraphics[width=\textwidth]{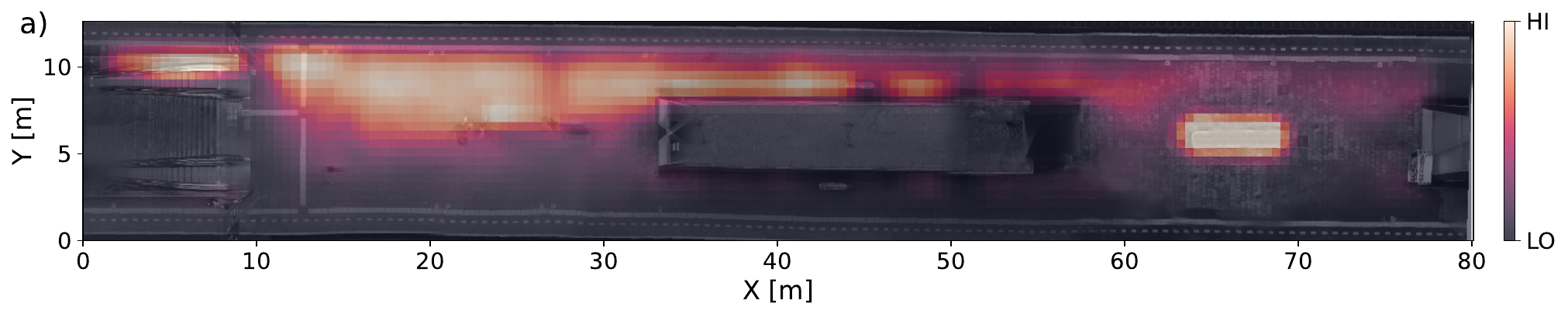}
    \end{subfigure}
    \begin{subfigure}[b]{\textwidth}
        \centering
        \includegraphics[width=\textwidth]{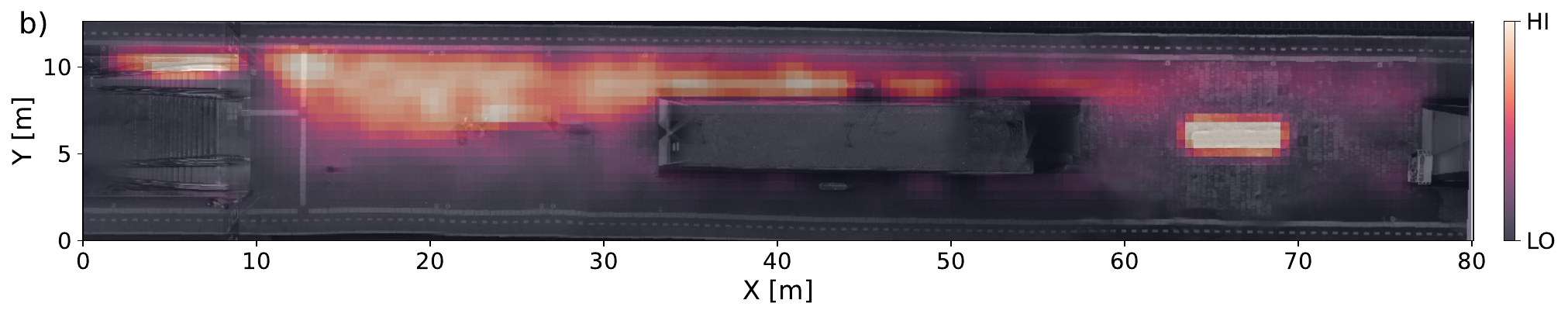}
    \end{subfigure}
    \caption{Estimated one-particle marginal densities $p_1$ from simulations of the (a) Strauss model and (b) DGS model. For comparison, a kernel smoothing is used with the same bandwidth ($0.5\,$m) as for the heatmap shown in Figure~\ref{fig:agg_heatmap}(a). Comparing with this figure, we see high agreement.}
    \label{fig:gpp_marginal}
\end{figure}

\subsection{Semiparametric pseudolikelihood estimation for Gibbs point processes}

\begin{figure}[p]
\centering
\begin{minipage}{\textwidth}
    \centering
    \begin{subfigure}[ht]{0.32\textwidth}
        \centering
        \includegraphics[width=\textwidth]{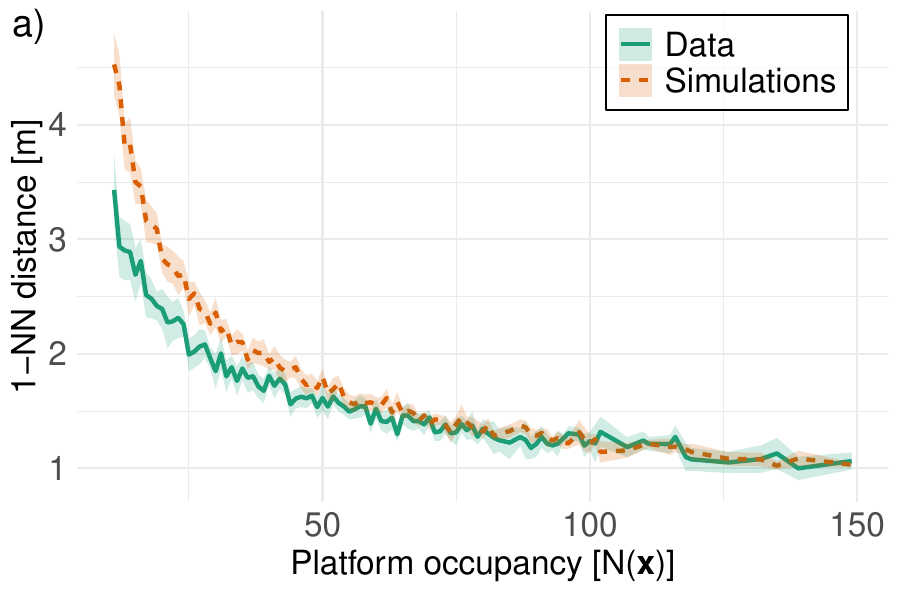}
    \end{subfigure}
    \begin{subfigure}[ht]{0.32\textwidth}
        \centering
        \includegraphics[width=\textwidth]{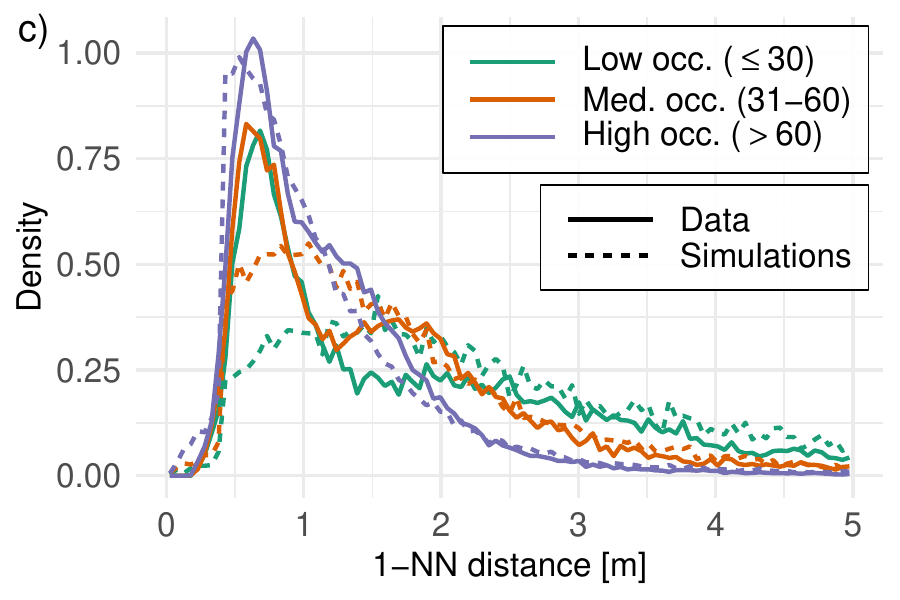}
    \end{subfigure}
    \begin{subfigure}[ht]{0.32\textwidth}
        \centering
        \includegraphics[width=\textwidth]{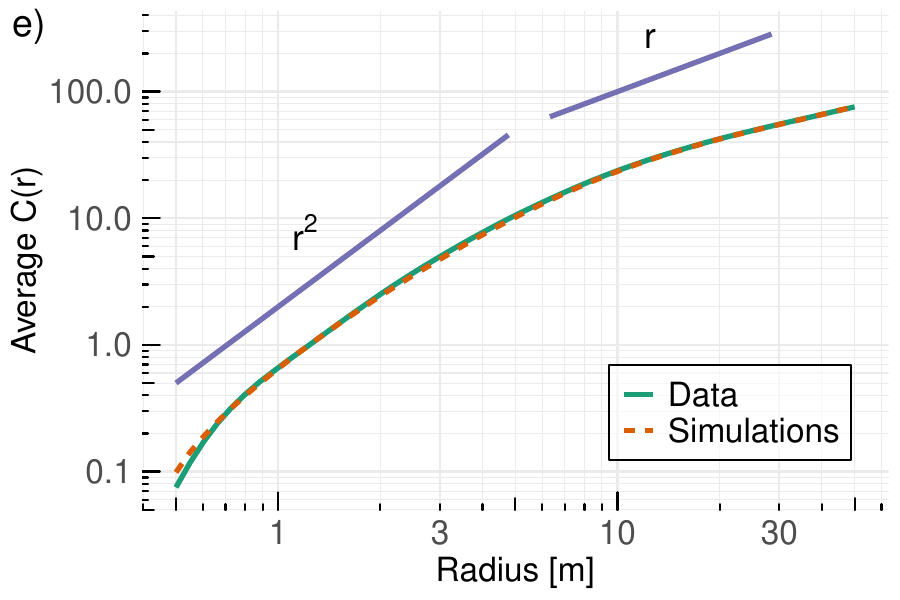}
    \end{subfigure}

    \begin{subfigure}[ht]{0.32\textwidth}
        \centering
        \includegraphics[width=\textwidth]{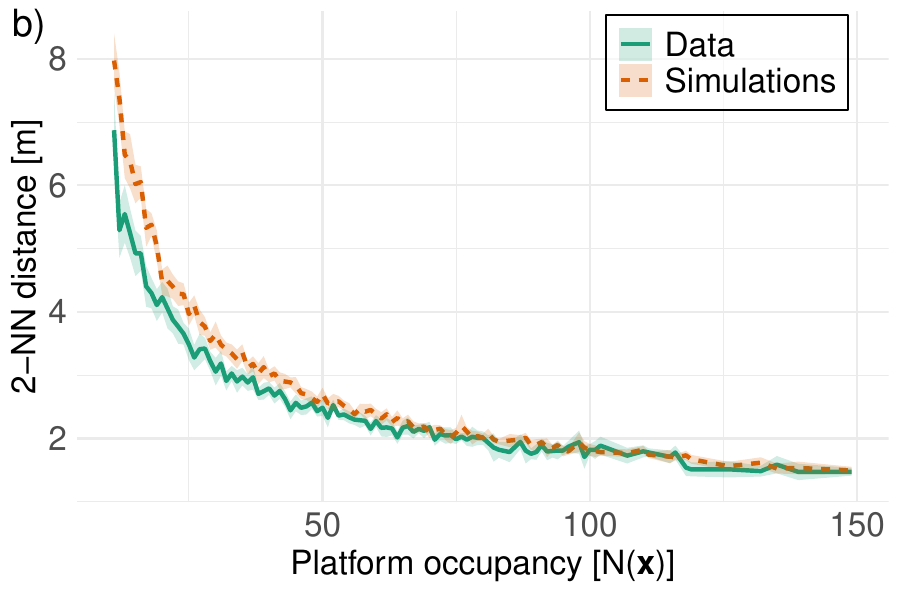}
    \end{subfigure}
    \begin{subfigure}[ht]{0.32\textwidth}
        \centering
        \includegraphics[width=\textwidth]{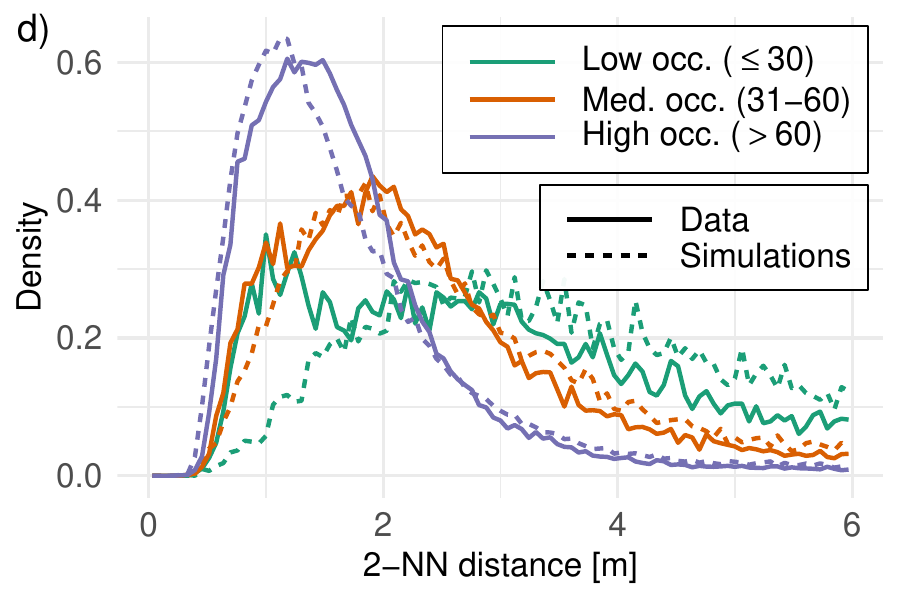}
    \end{subfigure}
    \begin{subfigure}[ht]{0.32\textwidth}
        \centering
        \includegraphics[width=\textwidth]{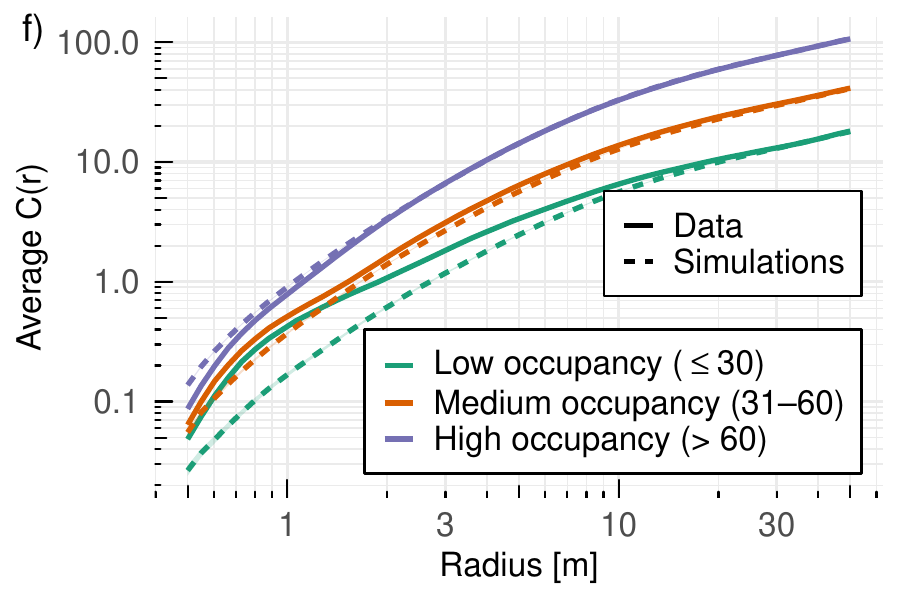}
    \end{subfigure}
    \caption{Comparison of \ref{item:D1}-\ref{item:D3} for Strauss simulations with $R=400, \vartheta=0.15$ vs. the dataset. (a) Avg. NN distance. (b) Avg. 2-NN distance.  (c) PDF of NN distances, separated by occupancy. (d) PDF of 2-NN distances. (e) Avg. $\mathcal{C}(r)$ over all snapshots in log-log scale.(f) Avg. $\mathcal{C}(r)$ separated by occupancy. We see quite strong agreement for most metrics, particularly at higher occupancy.}
    \label{fig:gpp_features_strauss}
\end{minipage}

\vspace{1em}

\begin{minipage}{\textwidth}
    \centering
    \begin{subfigure}[ht]{0.32\textwidth}
        \centering
        \includegraphics[width=\textwidth]{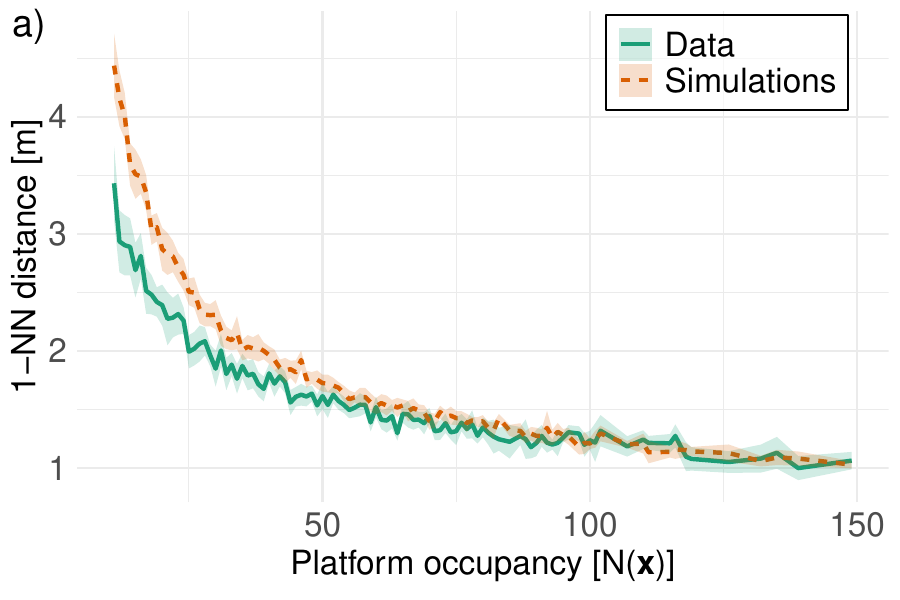}
    \end{subfigure}
    \begin{subfigure}[ht]{0.32\textwidth}
        \centering
        \includegraphics[width=\textwidth]{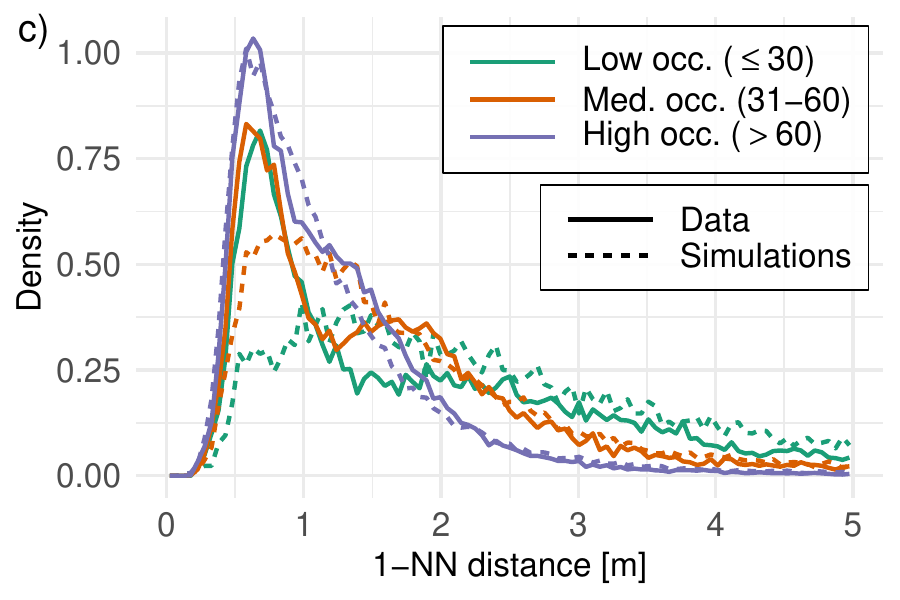}
    \end{subfigure}
    \begin{subfigure}[ht]{0.32\textwidth}
        \centering
        \includegraphics[width=\textwidth]{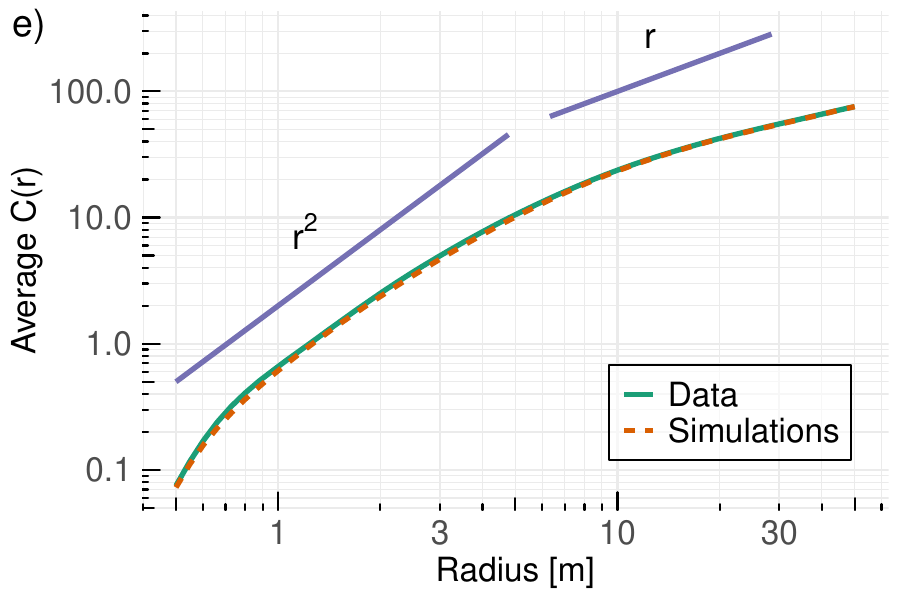}
    \end{subfigure}

    \begin{subfigure}[ht]{0.32\textwidth}
        \centering
        \includegraphics[width=\textwidth]{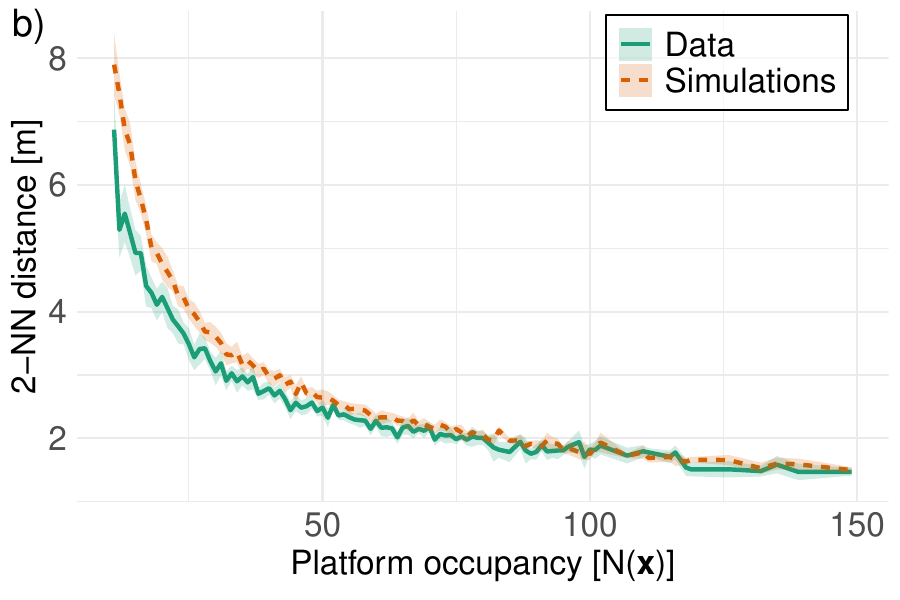}
    \end{subfigure}
    \begin{subfigure}[ht]{0.32\textwidth}
        \centering
        \includegraphics[width=\textwidth]{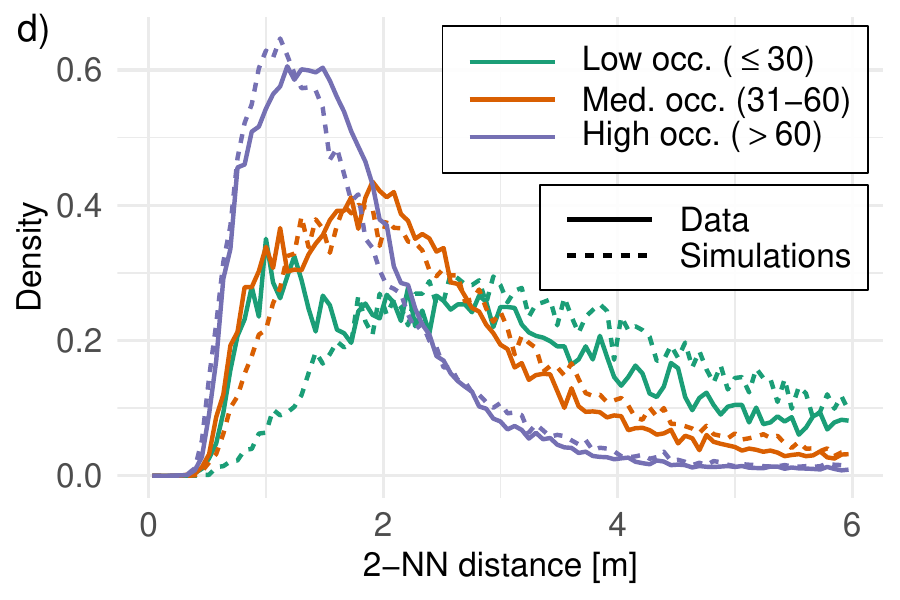}
    \end{subfigure}
    \begin{subfigure}[ht]{0.32\textwidth}
        \centering
        \includegraphics[width=\textwidth]{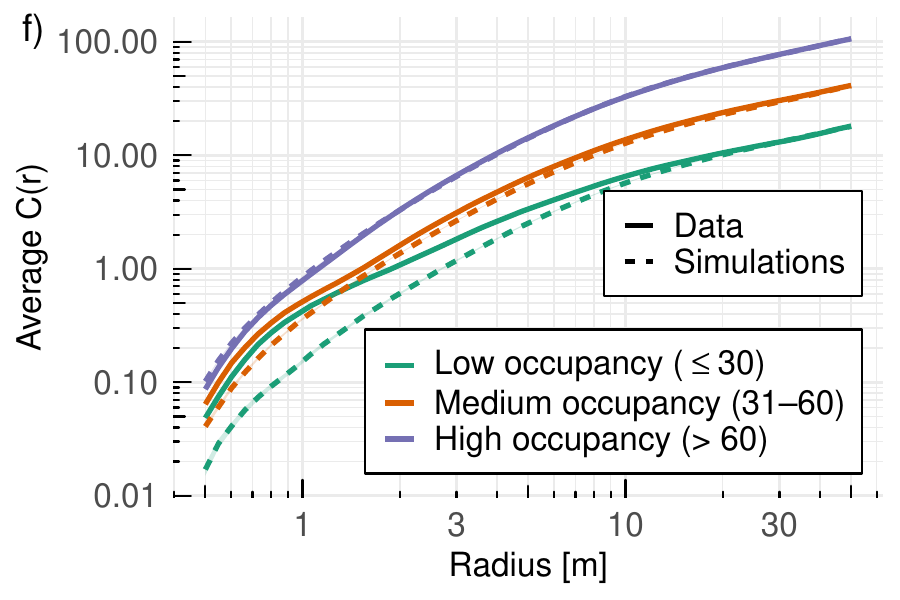}
    \end{subfigure}
    \caption{\ref{item:D1}-\ref{item:D3} for DGS simulations with $R=600, \alpha=2.7$ vs. the dataset. (a) Avg. NN distance. (b) Avg. 2-NN distance.  (c) PDF of NN distances, separated by occupancy. (d) PDF of 2-NN distances. (e) Avg. $\mathcal{C}(r)$ over all snapshots in log-log scale.(f) Avg. $\mathcal{C}(r)$ separated by occupancy. As in Fig.~\ref{fig:gpp_features_strauss}, most metrics coincide at high occupancy levels, with notably high agreement for (e).}
    \label{fig:gpp_features_dgs}
\end{minipage}

\vspace{1em}

\begin{minipage}{\textwidth}
    \centering
    \begin{subfigure}[ht]{0.49\textwidth}
        \centering
        \includegraphics[width=\textwidth]{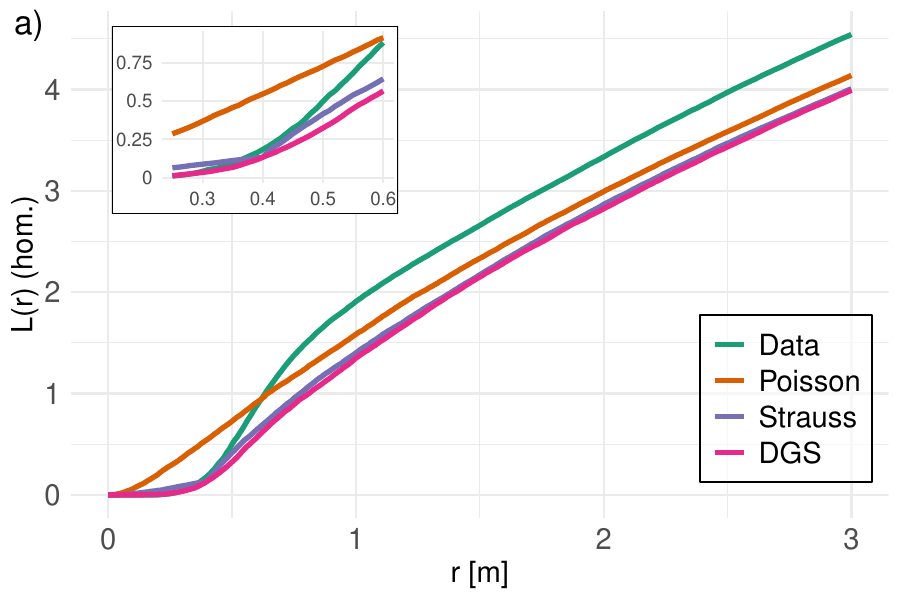}
    \end{subfigure}
    \begin{subfigure}[ht]{0.49\textwidth}
        \centering
        \includegraphics[width=\textwidth]{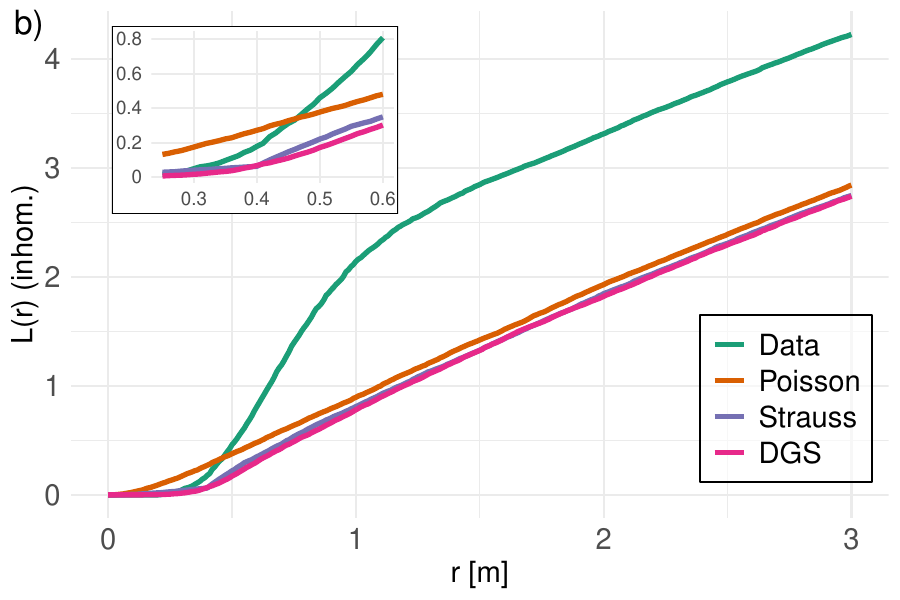}
    \end{subfigure}
    \caption{L-functions (\ref{item:D4}) for simulations from Strauss and DGS processes compared to data and Poisson. (a) homogeneous version, (b) inhomogeneous version. Both models match the data similarly well, although performance is worse under the inhomogeneity correction. In both cases, initial repulsion is comparable, but both Gibbs models show stronger repulsion at longer ranges. Insets highlight that DGS has a smoother repulsion mechanism than Strauss. Inhomogeneous curves for the data differ from those in Figure~\ref{fig:data_features} due to different intensity estimates being used for correction.}
    \label{fig:L_gpp}
\end{minipage}
\end{figure}

Running the inference for the Strauss model, we obtain parameter estimates of $\hat{R}=400\,$mm and $\hat{\vartheta}=0.15$, as well as a bandwidth estimate of $h=296\,$mm for the Gaussian kernel estimate of the intensity. The grid search for the interaction parameters is run over the entire $\vartheta$ grid for each choice of $R$, after which the pair with the maximum pseudolikelihood is chosen. The $R$ grid ranges from $100\,$mm to $1000\,$mm in steps of $50\,$mm, and the $\vartheta$ grid ranges from $0.05$ to $1$ in $0.05$ steps. Next, we simulate configurations from the resulting model, and compare these to the data. Results can be seen in Figures~\ref{fig:gpp_marginal}(a),~\ref{fig:gpp_features_strauss}, and~\ref{fig:L_gpp}. For the DGS model, estimation yields an interaction radius of $\hat{R}=600\,$mm and an interaction strength of $\hat{\alpha}=2.7$, as well as a bandwidth estimate of $297.5\,$mm. We use the same profile pseudolikelihood grid search as for the Strauss process, with the same ranges of possible $R$ values and an $\alpha$ grid ranging from $0$ to $10$ in $0.05$ steps. Results of the simulation versus data comparison can be seen in Figures~\ref{fig:gpp_marginal}(b),~\ref{fig:gpp_features_dgs}, and~\ref{fig:L_gpp}.

\begin{figure}[t]
\centering
\begin{minipage}{\textwidth}
    \centering
    \begin{subfigure}[ht]{0.32\textwidth}
        \centering
        \includegraphics[width=\textwidth]{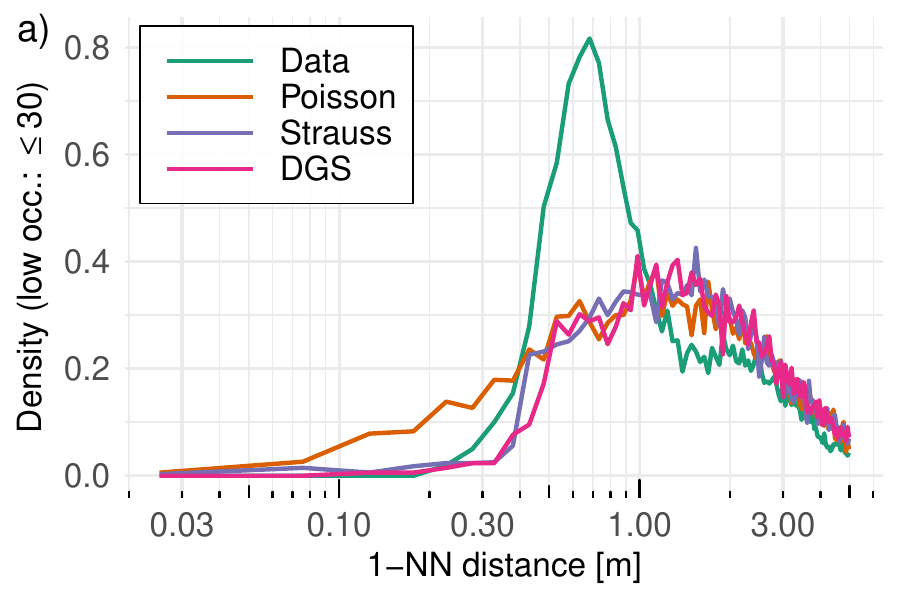}
    \end{subfigure}
    \begin{subfigure}[ht]{0.32\textwidth}
        \centering
        \includegraphics[width=\textwidth]{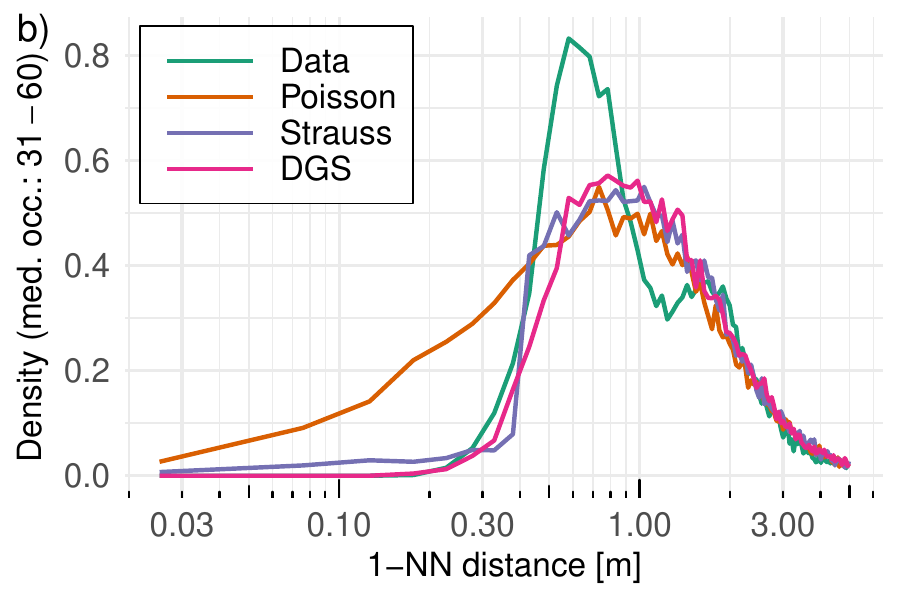}
    \end{subfigure}
    \begin{subfigure}[ht]{0.32\textwidth}
        \centering
        \includegraphics[width=\textwidth]{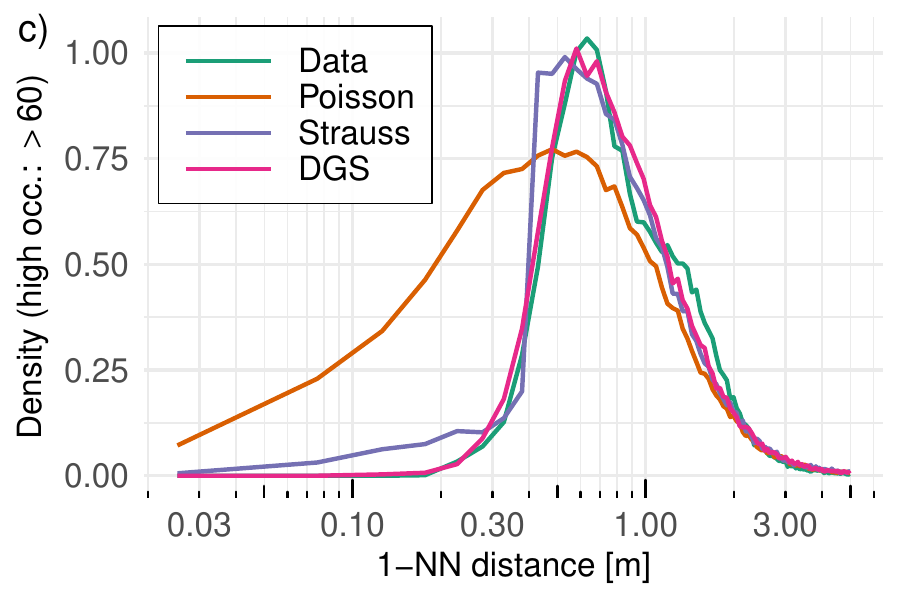}
    \end{subfigure}

    \begin{subfigure}[ht]{0.32\textwidth}
        \centering
        \includegraphics[width=\textwidth]{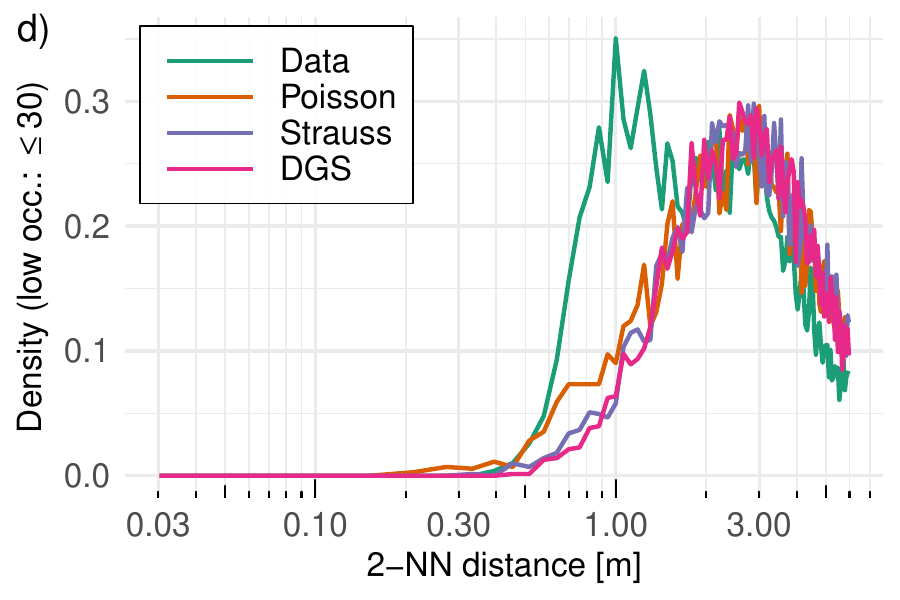}
    \end{subfigure}
    \begin{subfigure}[ht]{0.32\textwidth}
        \centering
        \includegraphics[width=\textwidth]{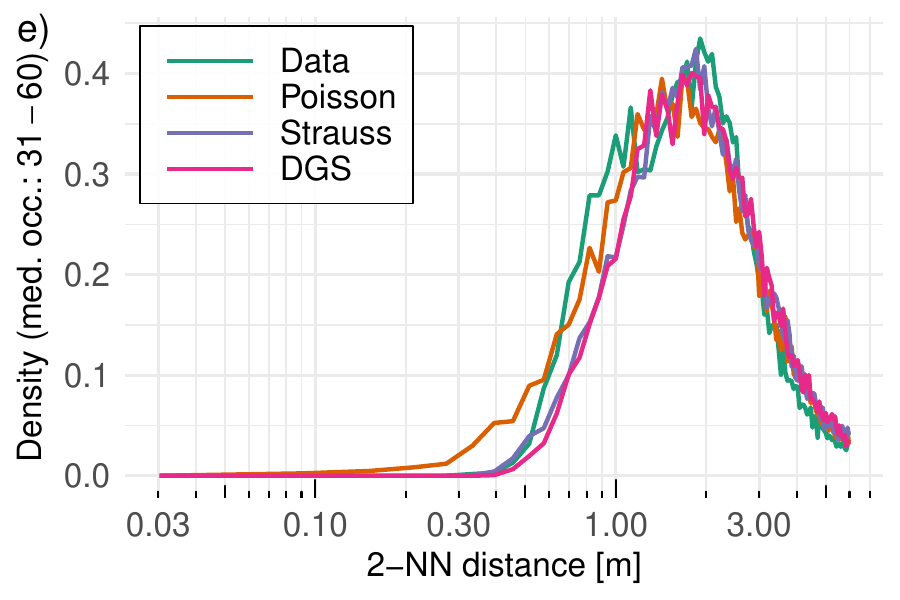}
    \end{subfigure}
    \begin{subfigure}[ht]{0.32\textwidth}
        \centering
        \includegraphics[width=\textwidth]{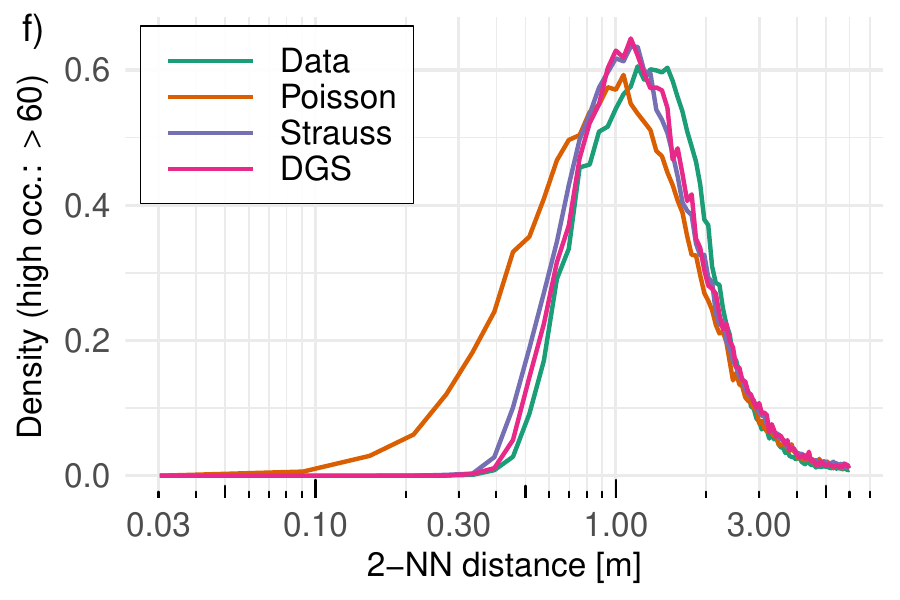}
    \end{subfigure}

    \begin{subfigure}[ht]{0.32\textwidth}
        \centering
        \includegraphics[width=\textwidth]{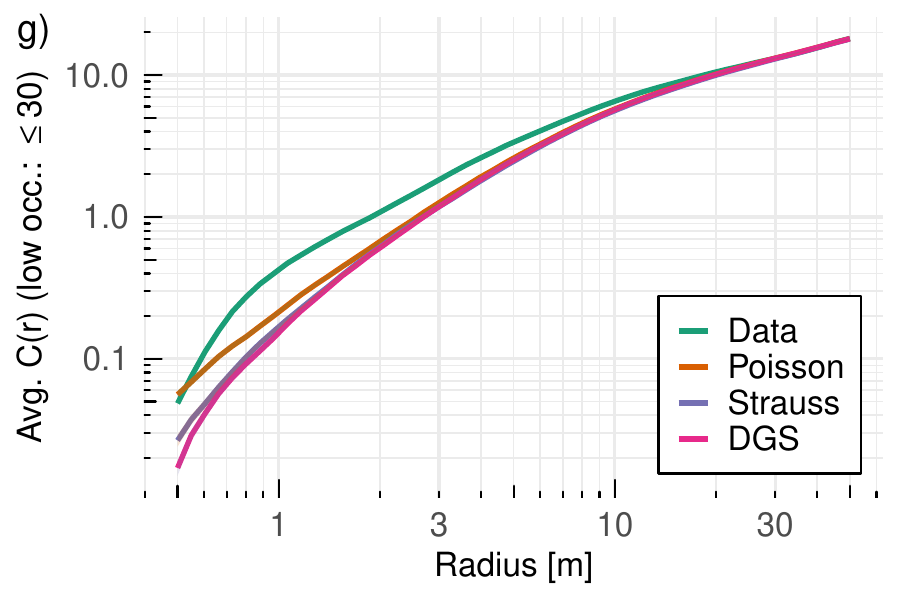}
    \end{subfigure}
    \begin{subfigure}[ht]{0.32\textwidth}
        \centering
        \includegraphics[width=\textwidth]{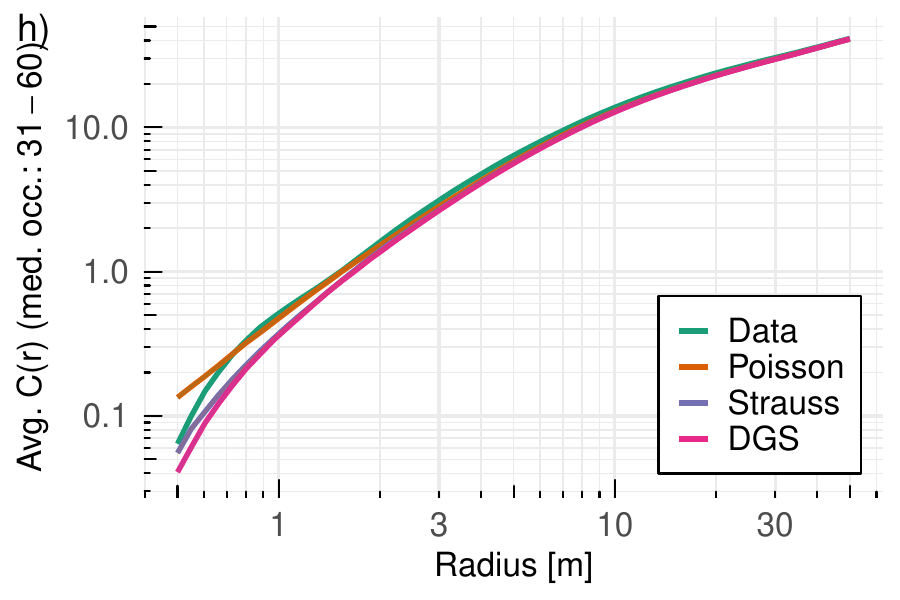}
    \end{subfigure}
    \begin{subfigure}[ht]{0.32\textwidth}
        \centering
        \includegraphics[width=\textwidth]{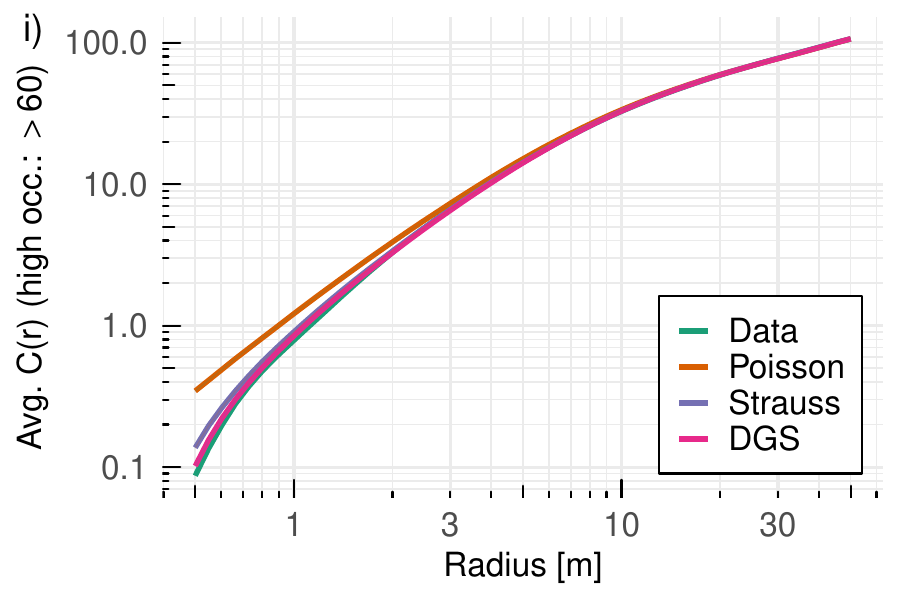}
    \end{subfigure}

    \caption{Side by side comparison of \ref{item:D2} and \ref{item:D3} for the Data, Poisson, Strauss, and DGS. Panels separated by occupancy, with NN and 2NN PDFs in logarithmic scale in $x$. (a) PDFs of NN distances at low occupancy. (b) PDFs of NN distances at medium occupancy. (c) PDFs of NN distances at high occupancy. (d) PDFs of 2-NN distances at low occupancy. (e) PDFs of 2-NN distances at medium occupancy. (f) PDFs of 2-NN distances at high occupancy. (g) Avg. $\mathcal{C}(r)$ at low occupancy. (h) Avg. $\mathcal{C}(r)$ at medium occupancy. (i) Avg. $\mathcal{C}(r)$ at high occupancy. In general, we see similar performance for both GPP processes, with the DGS model performing slightly better.}
    \label{fig:gpp_comparison}
\end{minipage}
\end{figure}

As expected when from the proposed method, the estimated one-particle marginal densities for both models agree well for the simulations and the data (Figure~\ref{fig:gpp_marginal}). The Strauss process is able to model the interaction seen in the data effectively at higher levels of platform occupancy, but it falls short for lower densities. Considering Figure~\ref{fig:gpp_features_strauss}(a-b), the NN and 2-NN distances agree quite well when occupancy exceeds around $50$ pedestrians, but both are overestimated when fewer people are present. Similarly, in Figure~\ref{fig:gpp_features_strauss}(c), peaks in the empirical distributions of NN distances agree at high occupancy, but again show an overestimation of the repulsion at medium and low occupancy. For 2-NN distance distributions (Figure~\ref{fig:gpp_features_strauss}(d)), the curve at medium occupancy also agrees well, but at low occupancy the model falls short, with the peak in the distribution being clearly shifted rightward in comparison with the data. This is also reflected in the average ball count, where overall we see very good agreement (Figure~\ref{fig:gpp_features_strauss}(e)), but when we separate the curves by occupancy level (Figure~\ref{fig:gpp_features_strauss}(f)), we again see similar issues. Considering the L-functions (see Figure~\ref{fig:L_gpp}(a) and (b)), we see that the repulsion visible at close ranges is well-modeled by the Strauss process. However, at longer ranges, the L-function indicates a more regular pattern than the data.

The DGS model shows similarly good agreement with the data as the Strauss process when it comes to the key metrics we consider. However, it also mirrors its shortcomings. As with the Strauss model, performance is quite good at medium and high levels of occupancy, but low occupancy levels are not generally well modeled. As before, it seems that the repulsion is overestimated at these regimes, visible in nearly all panels of Figure~\ref{fig:gpp_features_dgs}, except panel~(e), where the DGS process shows nearly perfect alignment with the data curve for $\mathcal{C}(r)$ averaged over all snapshots. Regarding the other plots, one example of similar limitations as the Strauss process can be seen in the peaks of the NN and 2-NN neighbor distance distributions Figure~\ref{fig:gpp_features_dgs}(c-d) are shifted to the right significantly for the simulation at low occupancy. Performance in terms of the L-function, visible in Figure~\ref{fig:L_gpp}(a) and (b), is again very similar to that of the Strauss model, with the shape of the curve for the simulations matching that of the data only slightly better at the first ``bend'' around $0.5\,$m, due to the interaction function being continuous. The DGS model also captures the initial repulsion at close ranges quite well, but the curve for the simulations then remains solidly below that of the data. Whether exactly this indicates an overestimation of the repulsion is unclear, as both the Strauss and the DGS models have no interaction past a certain range. In fact, this may indicate that in the real-world data there is an attractive effect at a certain range. This may be due to a variety of social factors, such as some pedestrians belonging to groups.

A clear picture of the differences between the two Gibbs models is given in Figure~\ref{fig:gpp_comparison}. Panels (a)-(c) compare the NN distance distributions. The logarithmic scale in the $x$ coordinate highlights the fact that the Poisson misses the repulsion at close ranges clearly visible in the data, and that both the Strauss and DGS models do a better job, particularly at higher occupancy. At this occupancy level, the DGS model has a clear advantage over the Strauss, very closely matching the distribution of the data, while the Strauss process still misses some repulsion due to its simpler interaction mechanism. In addition, the logarithmic plots of the NN distance distributions show a clear peak in around $0.5\,$m, particularly at lower occupancy levels, which can most likely be attributed to the influence of groups. Considering panels (d)-(f), it becomes clear that especially pairs of pedestrians may skew the distribution strongly at low and medium occupancy, as this peak is less pronounced (and even disappears from medium occupancy onwards) for 2-NN distances. As before, both Gibbs models are able to capture repulsion the Poisson process misses, with only very slight differences in performance between the two. When considering $\mathcal{C}(r)$ in panels (g)-(i), this picture is reinforced, with again only slightly better performance of the DGS at high occupancy.

\section{Discussion} \label{sec:discussion}

In this article, we made two closely connected contributions. First, we proposed a statistical inference procedure for semiparametric spatial point process models in the setting of replicated spatial patterns. Second, we showed that this framework provides a natural and effective description of waiting pedestrian crowds, where collective behavior reflects the combined influence of inhomogeneous intensity, driven by location attractiveness and repulsive interactions between pedestrians. In combination, these results demonstrate that waiting pedestrian configurations can be modeled quantitatively at the granular level of individual pedestrian locations, rather than only at an ensemble level. 

The methodological and applied contributions in this article are tightly intertwined. Replication of samples is uncommon in spatial point process inference, but the pedestrian setting considered here provides such data through repeated snapshots of comparable waiting scenarios at a train station. This enables the adaptation of classical inference ideas to replicated spatial patterns, as well as the quantitative assessment of semiparametric models. In parallel, the waiting crowd application highlights a central challenge of spatial point process modeling: disentangling the effect of the inhomogeneous background intensity from interpoint interaction.

Our analysis confirmed that spatial configurations of waiting pedestrians exhibit repulsive interactions, which cannot be adequately captured by Poisson models. Among the interactive models considered here, we showed that determinantal point processes offer mathematically tractable inference and sampling, together with a natural mechanism for modeling interaction. However, they appear too limited to reproduce the combination of strong short-range repulsion and weaker long-range interactions observed in the data. By contrast, Gibbs point processes provide the flexibility required to represent such behavior. In particular, both the Strauss model and a custom model based on a modified Diggle-Gates-Stibbard interaction function performed well in reproducing quantitative features of the train station data, especially at the higher occupancy levels that are most relevant in transportation and urban planning contexts.

\begin{figure}[t]
    \centering
    \begin{subfigure}[ht]{0.32\textwidth}
        \centering
        \includegraphics[width=\textwidth]{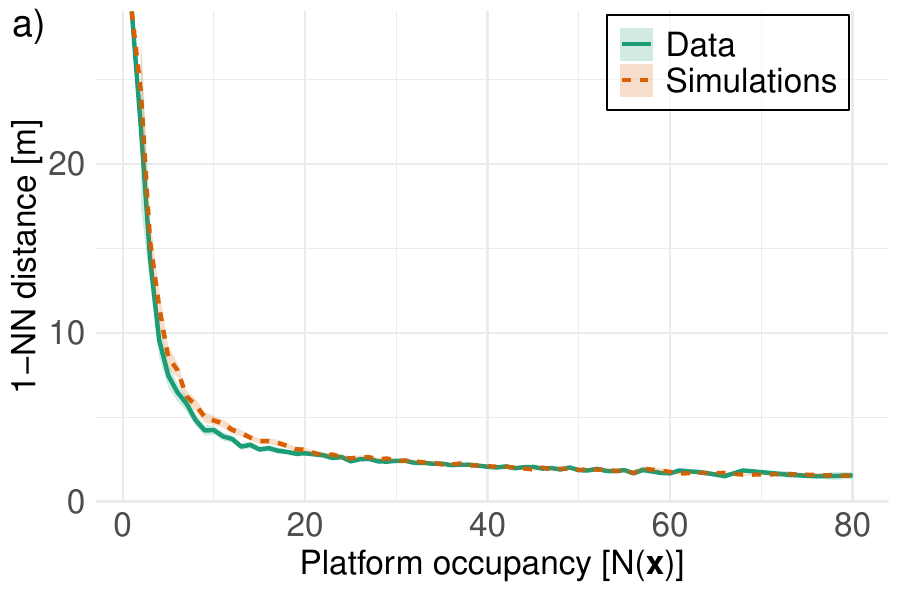}
    \end{subfigure}
    \begin{subfigure}[ht]{0.32\textwidth}
        \centering
        \includegraphics[width=\textwidth]{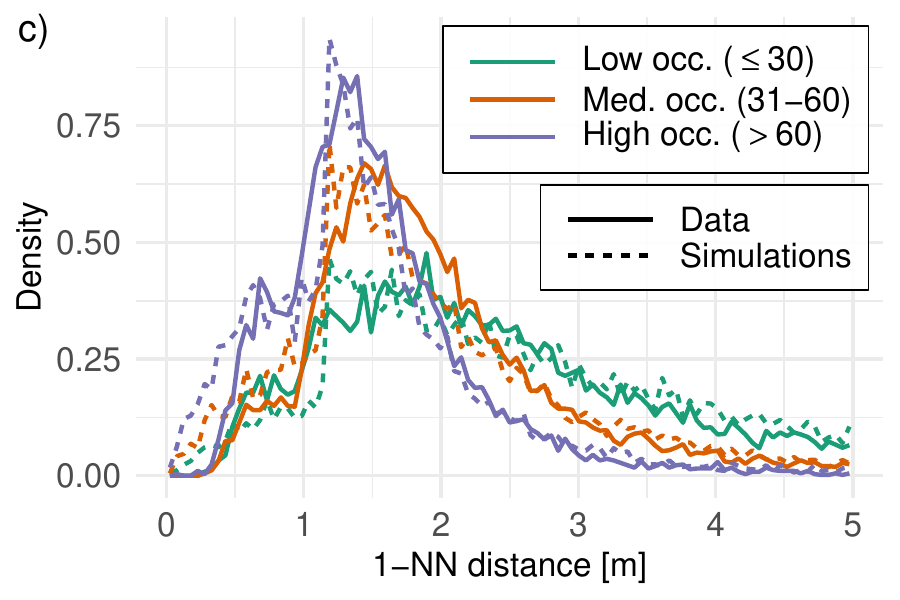}
    \end{subfigure}
    \begin{subfigure}[ht]{0.32\textwidth}
        \centering
        \includegraphics[width=\textwidth]{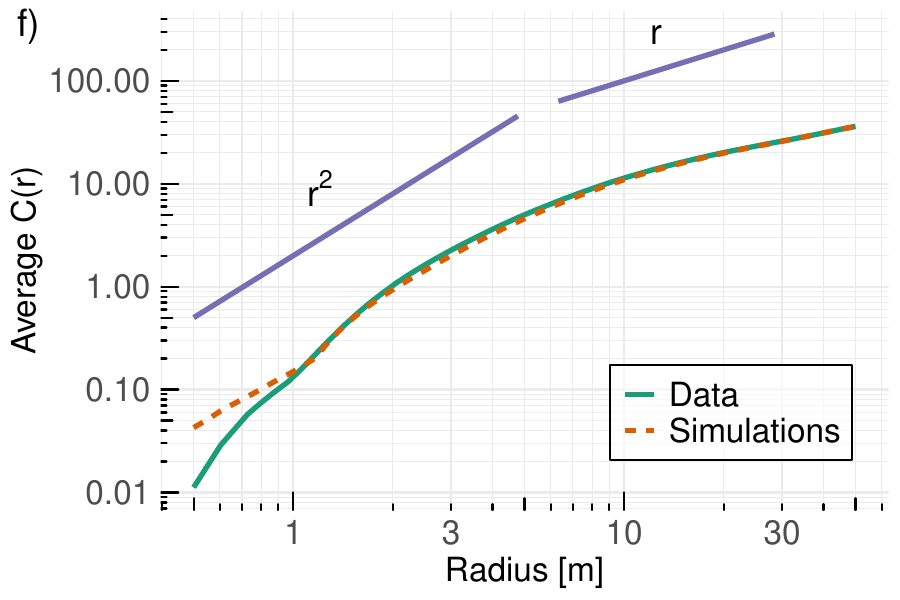}
    \end{subfigure}

    \begin{subfigure}[ht]{0.32\textwidth}
        \centering
        \includegraphics[width=\textwidth]{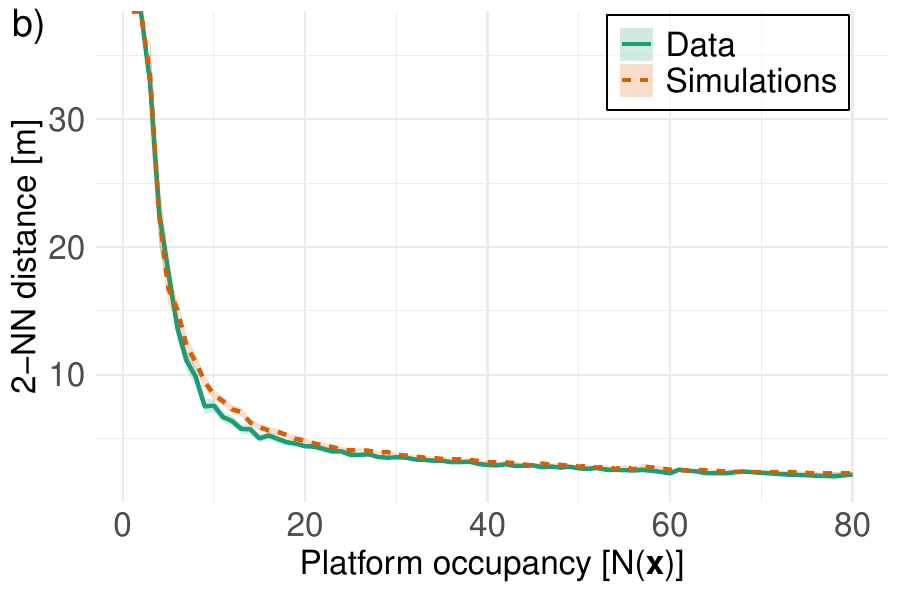}
    \end{subfigure}
    \begin{subfigure}[ht]{0.32\textwidth}
        \centering
        \includegraphics[width=\textwidth]{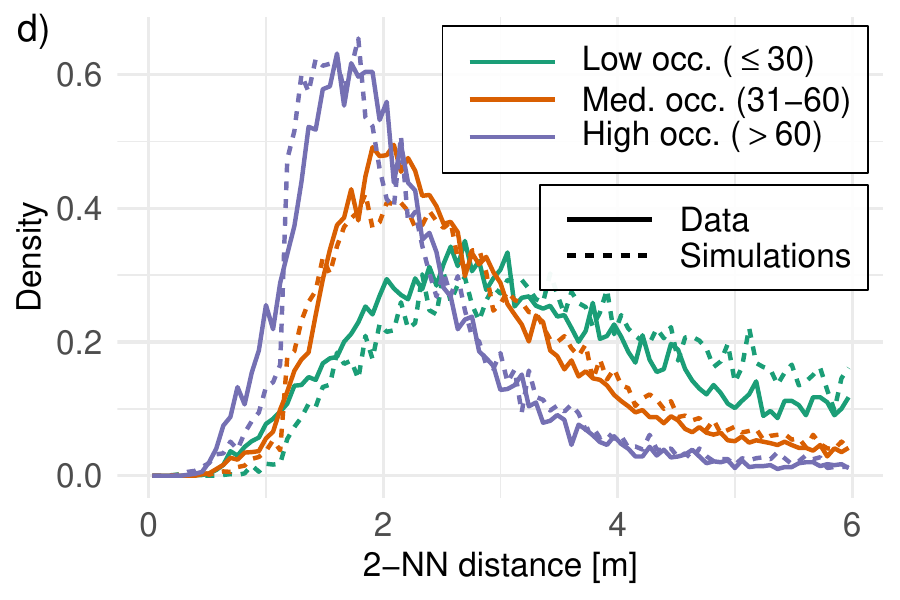}
    \end{subfigure}
    \begin{subfigure}[ht]{0.32\textwidth}
        \centering
        \includegraphics[width=\textwidth]{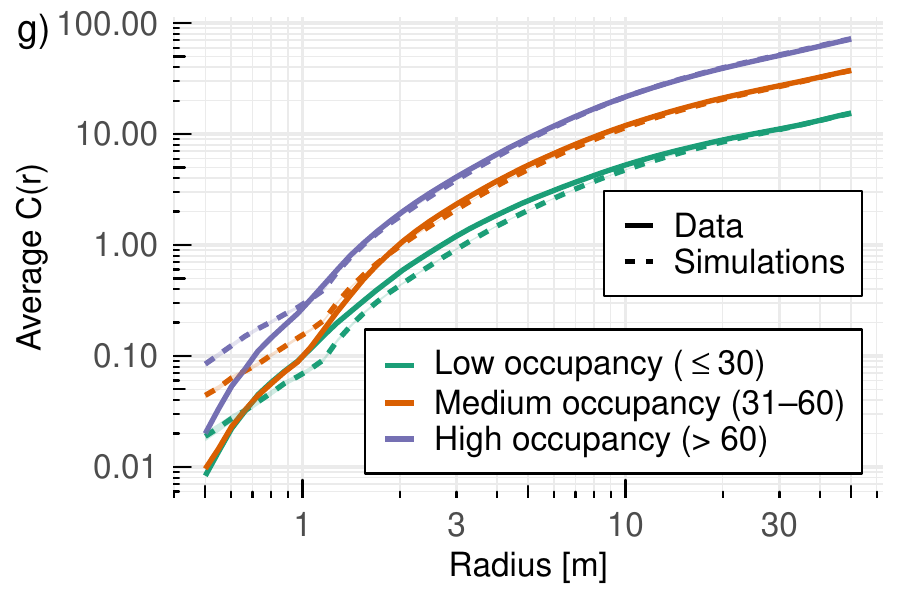}
    \end{subfigure}

    \begin{minipage}[t]{0.31\textwidth}
            \centering
            \vspace{0pt}
    \end{minipage}
    \begin{subfigure}[ht]{0.32\textwidth}
        \centering
        \includegraphics[width=\textwidth]{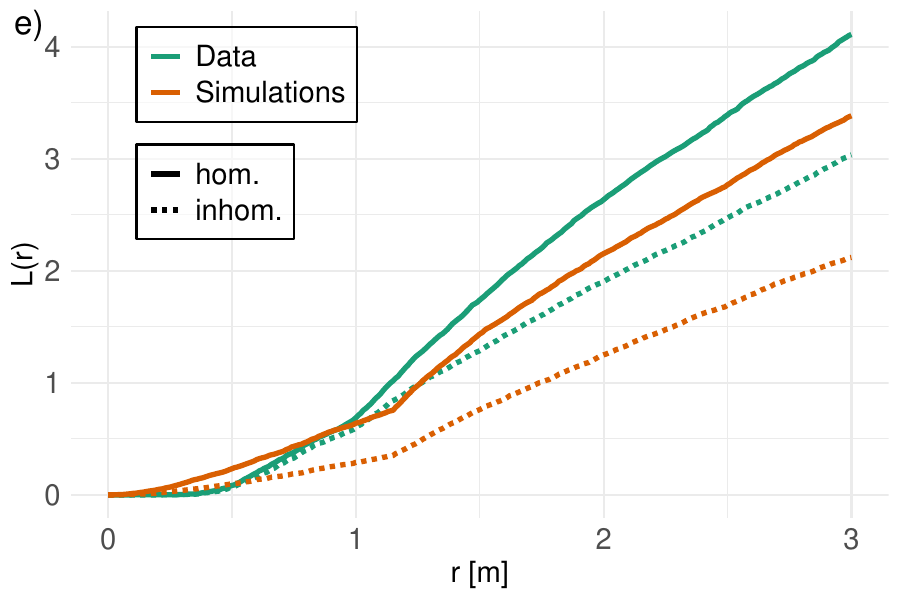}
    \end{subfigure}
    \begin{minipage}[t]{0.31\textwidth}
            \centering
            \vspace{0pt}
    \end{minipage}
    \caption{Comparison of distance-based statistics~\ref{item:D1}-\ref{item:D4} of Strauss simulations from the model taking groups of pedestrians into account versus the corresponding data (i.e. snapshots consisting of ``macro-pedestrians''). (a) Avg. NN distance. (b) Avg. 2-NN distance. (c) PDF of nearest neighbor distances, separated by occupancy. (d) PDF of 2-NN distances, same smoothing rule as for previous panel. (e) L-functions for simulations vs. data (including correction for inhomogeneity). (f) Avg. $\mathcal{C}(r)$ over all snapshots in log-log scale. (g) Avg. $\mathcal{C}(r)$ separated by occupancy. Overall, these results indicate that interaction, and particularly the radius of interaction, is slightly overestimated, while the repulsion within this large radius seems to be underestimated at close ranges.}
    \label{fig:grouped_strauss}
\end{figure}

A persistent difficulty across all considered models, however, is the role of spatial inhomogeneity during inference. Separating the contribution of the inhomogeneous intensity from that of the interaction remains fundamentally challenging. Our procedure relies on estimating the intensity via the one-particle marginal probability density $p_1$, but this necessarily introduces bias, since $p_1$ is itself affected by interactions. In particular, repulsive interactions regularize spatial patterns, and can therefore obscure evidence of the intensity. This issue is not specific to the present application, but is a more general obstacle in statistical inference for interactive spatial point processes~\cite{Baddeley_2015,Waagepetersen_2009}.

A further natural question in waiting crowd modeling is how strongly social groups affect the inferred interaction structure. Figure~\ref{fig:data_features}, and even more clearly Figure~\ref{fig:gpp_comparison} and the accompanying analysis, show that groups, and in particular pairs, have a significant influence on the distributions of nearest-neighbor distances, and thus on the entire interaction structure. This introduces an additional layer of complexity that is not easily accommodated in the present framework. Since graph-based group detection methods are already available in the pedestrian dynamics literature~\cite{Kupper_2023b,Pouw_2020}, we performed a preliminary analysis in which snapshots containing no groups larger than $5$ were retained, and each detected group was replaced by an average position, yielding a ``macro-pedestrian.'' Applying the inference methodology of Algorithm~\ref{alg:cv} to a Strauss model then not only led to different parameter estimates, with $\hat{R} = 1150\,$mm and $\hat{\vartheta} = 0.35$, but also resulted in visibly distinct phenomenology, as illustrated by the comparison with the data in Figure~\ref{fig:grouped_strauss}. While the interaction radius increased dramatically, the repulsion remained quite strong, suggesting an overestimation of one or both of these parameters. A likely explanation is that averaging group members' positions shifts nearest-neighbor distances upward, thereby inducing artificially larger interaction scales. This would constitute a reversal of the effect we see when we do not control for groups (cf. Section~\ref{sec:application}), where uncharacteristically close nearest-neighbor distances between group members could cause estimates to skew towards shorter interaction ranges. The macro-pedestrian approach is thus an adjustment that keeps the model within the current framework and addresses some biases seen when groups are disregarded, but it does so at the cost of discarding the shape and the internal configuration of groups, treating them as individuals. We investigate this approach further in~\cite{Sickert_2026b} as a first step towards accounting for social groups. Nonetheless, a more complete treatment of social grouping will require extending the current framework to e.g. marked point processes, by assigning different effective sizes or marks to individuals and groups.

Several further directions for future research emerge from these findings. On the methodological side, an important next step is to improve statistical inference in the presence of strong confounding effects between inhomogeneous intensity and interaction. One promising avenue is to incorporate additional information into the inference procedure that helps control the effects of the intensity. For example, temporal information on the order of point ``arrivals,'' which is at least partially available in many datasets, including the one studied here~\cite{Pouw_2024b}, could be used to formulate sequential models. These could reduce the bias introduced by relying on the one-particle marginal density $p_1$. Better intensity estimates could in tun lead to more reliable estimates of interaction parameters. On the modeling side, the present work focuses exclusively on repulsive interaction. In practice, pedestrian behavior may also involve attractive effects, for instance due to social clustering~\cite{Kupper_2023b,Pouw_2020,Moussaid_2010}, and mixtures of attraction and repulsion may be relevant in certain settings. More generally, multiple scales of interaction could be represented through richer multiscale interaction functions beyond the Strauss and DGS models considered here.

Overall, waiting pedestrian crowds provide a rich case study for statistical inference in spatial point processes, especially in the valuable setting of replicated spatial patterns. This work contributes to bridging the gap between inference in complex interactive point process models and the applied study of pedestrian behavior.

\ack

We thank the editor and reviewers for their valuable feedback, which greatly improved this work. Research funded by the Eindhoven Artificial Intelligence Systems Institute (EIASI) Exploratory Multidisciplinary Artificial Intelligence Research Program (EMDAIR) project \emph{AICrowd}.

During the preparation of this article, the authors used OpenAI's GPT 5.5 to support language editing, brainstorming, and debugging code. The authors reviewed, validated and critically assessed all AI-assisted outputs and take full responsibility for the content of the manuscript.

\appendix

\section{Implementation details}\label{app:implementation}

Data preparation and analysis is done primarily in \texttt{python}. Statistical inference for both DPPs and GPPs is also done in \texttt{python} using custom implementations. DPPs are simulated using the \texttt{dppy} package~\cite{Gautier_2019}. Poisson inference and simulation, as well as GPP simulation is done in \texttt{R} using the \texttt{spatstat} package~\cite{Baddeley_2015}, which is also used for evaluation of simulation results. A demo of our implementation can be found at \url{github.com/LSickertK/spatial_pps_for_waiting}.

\section{Dataset preparation details}\label{app:data_prep}

\begin{figure}[ht]
    \centering
    \includegraphics[width=\textwidth]{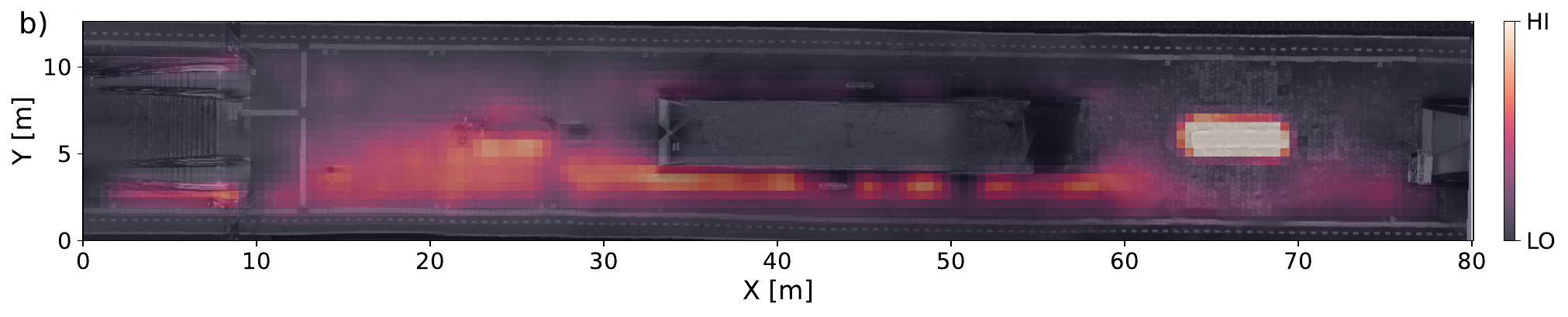}
    \caption{Estimate of the one-particle marginal density $p_1$ of pedestrians derived by aggregating snapshots where the majority of points is in the bottom half of the platform (i.e. waiting at track 4). Comparing to Figure~\ref{fig:agg_heatmap}(a), this clearly indicates a different underlying intensity.}
    \label{fig:bottom_heatmap}
\end{figure}

For details on the creation of the raw dataset of pedestrian trajectories on a platform at Eindhoven Central train station, see~\cite{Corbetta_2018,Pouw_2024b}. First, as described in Section~\ref{sec:data} the data is split into two halves by comparing the ratio of points in the top half of the platform versus the bottom half (as viewed from above). The intensity is greatly affected by this, as different locations are more attractive to people waiting for a train arriving at one track than the other. Consider Figure~\ref{fig:bottom_heatmap} and compare to Figure~\ref{fig:agg_heatmap}(a). To filter the dataset to contain only waiting pedestrians, we begin by calculating the velocities of individual trajectories using a Savitzky-Golay smoothing~\cite{Savitzky_1964}. Then, a rolling average over $4$ seconds is calculated, and pedestrians for which the average is below $0.4\,$m/s are flagged as ``waiting.'' This choice is based on finding in~\cite{Kupper_2023b}, in which this threshold is found to be the local minimum of the velocity distribution, which has two peaks corresponding to waiting and passing pedestrians. Using this flag, passing pedestrians are removed, while the fraction of waiting versus passing pedestrians is retained. Next, we check the dataset for duplicates arising from measurement errors. After calculating pairwise distances for each snapshot, pairs of points closer than $20\,$cm are classified as duplicates, as this is an unrealistic difference for real-world pedestrian measurements, and can be attributed to errors in the recording and processing of positional data. The average position of the two points is then retained. These steps are summarized in algorithm~\ref{alg:dataset} below, which also contains the steps for preparing the final dataset of snapshots described in Section~\ref{sec:data}.

\begin{algorithm}\label{alg:dataset}
    \begin{enumerate}
        \item Calculate velocities of pedestrian trajectories
        \item Calculate a $4$-second rolling average velocity for each trajectory at each time step
        \item Flag trajectories with a rolling average below $0.4\,$m/s as ``waiting''
        \item Calculate ratio of waiting versus passing pedestrians at each time step
        \item Remove duplicates (points closer than $20\,$cm) by taking average their position
        \item Detect peaks in the number of pedestrians
        \item Keep only data $1$ min before each peak
        \item Remove passing pedestrians
        \item Add flag for the majority of points lying above the midline of the platform for each snapshot (i.e. each peak) and separate the dataset into two based on this flag
    \end{enumerate}
\end{algorithm}

\section{On the stability of Algorithm~\ref{alg:cv}}\label{app:stability}

\begin{wrapfigure}{r}{0.35\textwidth}
    \centering
    \includegraphics[width=0.35\textwidth]{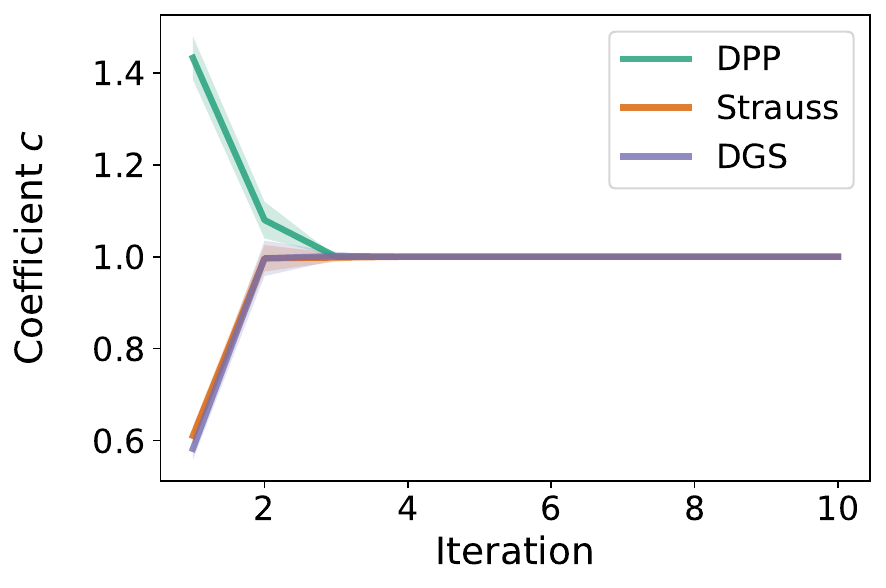}
    \caption{Coefficients of the bandwidth estimated in Algorithm~\ref{alg:cv} for $10$ iterations. Convergence occurs similarly quickly for all models.}
    \label{fig:algo_convergence}
\end{wrapfigure}

In this appendix we provide heuristic justification of the of Algorithm~\ref{alg:cv}, supported by numerical evidence. Both the bandwidth selection and parameter estimation steps optimize the same criterion, maximizing likelihood (or a suitable surrogate). More specifically, the likelihoods of both DPPs and GPPs contain the no-interaction model as a special case. The algorithm can therefore be read as block-coordinate ascent, suggesting that the attained likelihood is non-decreasing across iterations, and therefore the estimates approach a local maximum.

For each of the three models, running the algorithm for $10$ iterations on batches of $100$ synthetic snapshots (see Section~\ref{sec:inference}) with $30$ different randomized splits of $5$ folds gives the results seen in Figure~\ref{fig:algo_convergence}. On average, the algorithm converges for DPP after $2.8$ iterations, $2.4$ for Strauss, and $2.6$ for DGS.

\section{Biases in DPP estimation caused by inhomogeneous intensities}\label{app:dpp_just}

\begin{wrapfigure}{r}{0.35\textwidth}
    \centering
    \includegraphics[width=0.35\textwidth]{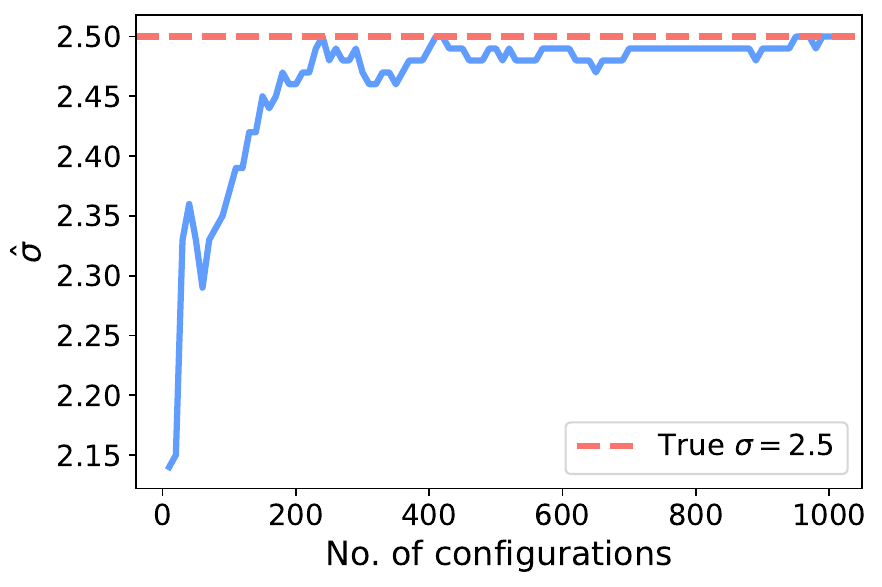}
    \caption{Consistency of $\sigma$ estimation when true intensity is known. Maximum likelihood estimation converges reasonably quickly.}
    \label{fig:dpp_consistency_known_intensity}
\end{wrapfigure}

As discussed in Section~\ref{subsec:numerics}, the estimation of the parameter $\sigma$ of a Gaussian interaction kernel for a discrete DPP is quite biased when the inhomogeneous background intensity is unknown. The approach taken in this article, namely estimating the intensity as a kernel smoothing estimate of the one-particle marginal probability density $p_1$, results in significant underestimation of the interaction parameter. The bias in this estimation is entirely caused by the mismatch between the estimated intensity and the true intensity. While the presence of spatial inhomogeneity already complicates the estimation of the parameter, when the true intensity is given, we are able to recover the true parameter, see Figure~\ref{fig:dpp_consistency_known_intensity}. However, when the intensity is estimated from the one-particle marginal density, we see that the procedure is no longer consistent, confer Figure~\ref{fig:dpp_consistency}.

\pagebreak
\section{A note on generalization}\label{app:generalization}

\begin{wrapfigure}{r}{0.35\textwidth}
    \centering
    \includegraphics[width=0.35\textwidth]{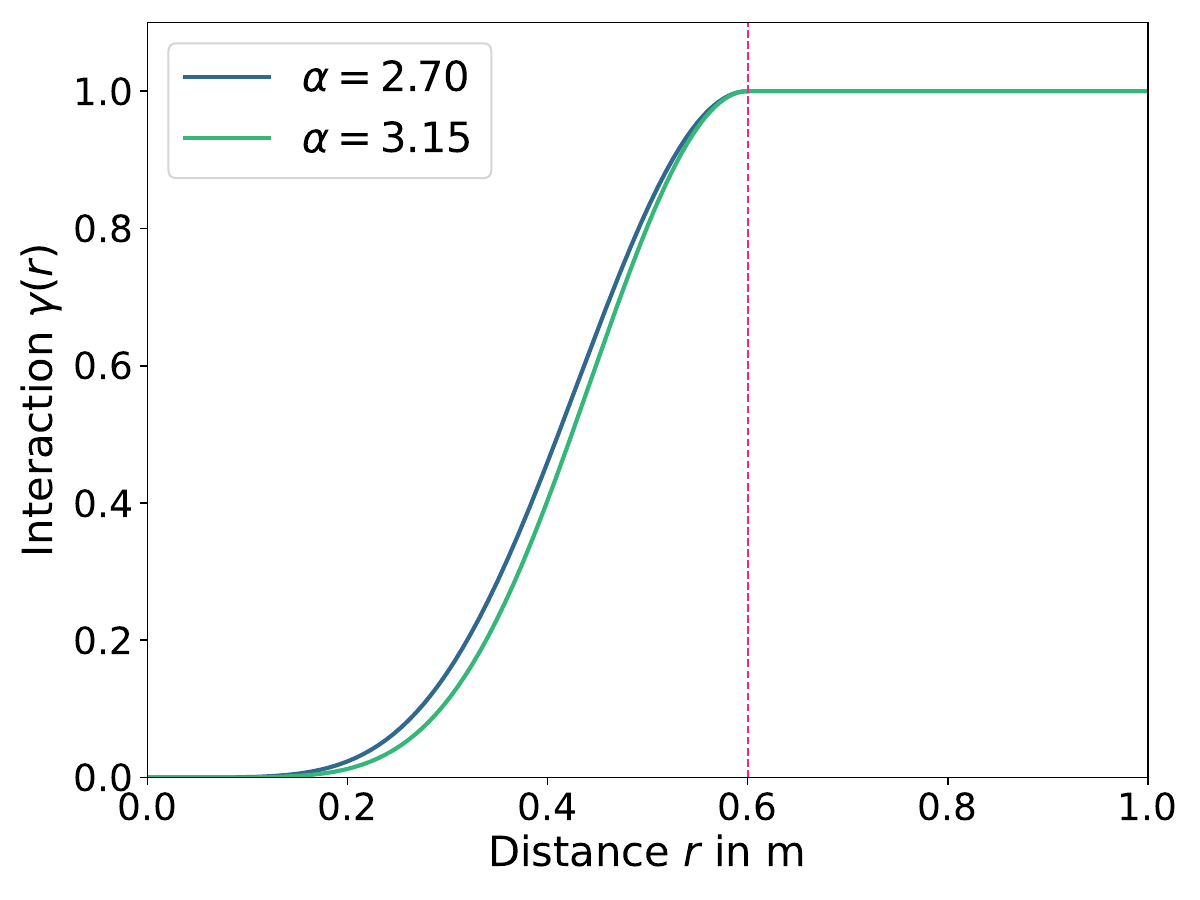}
    \caption{Comparison of the $\alpha$ estimate from Section~\ref{sec:application} and the estimate on the smaller observation area.}
    \label{fig:dgs_interaction_comparison}
\end{wrapfigure}

\begin{figure}[ht]
    \centering
    \includegraphics[width=0.75\textwidth]{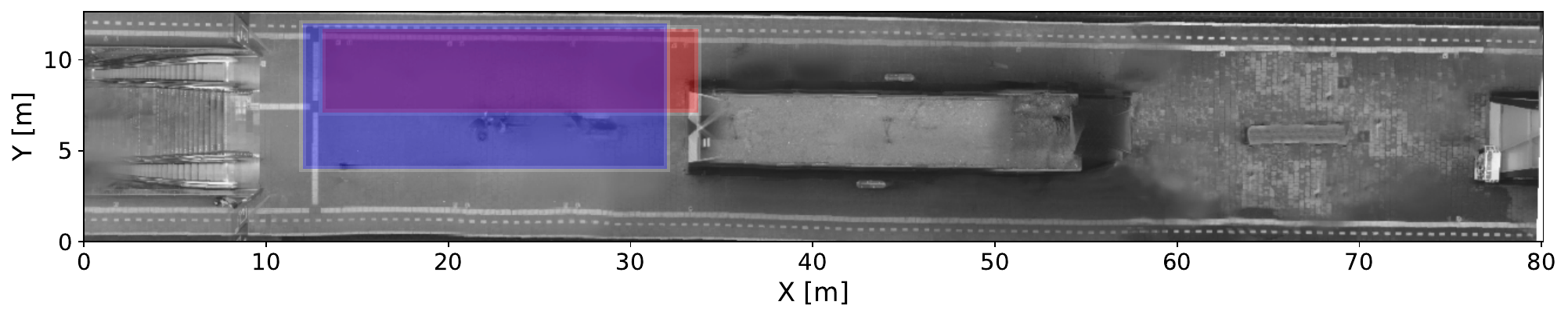}
    \caption{The original estimation area and the reduced area for comparison.}
    \label{fig:two_rectangles}
\end{figure}

To assess parameter sensitivity to the choice of observation window, the inference was repeated on a smaller subregion of the platform, see Figure~\ref{fig:two_rectangles}. For the GPPs, the reduction has little effect. For Strauss, estimates of $R$ and $\vartheta$ are identical, and for DGS $R$ remains unchanged and $\alpha$ is $17$\% larger, at $3.15$. While this may seem like a notable change, Figure~\ref{fig:dgs_interaction_comparison} shows the resulting interaction functions are very close. DPPs, in contrast, are more sensitive to the window: on the smaller rectangle, $\sigma$ is estimated as $0.901$. This suggests that the GPP models generalize better across observation windows than the DPP. This is expected, as DPPs are unable to simultaneously capture both strong repulsion at close ranges and independence at longer ranges, and thus benefit from smaller inference areas.

In addition, as described in Section~\ref{sec:application}, the data was restricted to scenarios with $\geq 10$ pedestrians. Tests on the full dataset resulted in the same parameter choices for both GPP models, and nearly no change for DPPs, at $\sigma = 0.712$.

\bibliographystyle{vancouver}
\bibliography{biblio}

\end{document}